\documentclass[]{aastex631}

\def\rr1#1{#1}

\graphicspath{{./}}

\begin{document}

\title{DECam Multi-Messenger Astrophysics Pipeline. I. from Raw Data to Single-Exposure Candidates}

\correspondingauthor{Shenming~Fu}
\rr1{\email{shenming.fu.astro@gmail.com}}

\author[0000-0001-5422-1958]{Shenming~Fu}
\affiliation{NSF National Optical-Infrared Astronomy Research Laboratory, 950 North Cherry Avenue, Tucson, AZ 85719, USA}

\author[0000-0001-6685-0479]{Thomas~Matheson}
\affiliation{NSF National Optical-Infrared Astronomy Research Laboratory, 950 North Cherry Avenue, Tucson, AZ 85719, USA}

\author[0000-0002-1125-7384]{Aaron~Meisner}
\affiliation{NSF National Optical-Infrared Astronomy Research Laboratory, 950 North Cherry Avenue, Tucson, AZ 85719, USA}

\author[0000-0001-5969-4631]{Yuanyuan~Zhang}
\affiliation{NSF National Optical-Infrared Astronomy Research Laboratory, 950 North Cherry Avenue, Tucson, AZ 85719, USA}

\author[0009-0009-7129-7538]{Sebasti\'an~Vicencio}
\affiliation{NSF National Optical-Infrared Astronomy Research Laboratory, 950 North Cherry Avenue, Tucson, AZ 85719, USA}

\author[0009-0007-0323-8929]{Destry~Saul}
\affiliation{NSF National Optical-Infrared Astronomy Research Laboratory, 950 North Cherry Avenue, Tucson, AZ 85719, USA}

\begin{abstract}

We introduce a pipeline that performs rapid image subtraction  and source selection to detect transients, with a focus on identifying gravitational wave optical counterparts using the Dark Energy Camera (DECam). 
In this work, we present the pipeline steps from processing raw data to identification of astrophysical transients on individual exposures. 
We process DECam data and build difference images using the Vera C. Rubin Observatory Legacy Survey of Space and Time (LSST) Science Pipelines software, and we use flags and principal component analysis to select transients on a per-exposure basis, without associating the results from different exposures. 
Those candidates will be sent to brokers for further classification and alert distribution.
We validate our pipeline using archival exposures that cover various types of objects, and the tested targets include a kilonova (GW170817), supernovae, stellar flares, variable stars (in a resolved galaxy or the Milky Way Bulge), and serendipitous objects. 
Overall, the data processing produces clean light curves that are comparable with published results, demonstrating the photometric quality of our pipeline. Real transients can be well selected by our pipeline when sufficiently bright (S/N $\gtrsim15$). 
This pipeline is intended to serve as a tool for the broader research community. 
Although this pipeline is designed for DECam, our method can be easily applied to other instruments and future LSST observations.

\end{abstract}


\section{Introduction} \label{sec:intro}

In 2015, the first detection of gravitational waves (GW) enabled a new probe of the Universe and verified general relativity from a new perspective~\citep{Abbott2016}.
Soon afterwards, in 2017, the successful observation of the electromagnetic (EM) counterpart of a GW source (a kilonova/KN corresponding to GW170817) opened up a new field: Multi-Messenger Astrophysics~\citep[MMA;][]{Abbott2017}.
In MMA, an astronomical source is detected by not only EM probes but also other physical manifestations, for example GW, high-energy neutrinos, or cosmic rays. Detection of these other messengers requires facilities quite different from traditional EM observatories, often with poor spatial resolution for the detections. 
This can make MMA more difficult than the regular multi-wavelength observations. 
Successful MMA observations provide information in more dimensions than traditional techniques, and thus enable a more thorough analysis and detailed modeling of the astronomical source. 
As the physical properties of the astronomical source may vary with time, its MMA observations can also be made at different times, which provide an extra dimension of measurement and enable the study of its dynamic processes in the time domain. 
MMA studies significantly open up new discovery space, e.g., the observations of GW and KN reveal the formation of heavy elements and the physics in extreme environments~\citep{Chornock2017,Kasen2017}. 
Additionally, combining the GW information and the redshift of the EM counterpart provides an independent measurement of $H_0$~\citep{Abbott2017b}.  

A primary goal of MMA is to find the optical counterpart, ideally in multiple wavelength bands, of e.g., a GW source. 
As the number of direct GW detections grows and the sensitivity of GW observatories improves, it becomes urgent to develop a new system that rapidly searches for and identifies the EM counterparts. This system should include instruments that perform fast and wide-field follow-up observations, and standardized data processing pipelines that carry out high-speed and consistent analyses. 
With the completion of the Dark Energy Survey~\citep[DES;][]{DES2016} observations in 2019 and the beginning of the Vera C. Rubin
Observatory Legacy Survey of Space and Time~\citep[LSST;][]{Ivezic2019} in the next few years\footnote{\rr1{
Typical $5\sigma$ depth of DES and LSST for point sources in 30 sec $r$-band exposures: $\sim23.5$ and $\sim24.5$, respectively. Each LSST exposure/visit consists of a pair of 15 sec snaps back to back~\citep{LSST2009,Ivezic2019}, and the DES depth is converted from its 90-sec exposure: $r=23.34$ for $S/N=10$~\citep{Morganson2018,Abbott2018}; the $r$ band has a balance between the seeing and the sky brightness. 
The depth grows with exposure time, and the MMA source detection ideally happens at single exposures. Note the goal of this work is not to make a comparison with LSST or other surveys but to present our pipeline. 
}}, one of the roles of the Dark Energy Camera~\citep[DECam;][]{Flaugher2015} will be searching for the optical counterparts of GW sources.
DECam is mounted on the 4-meter Blanco Telescope at the Cerro Tololo Inter-American Observatory (CTIO) in Chile (typical seeing $\sim1''$). 
As the instrument used by DES, DECam has 62 science CCDs (each has 2k$\times$4k pixels and a plate scale of $\sim0.263''$ per pixel) and a Field-of-View (FoV) of 2.2$^\circ$ (3 deg$^2$). Its wide aperture, large FoV, and sensitive CCDs facilitate efficient and deep observations in ultraviolet (UV), optical, and near-infrared (NIR) bands. DECam thus has high efficiency in detecting transients. 

DECam naturally has the potential for MMA studies, and it has proved its capacity in \rr1{the observations of GW EM counterparts by helping} to locate the exact source of GW170817\rr1{~\citep{SoaresSantos2017,Herner2020}}.
However, to be compatible with future wide-area surveys such as LSST, a \rr1{pipeline that can consistently process data from different instruments} is required. 
The open source LSST Science Pipelines software\footnote{\url{https://pipelines.lsst.io}}~\citep[][]{Juric2017,Bosch2019} includes state-of-the-art image processing and measurement algorithms developed for the future LSST observations.
The software is compatible with other instruments as well, such as DECam, Hyper-Suprime Cam (HSC), Canada–France–Hawaii Telescope (CFHT)/MegaCam -- the user only needs to set up the configuration at the beginning of the data processing; after detrending the raw images, the subsequent pipeline steps are almost identical. This compatibility improves the synergies between telescopes, and allows researchers to easily intercompare the results from different instruments. 
Though still undergoing rapid upgrade iterations, the LSST pipeline (and its custom versions) has been successfully applied to HSC and DECam observations for time-domain studies~\citep[e.g.,][]{Yasuda2019,Rawls2019}. 
We summarize how we apply the LSST pipeline to process DECam time-domain data in Section~\ref{sec:pipeline}. 

DECam can fill the gaps of the LSST \rr1{observations (e.g., cadence $\sim3$ days)} to catch transients that have fast brightness changes such as KN ($\sim1-2$ weeks, one of the optical manifestations of GW sources), stellar flares ($\sim1$ hour), and faint short-period variables ($\lesssim1$ day), in multiple bands\rr1{~\citep{Bianco2022,Alves2023}}. 
Although the LSST observing strategy has not been finalized yet\footnote{\rr1{LSST may spend $\leq 3\%$ of time for Target of Opportunity observations (ToO; \url{https://survey-strategy.lsst.io/baseline/too.html} and references therein). Under a similar time scale, LSST may discover counterparts of $\sim10$ binary neutron star mergers per year~\citep{Andreoni2022}. DECam can follow GW detections outside that $3\%$ LSST time. }
}, as the number of GW detections increases, LSST could be limited by its capacity to make follow-up observations, and DECam can always serve as a valuable complement to the LSST observations for MMA studies. 
Searching for KN requires rapid observation and data analysis covering large sky regions, which means the number of exposures (per unit sky area) would be small, and the candidate selection would preferably happen at \textit{individual} exposures, rather than waiting for a sequence of exposures and combining sources from different exposures/external catalogs (which is a more commonly used method).  
There have been efforts to determine the optimal telescope pointing strategies for observing the EM counterparts after GW detections\rr1{~\citep[e.g.,][]{SoaresSantos2017,Coughlin2018,Bom2024}}, and efforts to thoroughly classify transient candidates and to distribute alerts via brokers after multiple exposures~\citep[e.g., ANTARES;][]{Matheson2021}. 
In this work, we focus on the post-exposure and pre-broker steps of the MMA pipeline, including the image processing (detrending and image subtraction; Section~\ref{sec:pipeline},~\ref{sec:examples}) and preliminary source selection at individual exposures (Section~\ref{sec:rb}). 
This work represents an augmentation of the current DECam capabilities for GW optical counterpart searches and is intended to serve as an MMA follow-up tool for the entire astronomy community. 

To obtain candidates in individual exposures, we first need to build a source catalog for each exposure. 
The source catalog usually comes from two approaches: 
(1) a catalog detected from a difference image, which is the difference between a science image and a template image, or 
(2) a comparison between the catalog of a new science exposure and an existing reference catalog. The first method requires clean difference images and high-quality point-spread-function (PSF) modeling, while the second  method is more often used in high density
regions as the blending between objects can affect the image subtraction and PSF measurement. 
Here we choose the first approach because the sky area to be studied usually has \rr1{low source density (compared to high density regions such as the bulge of a nearby galaxy or the Milky Way)}. Therefore, the blending effect is small, and the difference images and catalogs are relatively clean, as the background subtraction and PSF modeling are less affected than in high stellar density regions. 
We revisit this topic when analyzing Galactic Bulge data (Section \ref{sec:bulge}). 

Typical image subtraction algorithms include Alard-Lupton~\citep[AL;][]{Alard1998} and Zackay-Ofek-Gal-Yam~\citep[ZOGY;][]{Zackay2016}. 
To obtain a difference image in the AL algorithm, a template (reference) image is convolved with a kernel model (approximated by basis functions) to match a science image, whereas in the ZOGY algorithm, the science image and the template are convolved with the template's PSF and the science image's PSF, respectively, with a decorrelation of noise.
Both algorithms are available in the LSST Science Pipelines software.  
\rr1{
In this work we use AL because it is the default method in the software. 
We select exposures for building the template based on their seeing and depth, because AL has better performance when the template PSF is sharper than that of the science image. In the future we will also test ZOGY, and we will explore more recent algorithms~\citep[e.g.,][]{Hu2022}.  
}

After performing source detection on each difference image, we classify the sources and select out candidates that likely correspond to real objects -- the so called ``Real/Bogus classification'' \citep[R/B;][]{Bloom2012}, as a first pass of source selection to reduce the burden placed on brokers downstream. 
Various types of artifacts and noise can contaminate the sources detected in difference images. In recent years, researchers have been using machine learning (ML) algorithms to remove bogus detections, because those ML algorithms are able to perform R/B based on multiple features extracted from observations, without exactly knowing the causes of artifacts and the correlations between those artifacts. As a result, those algorithms can efficiently reduce the data dimensionality and the human effort required for carrying out R/B. 
But note, most of those algorithms require images and/or catalogs from \emph{multiple} exposures, and ideally a labeled dataset for training/learning before the actual classification.

The ML algorithms can run at the image level or the catalog level. They can be divided into supervised and unsupervised methods depending on whether \rr1{the labeled} dataset is required and provided. 
At the catalog level, a common supervised algorithm is Random Forest (RF), which takes features in the catalog as an input, and uses an ensemble of decision trees to perform R/B; 
RF uses randomly sampled training data and features to improve stability. 
Example applications of RF include datasets from DES~\citep{Goldstein2015}, Zwicky Transient Facility~\citep[ZTF;][]{Bellm2019,Mahabal2019}, and Sloan Digital Sky Survey~\citep[SDSS;][]{duBuisson2015}, where RF performed best among various algorithms. 
At the image level, a typical supervised algorithm is Convolutional Neural Network (CNN), which is sensitive to graphical structures and can be trained to detect specific patterns in an image rather than its corresponding catalog. CNN has been applied to the exposures of, e.g., ZTF~\citep{Duev2019}, Gravitational-Wave Optical Transient Observer~\citep[GOTO;][]{Killestein2021}, DES\rr1{~\citep{Ayyar2022,Shandonay2022}}, and DECam Deep Drilling Fields~\citep[DDF;][]{Graham2023}. 

Most ML methods applied in recent R/B studies used supervised learning, which needs a training set -- a group of real objects and/or a group of defects that are both known, and also requires the human effort to build this training set. Furthermore, in real observations the objects identified by a supervised learning method could be limited to the ones similar to the training set. In contrast, an unsupervised method is able to select out targets that are not covered by the training set, which expands its detection capability to special/unusual targets. Also, an unsupervised method can avoid the reliance on a single and fixed training set, especially when the feature distributions per observation/measurement vary (e.g., if we consider sources detected in independent/non-overlapped exposures, or in exposures that are strongly affected by weather conditions).  For MMA, it is common to have limited comparable observations available for training.

Our DECam MMA pipeline requires a fast and simple R/B algorithm for single exposures, since a KN changes its brightness quickly and the search area (for DECam follow-up observations) derived by the GW source probability sky map is usually large\rr1{~\citep[tens to hundreds of DECam FoV depending on the source distance;][]{Abbott2018b,Petrov2022}}.
We find that the LSST Science Pipelines software is able to efficiently produce clean difference images and light curves for DECam exposures (Section \ref{sec:examples}).
The software also provides a wide range of flags that can help clean the catalog -- \rr1{those flags are made by the software during the image processing and the measurements afterwards, and} recent studies~\citep[e.g.,][]{Liu2024}  
showed that the flags could be used to remove most of the artifacts. 
Therefore, \rr1{for each exposure} we first filter \rr1{the difference image} catalog by those flags \rr1{(Section~\ref{sec:flags}, Table~\ref{tab:flags})}, and then we
carry out R/B using an unsupervised method. 

Compared to spurious sources, a real source on one difference image is expected to have a higher signal-to-noise ratio (S/N), a rounder shape, and a reasonable size, etc.\footnote{Cosmic rays may also have high S/N, but they are generally removed by the LSST Science Pipelines and have special shapes. The detection of cosmic rays is not the main topic of this work, though they may come from some object that generates MMA signals. } Thus, we first \rr1{select a group of features} based on those characteristics, \rr1{
and the real sources are expected to have greater feature values than the} artifacts/noise \rr1{(Section~\ref{sec:features}, Table~\ref{tab:features})}.  
Then we use Principal Component Analysis (PCA), an unsupervised algorithm  that has a straightforward mathematical derivation, to divide the sources based on those features. 
\rr1{
Finally we select sources that have significantly larger values than the others in the transformed space as per-exposure candidates, and we assign weights to the candidates based on the PCA results and the weather condition (Figure~\ref{fig:flow_chart}, Section~\ref{sec:weight_count}). Here, for the transformed space we consider the first principal component. For the PCA results, we consider the explained variance ratio of the first component, and the rank of the first component value of each source; for the weather condition, we consider the effective exposure time~\citep{Morganson2018}. 
}
Those candidates will later be sent to brokers for further classification and alert distribution. 
We have tested this method on archival exposures, and we found that real objects generally ``stood out'', regardless of their stellar types and local environments. 
We note that PCA has been used to find the main components of the source images for feature extraction~\citep{duBuisson2015} or to reduce the dimensionality of alert data~\citep{Graham2023},  
but it has not been used on source catalogs for R/B, which is studied in this work. 
We give more details of our R/B algorithm and its performance in Section \ref{sec:rb}.

In this paper, we test various types of objects located in different environments using \textit{archival} data, to assess the flexibility of the LSST Science Pipelines software and our algorithm. 
Those targets span a wide range of time scale (from hours to months), and the software is able to produce clean results even if the transient has fast brightness changes (Section \ref{sec:examples}). 
Although we have built individual templates for those targets, in the future we will explore methods to build a uniform template using the archival data of large-footprint DECam surveys, e.g., the Dark Energy Camera Legacy Survey~\citep[DECaLS;][]{Dey2019}\footnote{\url{https://www.legacysurvey.org/decamls/}}, which will reduce the burden of building templates for new observations.

This paper mainly consists of two parts: (1) our tests of the time-domain photometry from the LSST Science Pipelines for various types of targets captured by DECam; and (2) our Real/Bogus classification algorithm for selecting per-exposure candidates based on PCA. 
In Section \ref{sec:pipeline}, we show the framework of our pipeline. 
Then in Section \ref{sec:examples}, we present light curves from our photometry tests. 
In Section \ref{sec:rb}, we describe the details of single-frame R/B. 
We discuss our methods and results in Section \ref{sec:discussion}, and summarize the paper in Section \ref{sec:summary}.

\section{Pipeline} \label{sec:pipeline}

We develop our MMA pipeline based on the LSST Science Pipelines software.
The \rr1{LSST Science Pipelines} software incorporates two pipelines: the Alert/Prompt Production Pipeline (AP) and the Data Release Pipeline (DRP). 
The AP produces results of image subtraction and source association for rapid time-domain studies.  
The DRP carries out coaddition and performs measurements on coadded images primarily for static science studies, and it works on long-term time-domain analysis as well; in this paper we only use the static analysis part of DRP. 
The raw images  are  detrended in the same way in both pipelines, and the coadded image built by DRP can serve as a template for AP. 
The algorithms and functionalities of AP and DRP have been presented by~\citet{Yasuda2019} and~\citet{Bosch2018} respectively. 

\begin{figure*}[htb]
    \plotone{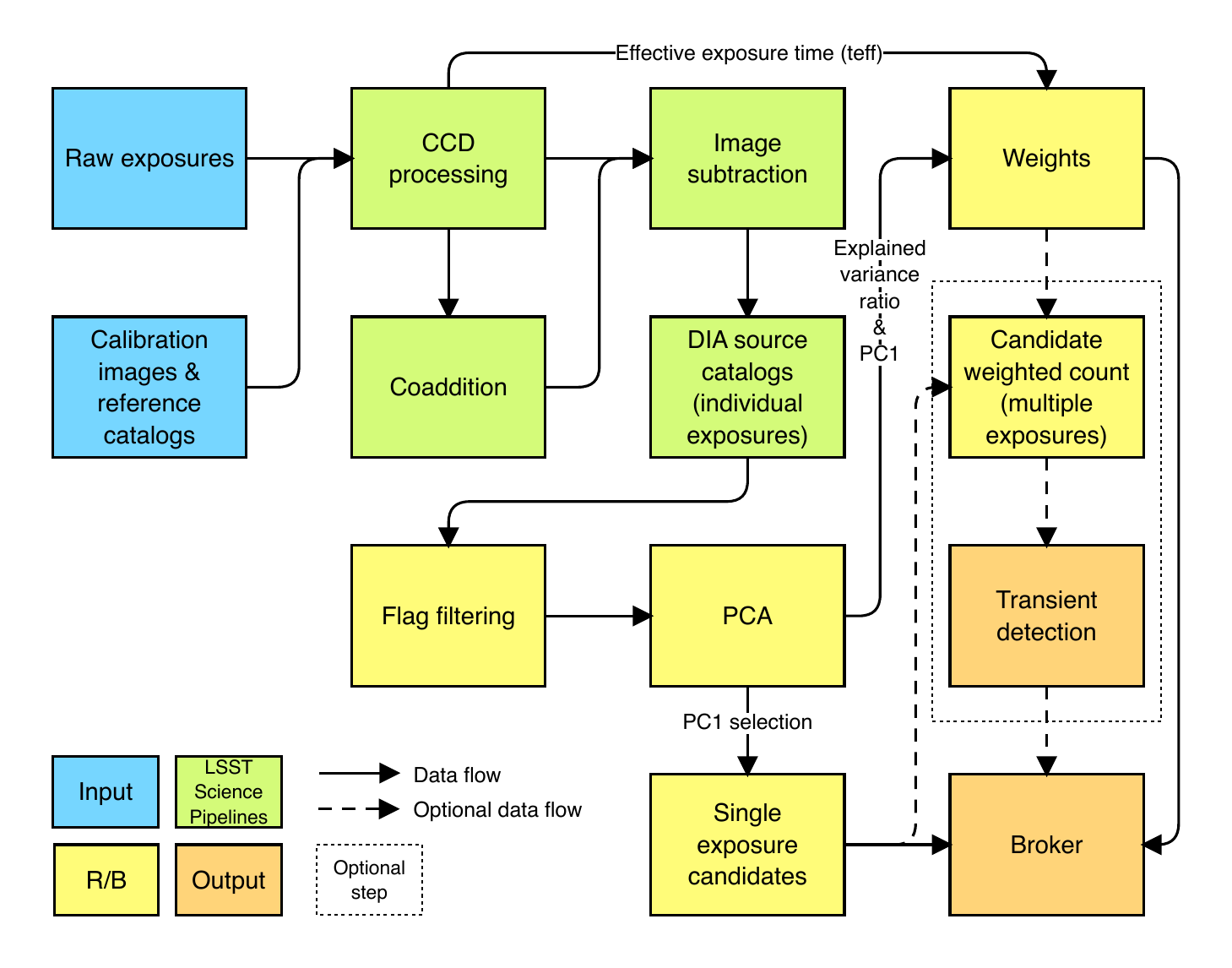}
    \caption{Flow chart of the MMA pipeline steps described in this paper. The arrows show the data flow. 
    The blue rectangles describe the input data.
    The green rectangles show the pipeline steps from the LSST Science Pipelines software. 
    The yellow rectangles give the steps for Real/Bogus classification (R/B) to generate per-exposure candidates and weights. 
    The orange rectangles show the output (the places that our data product can be fed into). 
    The dashed arrows and the steps enclosed by the dashed rectangle are optional (Section~\ref{sec:weight_count}).  
    }
    \label{fig:flow_chart}
\end{figure*}

In our MMA pipeline, we use the above functionalities to process DECam images and run image subtraction, and then use our own scripts to obtain transient candidates. Those steps require raw images, valid templates, and efficient source selection. 
In this section, we give the details of our pipeline -- the essential steps are as follows, and we use  Figure~\ref{fig:flow_chart} to summarize the pipeline details. 

(1) Pre-processing and detrending: selecting calibration images and reference catalogs, and processing raw images. (Section~\ref{sec:prep_detr})

(2) Building templates for image subtraction. (Section~\ref{sec:template}) 
		
(3) Image subtraction and source generation. (Section~\ref{sec:img_sub})

(4) Real/Bogus classification. (Section~\ref{sec:rb_s})

Note, the LSST Science Pipelines software can associate the sources from individual difference images and run forced photometry, and this does not require the R/B step above. We present the light curves generated by the forced photometry in Section~\ref{sec:examples} as a test of the software performance of image subtraction and photometry. However, when we run the MMA pipeline in real time to produce MMA candidates, the source association and forced photometry are not required. 

In this work, we mainly use the LSST Science Pipelines software version 19.0.0, which was an available stable release with sufficient functionality when this project started. 
We also tested the functionality of more recent versions and did not find significant changes in the results. 
Additionally, our R/B algorithm is insensitive to the software version and can be applied to the outputs of other data reduction pipelines. 

The typical costs of processing time and disk space are as follows.  
Processing of one DECam CCD chip from raw (\texttt{processCcd}) takes $\sim1-2$~min while the disk space used is $\sim50-100$~MB.
Difference imaging of one CCD typically takes $<1$~min with disk space usage of $\sim50$~MB. 
The cost of time and memory scales with star/galaxy number density (much larger in the Galactic Bulge region or nearby galaxies; Section~\ref{sec:bulge}).
The cost of coaddition scales with the number of exposures that are included ($\sim1-2$ times the \texttt{processCcd} disk space). 
R/B takes $\sim10$~sec per CCD. 
\rr1{Those timings are similar to the scale of exposure time plus overheads, and the data can be processed in parallel (one core per CCD). }

\subsection{Preprocessing and detrending}\label{sec:prep_detr}
The pipeline starts with determining the exposures for detection/measurement, selecting the corresponding calibration frames and reference catalogs for removing the instrumental signatures, and calibrating the astrometry and photometry. 
We adopt a strategy similar to the study of~\citet{Fu2022}, which requires processing a large number of DECam exposures as well, to select the calibration frames and catalogs. 
We use a database of high-quality master bias/flat frames ($\sim$ one per month), instead of nightly calibration frames, for detrending the raws, as the frames are generally stable in the observation window of our targets\rr1{~\citep{Morganson2018}}; this reduces the requisite disk space and processing time. 
For the astrometric calibration, we take the high-precision Gaia  catalogs~\citep{Gaia2016a,Gaia2016b,Gaia2018,Gaia-DB,Gaia-DB2}, and we use the Data Release 2 (DR2) unless otherwise specified. 
For the photometric calibration, we use the deep catalogs of Pan-STARRS1 (PS1) DR1~\citep{Chambers2016,PS1-DB} and SkyMapper~\citep{Wolf2018,smdr1,Onken2019,smdr2} in the sky regions of $\delta>-30^\circ$ and $<-30^\circ$, respectively; for SkyMapper, we use DR2 unless otherwise specified. For $u$-band exposures in the northern sky, SDSS DR12~\citep{Alam2015} can be used as a reference catalog. 
We generally use the external (and independent) reference catalogs described above to consistently calibrate the astrometry and photometry; for the objects covered by DES, we also test using the DES DR1 catalog~\citep{Abbott2018} as the reference.

The science exposures come from archival observations of test targets (Section~\ref{sec:examples}) but our methods should be broadly applicable to future DECam observations. 
After downloading the raw exposures from the NOIRLab Astro Data Archive\footnote{\url{https://astroarchive.noirlab.edu}}, we process those raw images using the LSST Science Pipelines software (\texttt{processCcd}), which produces a detrended and calibrated exposure image (\texttt{calexp}) and a source catalog (\texttt{src}) for each CCD. 
Then the pipeline \rr1{resamples} the CCD images onto a grid of sky patches (4k$\times$4k pix each) and stacks them. Transient artifacts are generally filtered out during the coaddition~\citep{Aihara2019}.
The processing of individual CCDs and the coaddition of individual patches can both run in parallel to save time. 
More details of CCD processing and stacking in the LSST Science Pipelines software have been described by~\citet{Bosch2018}. 
We skip the further coadd measurements in the LSST Science Pipelines software (DRP) in this work.

\subsection{Template generation}\label{sec:template}

There are different ways of producing template/reference images for image subtraction.  The best approach is subject to  debate -- it usually depends on the weather conditions, the image quality, the image subtraction algorithm, etc. 
In real observations, the template can come from the following places.

(a) Pre-made \rr1{coadded} images based on archival data; 

(b) Dynamically updated \rr1{coadded} images based on new high-quality exposures (e.g., with sufficient exposure time and satisfactory weather conditions); 

(c) Synthetic images. An example is the model images of stars and galaxies derived from observations in the DESI Legacy Imaging Surveys~\citep{Lang2016,Dey2019}. However, the modeling of the complex morphologies of nearby galaxies and in dense regions can be difficult for MMA purposes. Also, the model image needs to be scaled carefully (with the FITS header being properly set), so that it is usable for image subtraction in the data processing software. 

In this paper, we mainly focus on method  (a) and also discuss method (b). We will explore the method (c) in future work. 
As mentioned in Section~\ref{sec:prep_detr}, we need to use the first few steps of DRP to make a \rr1{coadded} image as a template, starting from raw images. The template can be made from archival exposures, or recent exposures that include the transient. The latter case may be common in MMA observations.  
We have considered using archival public exposures to build a large footprint template in the future, but it is possible that some sky regions are still not covered. 
\rr1{When the numbers of both archival and new exposures are limited in MMA studies, we could use a single exposure to build a ``coadded'' image as the template, supposing that the selected exposure has good seeing and depth and there is no significant dithering between telescope pointings. In the version of the LSST Science Pipelines software we are using, it even allows a processed CCD image (\texttt{calexp}) to be directly used as a template, without the later coaddition step (which resamples the \texttt{calexp}  onto patches). However, this functionality is not supported in newer versions of the software. 
Note, a \texttt{calexp} image and a coadded image are separate items in the software. 
}
We present examples \rr1{of template images} in Section~\ref{sec:examples}. 

To determine high-quality exposures for building templates, we use the seeing and the effective exposure time~\citep[\texttt{teff};][]{Neilsen2016,Morganson2018} as metrics.  
The \texttt{teff} metric considers seeing, sky brightness, and transparency, combined as follows:
\begin{equation}
    t_\textrm{eff} \propto \eta^2 \theta^{-2} b^{-1}\equiv\tau
    \label{eq:teff}
\end{equation}
where $\eta$ is the atmospheric transmission, $\theta$ is the seeing/PSF size (full width at half maximum; FWHM), and $b$ is the sky brightness. 
For one exposure, a high \texttt{teff} usually means a good weather condition. 
Near the detection limit, the S/N of a point source satisfies $S/N \propto f\sqrt{t\tau}$, where $f$ is the (true) source flux and $t$ is the exposure time~\citep{Neilsen2016}. 
The low S/N caused by sky transparency and brightness can be improved by coaddition, and therefore the total \texttt{teff} will increase after stacking. 
The \texttt{teff} of an exposure is usually normalized by some fiducial values of seeing, sky brightness, and transparency. By default, we adopt the \texttt{teff} measurements from CTIO inventory \texttt{qcInv} \rr1{available to DECam observers}.\footnote{\rr1{\url{https://noirlab.edu/science/programs/ctio/instruments/Dark-Energy-Camera/User-Guide/During-Night\#GODB}}}

The AL algorithm prefers the seeing of the template to be smaller than (or close to) that of the science exposures, and thus we select exposures with small PSF sizes, but also make sure that most stars are not saturated (which can happen when the seeing is too small and the exposure time is fixed). We note that even when the PSF size of the template is slightly larger than the direct \rr1{(science)} image, the image subtraction and photometry still work well in the LSST Science Pipelines software. 
The seeing, sky brightness, and transparency  values are available from \texttt{qcInv}.  
There is also information about 
transparency and seeing recorded in the header of the raw exposure, and sky brightness recorded in the calibrated exposure. 
The PSF size and shape may also be derived from the second moments of stars (used for PSF modeling) in the source catalog, or we can fit the stars on the processed image with Moffat profiles to get FWHM; their median values on each CCD can be used as quality indicators~\citep{Fu2022}.  
Here we use \texttt{teff} as a metric of selecting exposures for building templates, but it can also be used as a weight for assessing transient candidates (Section~\ref{sec:rb}).

\subsection{Image subtraction and source generation}\label{sec:img_sub}

We conduct image subtraction using the LSST Science Pipelines software, producing the difference between a detrended/calibrated exposure and a template (usually a coadded image). 
The image differencing between a deep coadded image and the template may be able to detect very faint transients; we will study that in future work.  


The image subtraction can run as a single task, but it is also a part of the AP pipeline in the LSST Science Pipelines software. 
When we generate the light curves in Section~\ref{sec:examples}, we run through the whole AP pipeline from detrending, image differencing, source detection and association, to database generation.  
The processing of each exposure (visit) generates a difference image (\texttt{diffexp}) and a Difference Image Analysis (DIA) source catalog (\texttt{diaSrc}); the default detection limit is $5\sigma$, where $\sigma$ is the per-pixel noise.
The AP pipeline can match the DIA sources in different visits using their positions (source association) and generate a DIA object catalog.
Also, it runs forced photometry (on both the direct and difference images of all visits) with fixed positions, even when the target is below the detection limit in some visits, which is useful for making light curves.
The catalogs are stored in FITS tables and an SQLite table (Alert Production Database, APDB; or Prompt Products Database, PPDB).
We use the forced photometry to validate the image subtraction by testing the light curves, but for the MMA pipeline we only use the sources detected on the difference images of single visits.

\subsection{Real/Bogus classification}\label{sec:rb_s}
We first use the flags in the source catalog of a difference image (\texttt{diaSrc}) to reduce artifacts. 
Next, before we feed features (e.g., the source flux and shape) into the ML classifier algorithm, we standardize those features by subtracting the mean and then dividing the standard deviation to scale them -- this makes the data consistent and simplifies the following analysis.  
After that, we carry out a two-component PCA on those scaled features to divide the sources. Then, using the PCA component values of those sources in the transformed space, we select a set of good candidates from the sources. Finally, the quality (or ``score'') of this division (or the selected candidates) can be described by the PCA-explained variance ratio, component values,  or \texttt{teff}. 
We give more details of our R/B algorithm in Section \ref{sec:rb}.
The single exposure candidates can then be directly sent to brokers for further classification and alert distribution, or the candidates from different exposures can be weighted by the score and then stacked for transient detection (Section~\ref{sec:weight_count}).

\section{Examples of difference imaging photometry: light curves} \label{sec:examples}

In this section, we examine the light curves as validations of the image subtraction and photometry of our MMA pipeline. The targets include various types of phenomena -- KN, supernova (SN), stellar flare, and variable star, and they reside in environments with different neighboring stellar densities (e.g., relatively compact host galaxies, resolved galaxies, and the Galactic Bulge) mimicking real-world use cases. 

We first run the AP pipeline on the DECam archival exposures of each target, then visualize the difference images, and plot the light curves of repeatedly detected sources (DIA objects) in the difference images -- the AP pipeline automatically associates sources in individual exposures and runs forced photometry on both direct images (\texttt{calexp}) and difference images (\texttt{diffexp}). Finally, we compare the light curves with previously published results.

The examples are summarized in Table~\ref{tab:example}, and the celestial coordinates of the targets  are given in Appendix~\ref{sec:target_list}. 
The tested variables have short periods ($\lesssim1$ day). Long-period variables such as Active Galactic Nuclei (AGN) are not included, and we will study them in follow up papers.  
We note that long-period variables, such as Asymptotic Giant Branch (AGB) stars, may show up on the difference images, when the new exposure and the exposures used for making the template have large time difference ($\sim$ months or years), and they may be selected as candidates after R/B (Section~\ref{sec:serendipitous}). However, those stars are more frequent in dense regions, such as the Galactic Bulge, instead of the low density regions that we focus on to search for extragalactic KN. In addition, the downstream broker will be able to filter out those local long-period variable stars. 
In addition to the examples above, we also test the pipeline in other fields. In particular, we processed archival exposures of DECaLS DR9 taken in the Cosmic Evolution Survey (COSMOS) field (using the LSST Science Pipelines software version 23.0.1 in Generation 3) and successfully obtained several SN in difference images that match with published results (Appendix~\ref{sec:cosmos_gen3}). 

We present our light curves, together with the corresponding exposure images and weather/observing condition metrics, in the following subsections. 
Note, the LSST Science Pipelines software does not use the weather information in our processing; we only use  the weather information when selecting exposures for making templates. 
We do notice the following signs of consistency: 
the light curve data points have relatively large uncertainties at low \texttt{teff} and have large fluctuations at large seeing (especially in \texttt{totFlux}). 
This emphasizes the critical significance of integrating weather information into transient source detection and measurement. 
In the light curve figures, we require the direct image flux (\texttt{totFlux}) to be greater than its error bar plus 100~nJy to filter out unphysical data points, and we remove outliers that have significantly large difference image flux ($|\texttt{psFlux}|\geq3\times10^6$~nJy) due to the weather.   
We do not bin the data points in order to show the results of individual exposures.

\begin{table*}[htb]
    \centering
    \begin{tabular}{l l l l}
    \hline
    Type & Field & Source density & Photometry \\
    \hline
    Kilonova & GW170817 & Low  & Sect.~\ref{sec:kn} \\
    Supernova & HiTS & Low  & Sect.~\ref{sec:hits} \\
    Supernova & DES & Low  & Sect.~\ref{sec:des} \\
    Supernova & DDF/COSMOS & Low & Sect.~\ref{sec:ddf} \\
    Stellar flare & DWF & Low & Sect.~\ref{sec:flare} \\
    Variable star  & Diffuse dwarf Crater-II & Low & Sect.~\ref{sec:craterii} \\
    Variable star  &  Galatic Bulge & High & Sect.~\ref{sec:bulge} \\
    \hline
    \end{tabular}
    \caption{Pipeline test bed data sets. For KN/SN, though they may apparently reside in/near some relatively compact host galaxies, the transient source density in observations is still much lower compared to the Bulge region. }
    \label{tab:example}
\end{table*}

\subsection{Kilonova in the DECam observation}\label{sec:kn}

We start with the well-known GW source that has an EM counterpart (kilonova/KN) captured by DECam -- GW170817~\citep{SoaresSantos2017}.  
The KN is fainter and more difficult to detect in bluer bands. 
We consider the $g$-band follow-up exposures.  
The number of exposures is limited and the KN drops in brightness quickly. 
We use their coadded image as the template to reach enough depth, and as the KN fades rapidly, the mean brightness is sufficiently lower than the peak brightness, which is good for the template generation.  
We use PS1 for photometric calibration. 

\begin{figure*}[htb]
    \plottwo{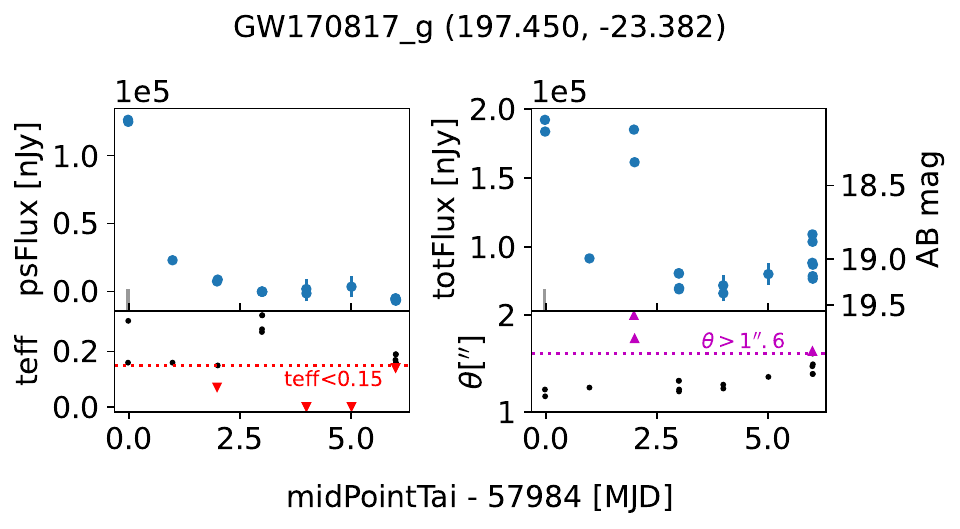}{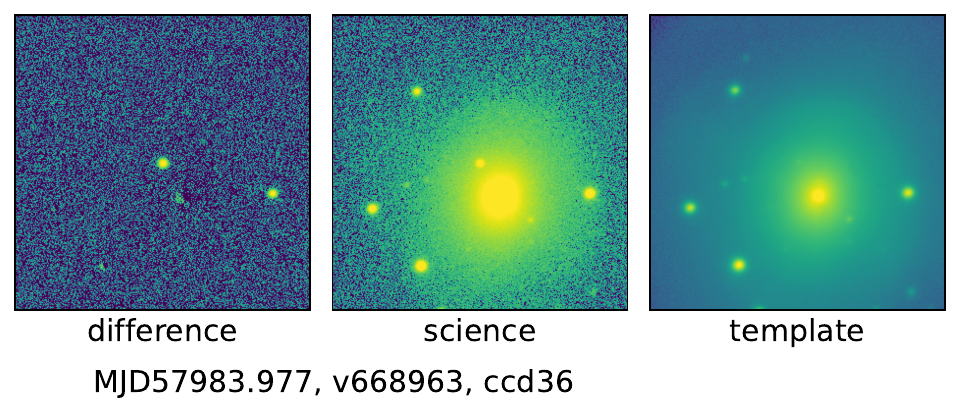}
    
    \caption{\textit{Left and Middle}: light curves of the KN of GW170817 in $g$ band from the forced photometry of the difference image and direct image, respectively. We give the celestial coordinates (RA, DEC) in the title. Here \texttt{psFlux} is the point source PSF flux from the difference image, while \texttt{totFlux} is from the direct (science) image. 
    The gray vertical tick in each light curve diagram shows the time of the difference/direct image on the \textit{right}. The ``midPointTai'' means the International Atomic Time  (Temps Atomique International in French, TAI) at the middle of the exposure. 
    Underneath we provide the weather metrics -- seeing (FWHM $\theta$ in arcsec) and the effective exposure time \texttt{teff} obtained from \texttt{qcInv} for reference. For the exposures we consider seeing has the greatest impact on \texttt{teff} compared to other observing conditions such as transparency, sky brightness, and airmass. The horizontal dashed lines give the thresholds of poor weather conditions ($\texttt{teff}=0.15$ and $\theta=1\farcs6$), and the triangle markers highlight the conditions worse than the thresholds.  
    \textit{Right}: the difference image from the first day (i.e., the flux peak; visit 668963 at CCD 36, MJD 57983.977; $300\times300$ pix; North on the top and East on the left; scaled by \texttt{arcsinh}), together with the \rr1{processed science (direct)} image and the template. The target (KN) is at the image center. The source near the right edge of the difference image is probably from a variable star. }
    \label{fig:kn_lc}
\end{figure*}

We present the light curve in Figure~\ref{fig:kn_lc} and also show the difference image to demonstrate the quality of the image subtraction. We note there is a possibly variable star nearby. 

\subsection{Supernovae in the HiTS data}\label{sec:hits}
\begin{figure*}[htb]
    \plottwo{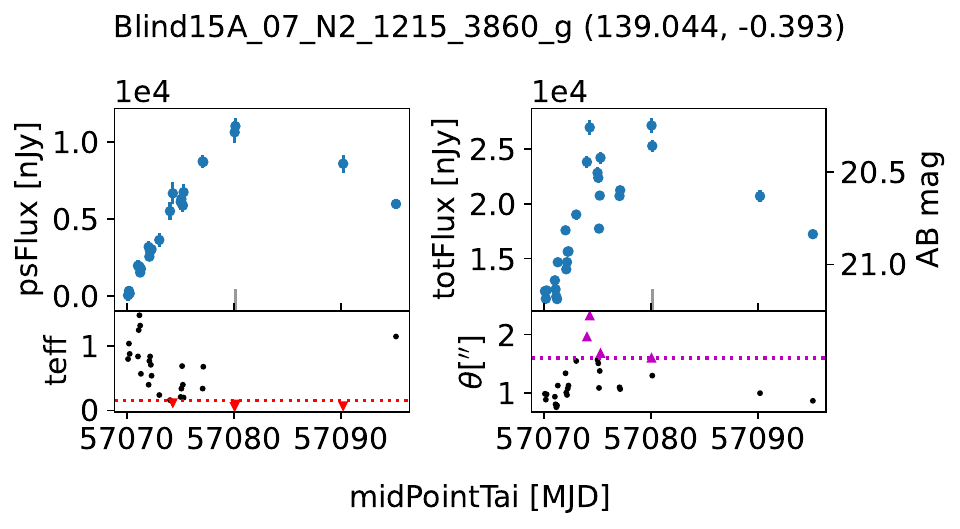}{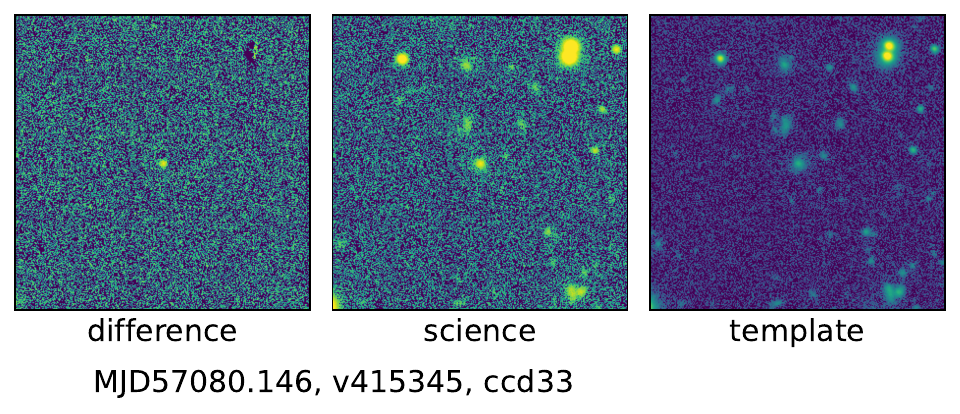}
    \plottwo{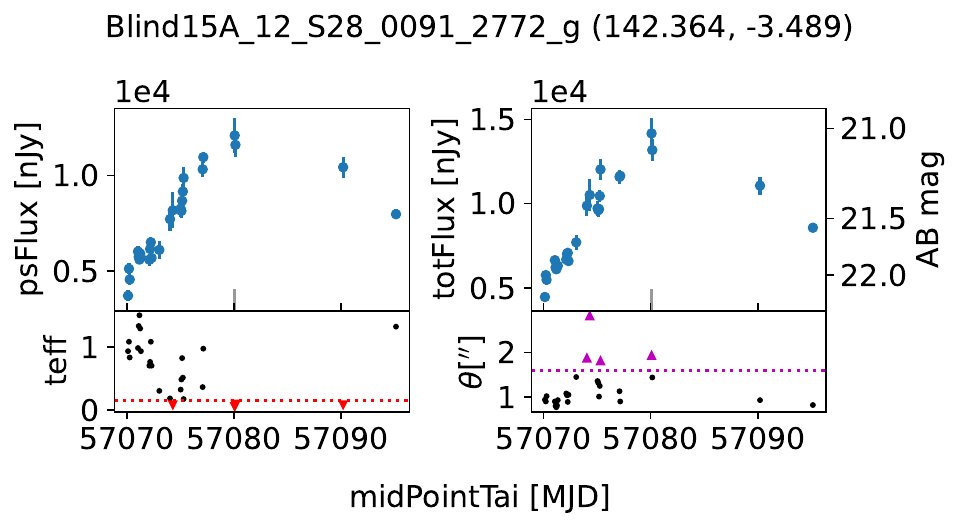}{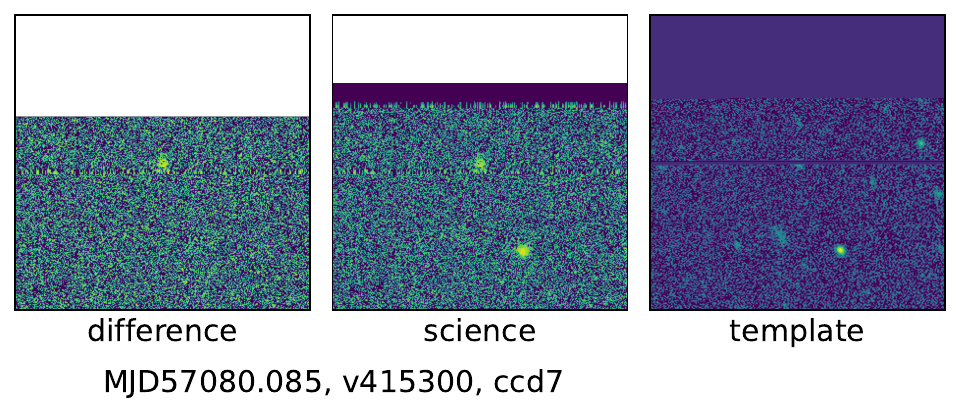}
    \caption{HiTS SN (both in $g$ band). \textit{Top}: Blind15A\_07\_N2\_1215\_3860 (SNHiTS15H). \textit{Bottom}: Blind15A\_12\_S28\_0091\_2772 (SNHiTS15bd). The blank regions are caused by CCD edges. Even though there is a bias artifact in the template, the pipeline still produces clean light curves. 
    These two objects are both from the HiTS labeled set for ML training and thus have high credibility. We titled them with their names in HiTS (year, field, CCD, and pixel coordinates). 
    The presentation and format here (and in the following figures) are the same as Figure~\ref{fig:kn_lc}. 
    }
    \label{fig:hits_lc}
\end{figure*}

We reprocess the DECam data of two SN candidates observed in the High Cadence \rr1{Transient} Survey~\citep[HiTS;][]{MartinezPalomera2018} and present their light curves in Figure \ref{fig:hits_lc}. Note those two objects happened to reach the peak brightness at almost the same time.  Other sources in the same fields do not show the same pattern, indicating that the photometry is not affected by any systematics such as weather.  
Here we use PS1 for photometric calibration. We use the first exposure as the template because of its sufficiently good observing conditions. 

\subsection{Supernovae in the DES data}\label{sec:des}
\begin{figure*}[htb]
    \plottwo{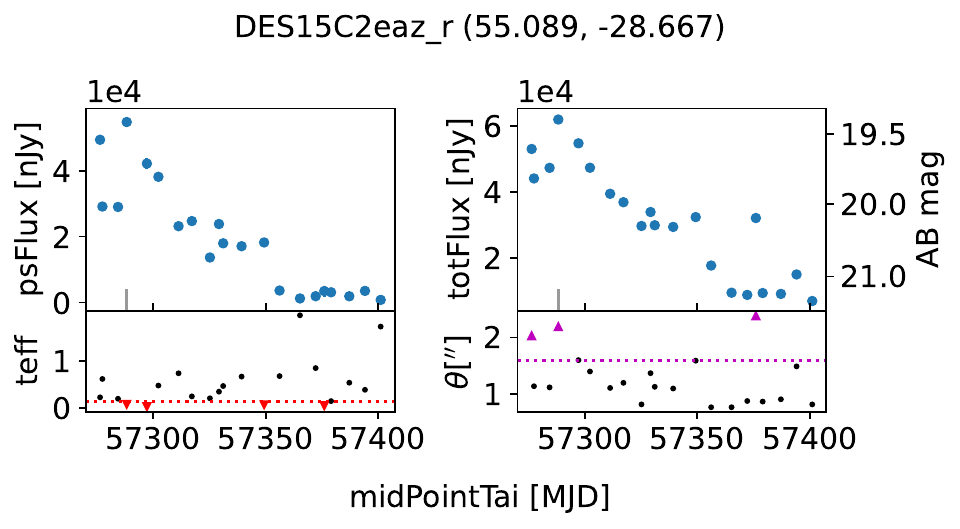}{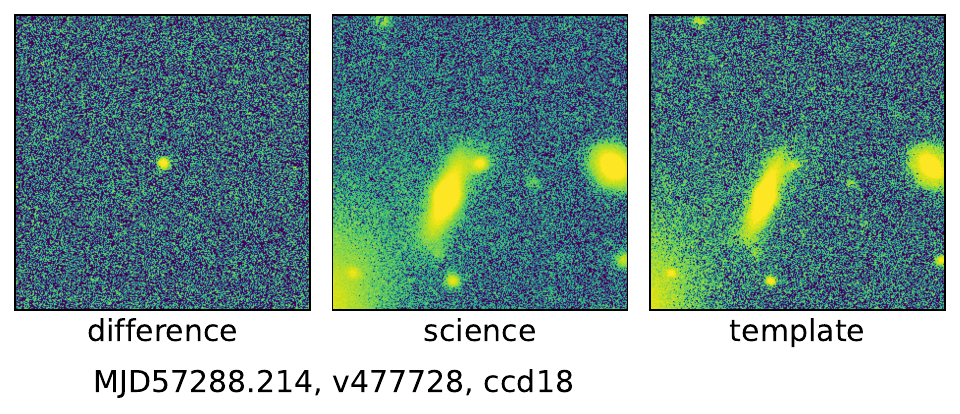}
    \plottwo{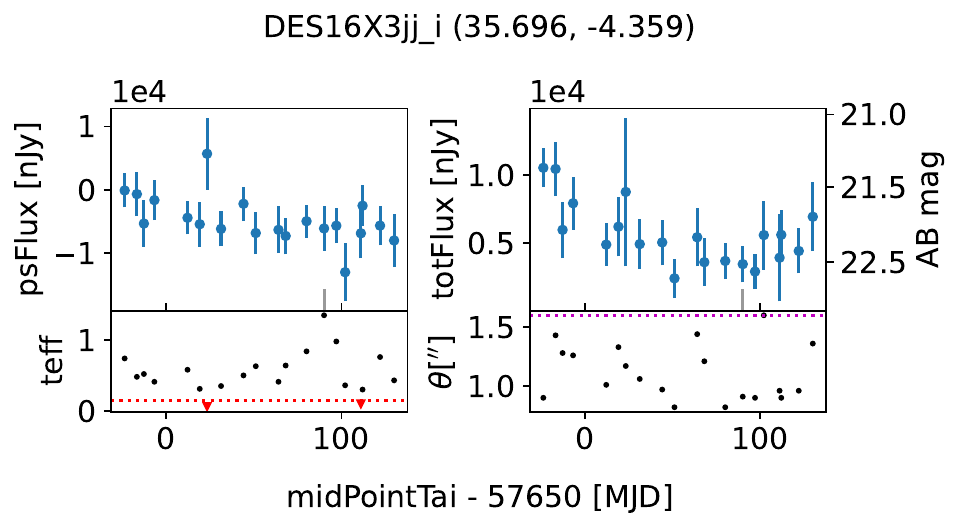}{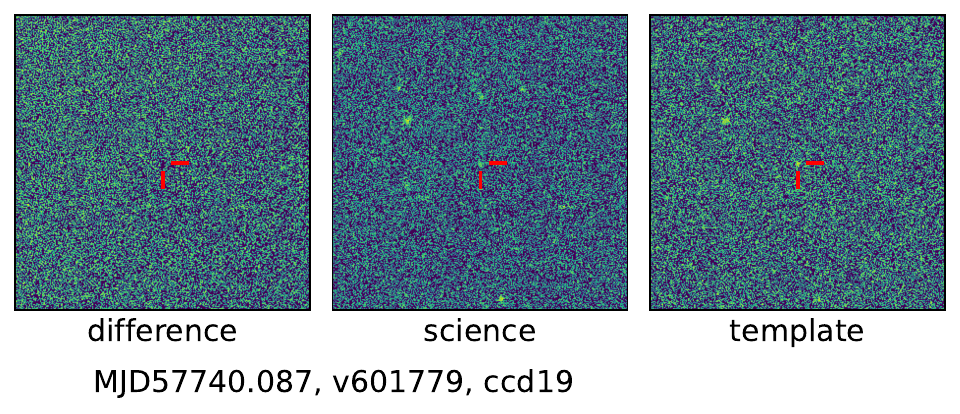}
    \caption{DES SN (Type II): 15C2eaz in $r$ band observed with 150-sec exposures (\textit{top}) and 16X3jj in $i$ band observed with 10-sec exposures (\textit{bottom}).}
    \label{fig:des_lc}
\end{figure*}

We test SN (Type II) from DES~\citep{deJaeger2020} and consider two very different cases -- 15C2eaz and 16X3jj. 
Here we use DES for photometric calibration, and use a calibrated exposure (\texttt{calexp}) right before the first visit in the light curve as the template. 
15C2eaz was observed with 150-sec exposures in $r$ band, and this SN is much brighter than the other one. 
16X3jj was observed with 10-sec exposures in $i$ band, and  we decrease the detection threshold (from the default $5\sigma$ to $1.5\sigma$) because of the short exposure time and low brightness of this object. 
Our results are shown in Figure~\ref{fig:des_lc}, which generally match the literature results. 

\subsection{Supernovae in the DECam DDF data}\label{sec:ddf}
\begin{figure*}[htb]
    \plottwo{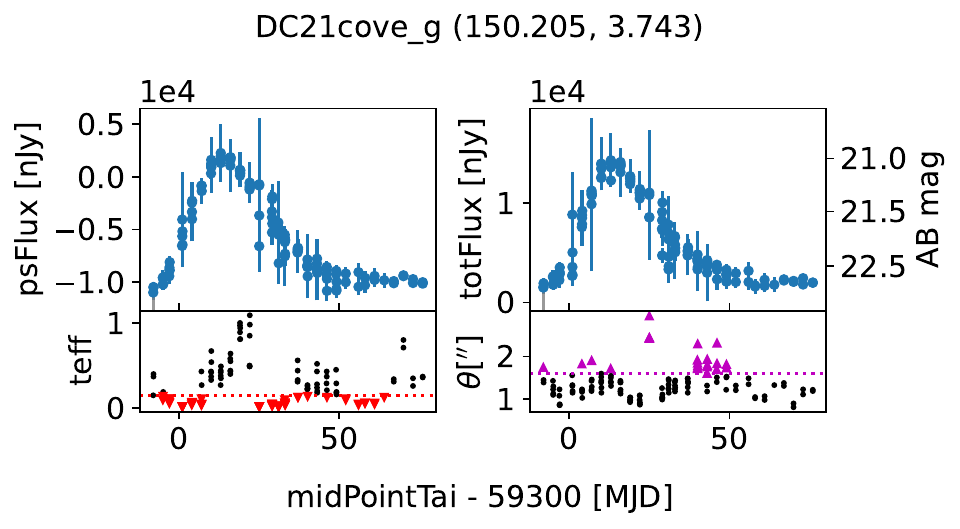}{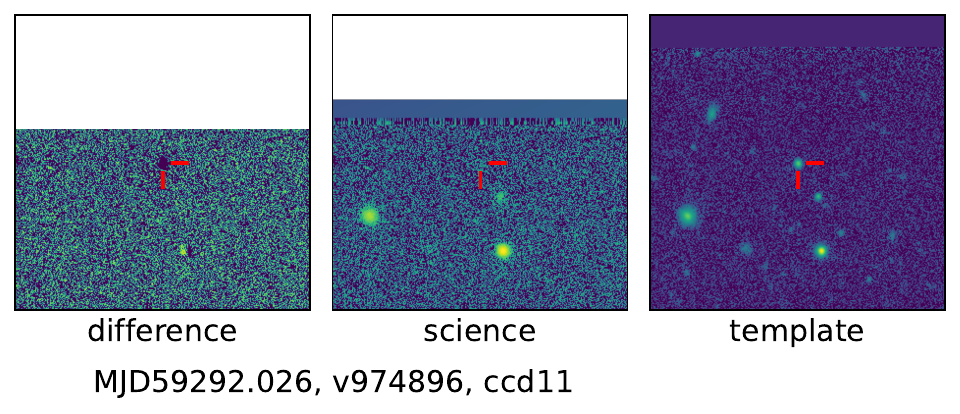}
    \plottwo{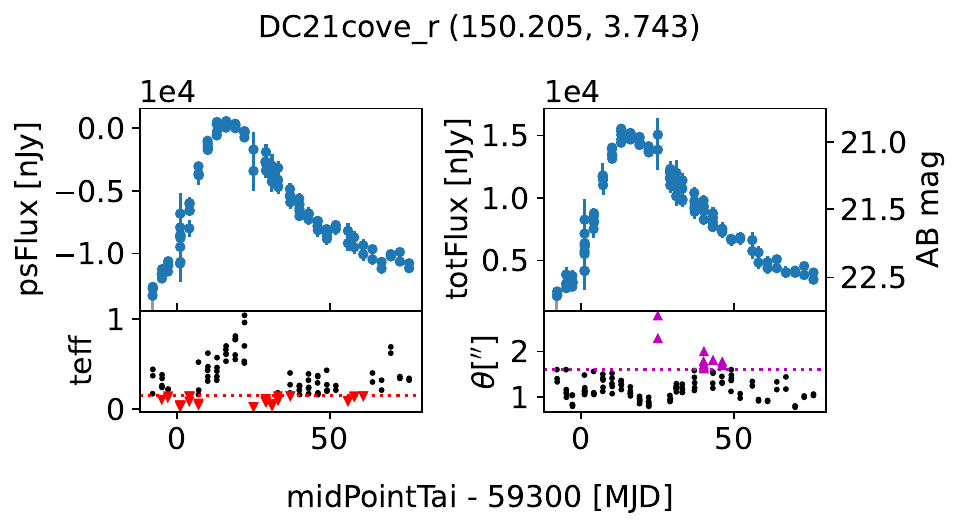}{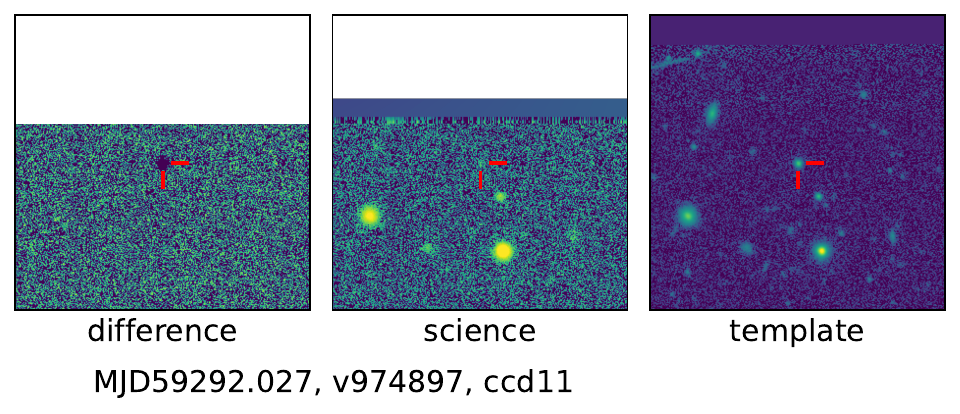}
    \plottwo{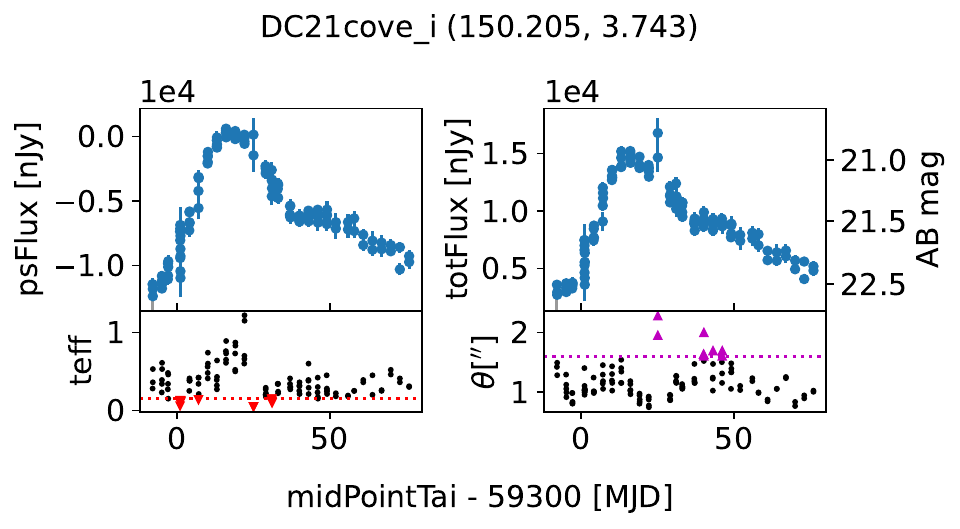}{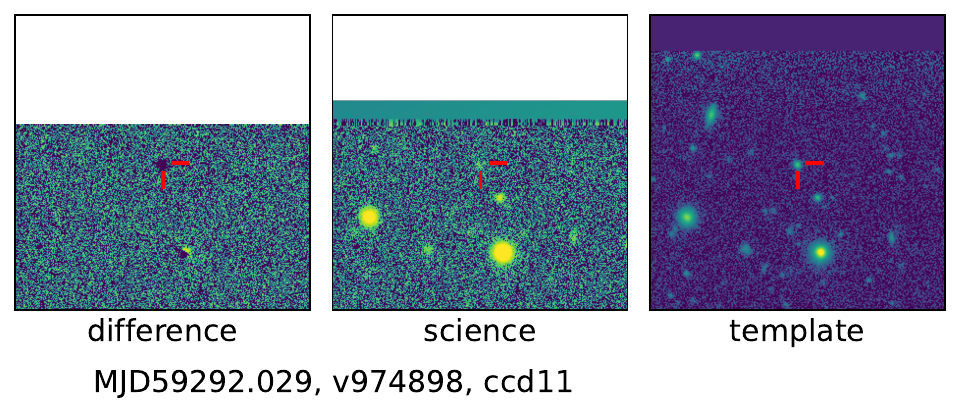}

    \caption{DECam DDF SN candidate DC21cove in $g,r,i$ bands. For clarity, we use individual diagrams for each band, instead of displaying all light curves on the same set of axes.   
    }
    \label{fig:decat_lc_cove}
\end{figure*}

\begin{figure*}[htb]
    \plottwo{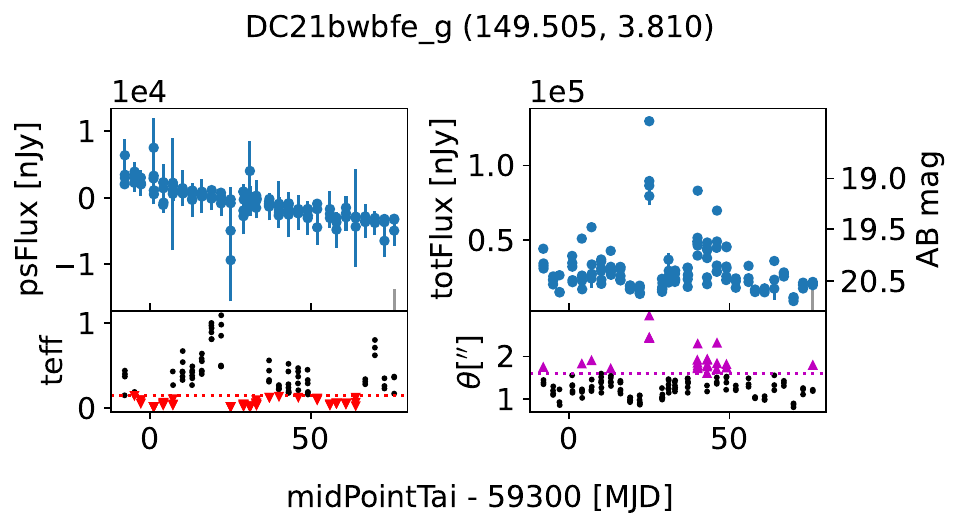}{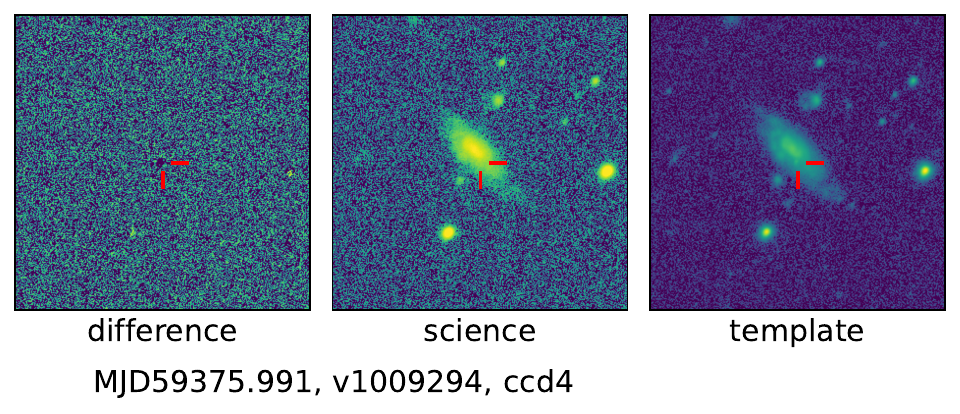}
    \plottwo{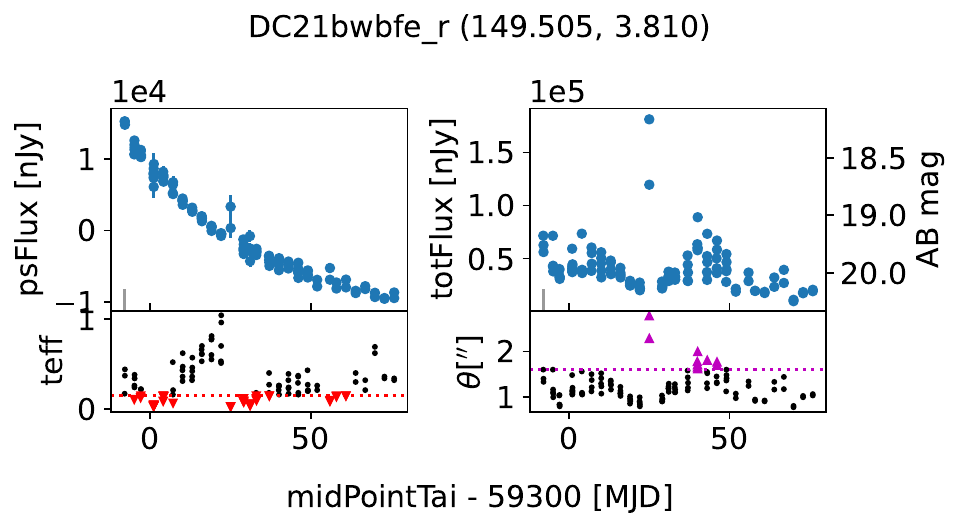}{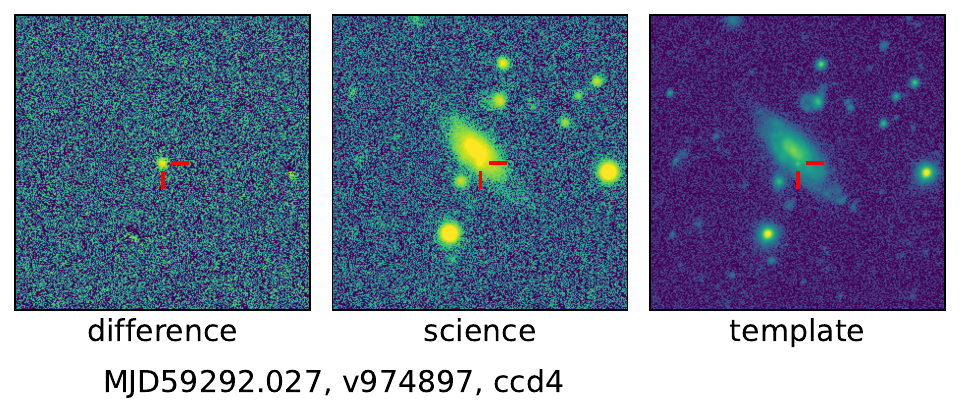}
    \plottwo{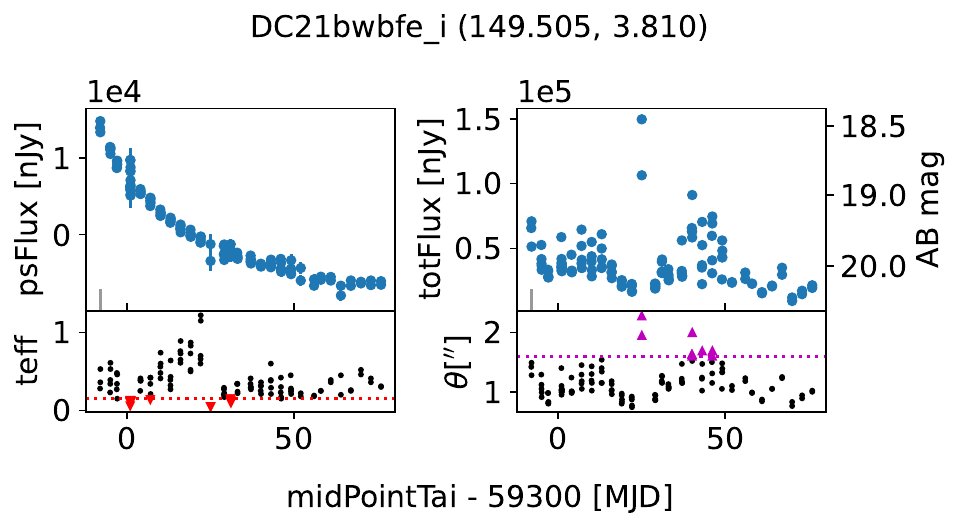}{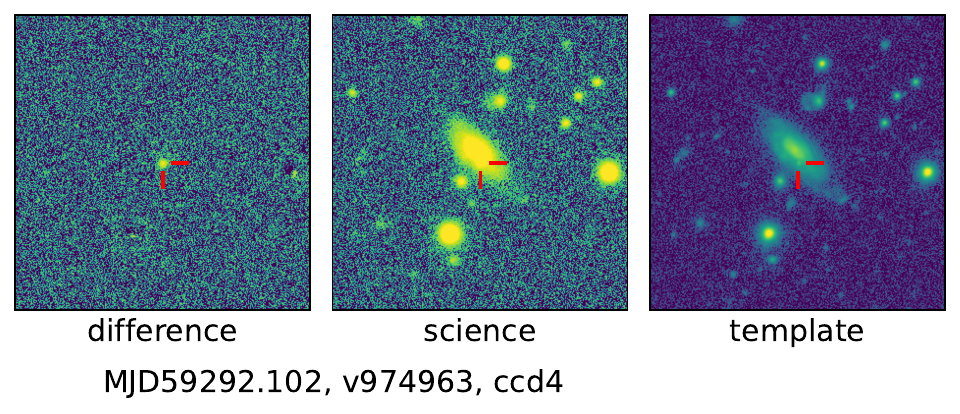}
    \caption{DECam DDF SN (Type Ia)  DC21bwbfe/SN2021bnv in $g,r,i$ bands. Note the host galaxy affects the light curves of the object on the direct images (\texttt{totFlux}), but the light curves from the difference images (\texttt{psFlux}) are much cleaner. 
    }
    \label{fig:decat_lc_bwbfe}
\end{figure*}

\begin{figure*}[htb]
    \plottwo{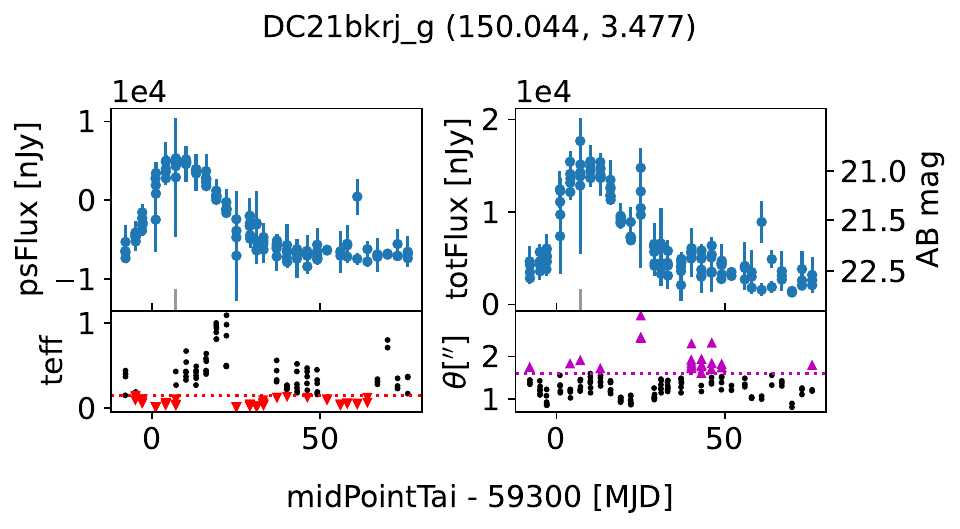}{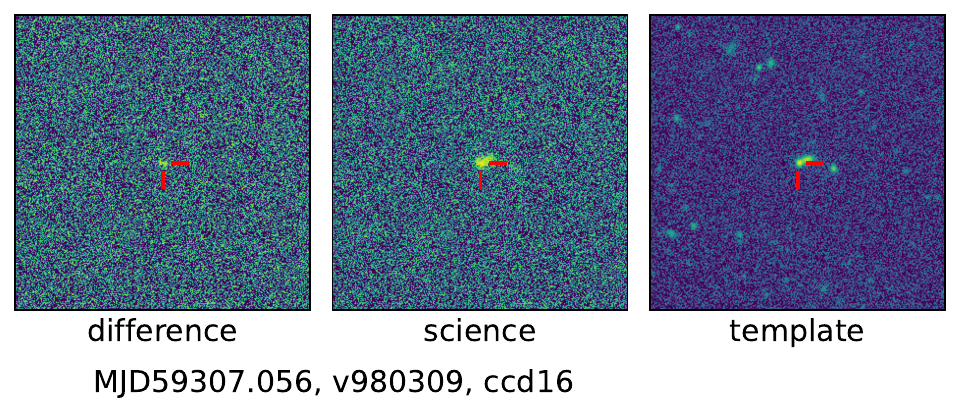}
    \plottwo{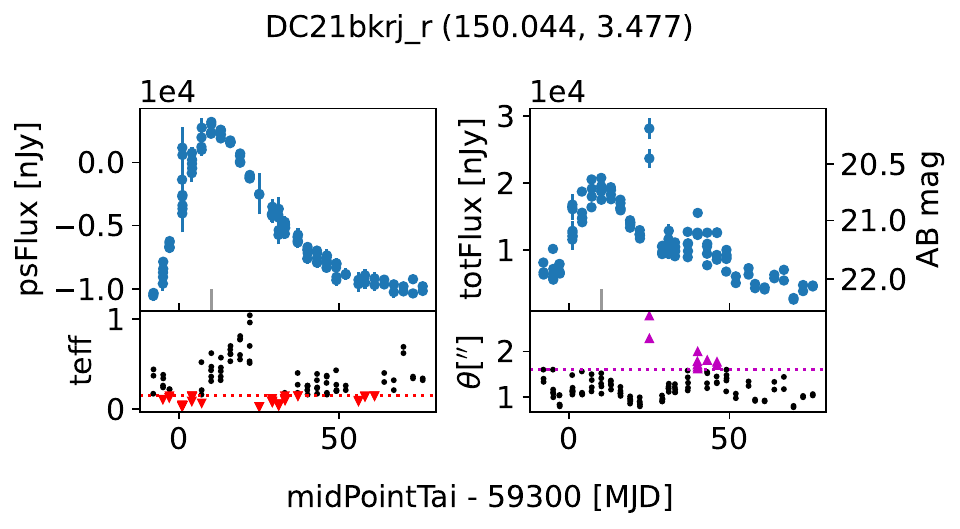}{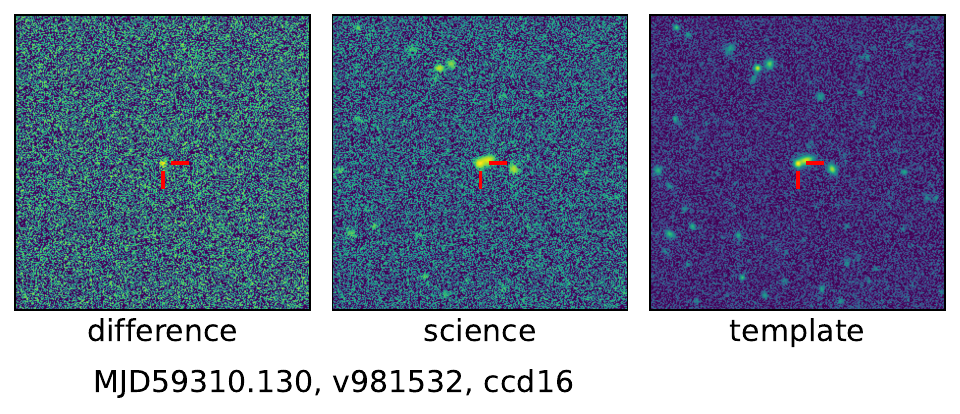}
    \plottwo{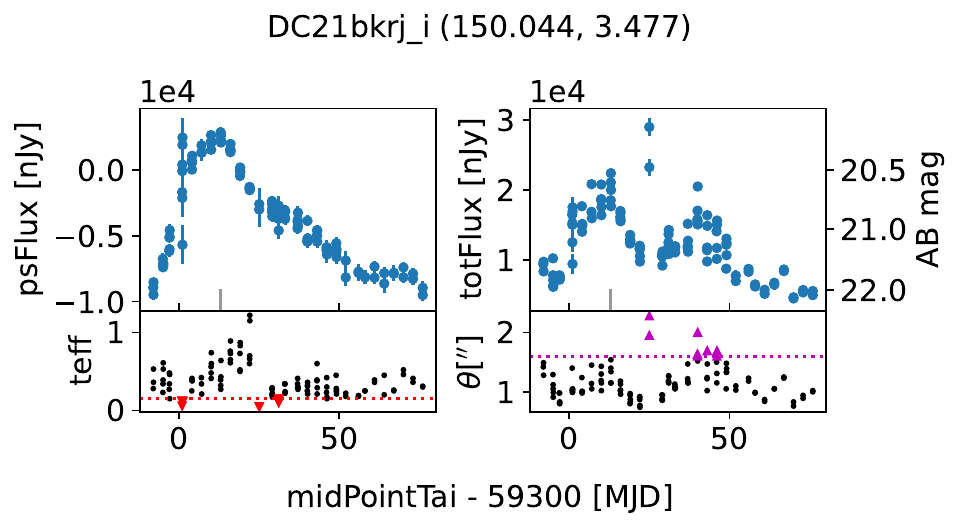}{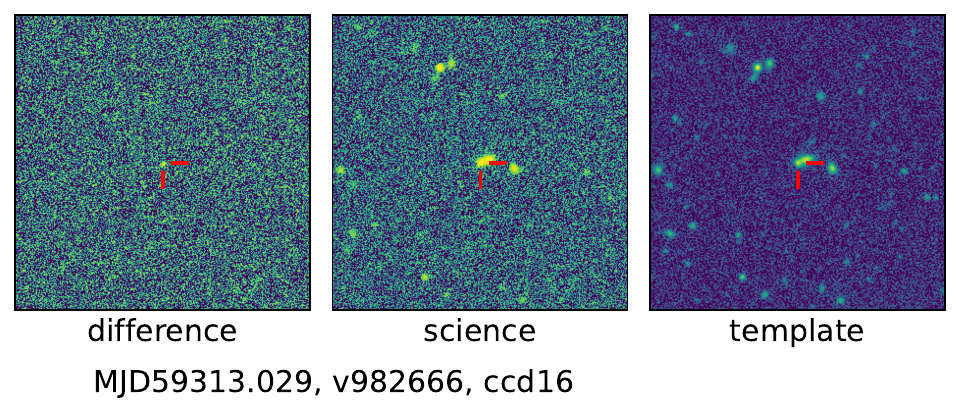}
    
    \caption{DECam DDF SN candidate  DC21bkrj in $g,r,i$ bands. 
    }
    \label{fig:decat_lc_bkrj}
\end{figure*}

We consider SN candidates in the COSMOS field observed during the DECam DDF survey~\citep{Graham2023}, which is part of the DECam Alliance for Transients (DECAT) program.\footnote{The DECAT program provides a candidate viewer: \url{https://decat-webap.lbl.gov/decatview.py/}. For example, DC21cove: \url{https://decat-webap.lbl.gov/decatview.py/cand/DC21cove}.  
} 
Here we use PS1 for photometric calibration. We select and stack high-quality exposures (six per band with good \texttt{teff} and seeing) for the templates in the $g,r,i$ bands respectively. 
Figure~\ref{fig:decat_lc_cove},~\ref{fig:decat_lc_bwbfe},~\ref{fig:decat_lc_bkrj} show the multi-band light curves of three SN candidates reprocessed by the LSST Science Pipelines software: DC21cove, DC21bwbfe/SN2021bnv\footnote{SN2021bnv is classified by spectrum as a Type Ia SN at redshift $z=0.08$~\citep{Fremling2021,Dahiwale2021} \url{https://www.wis-tns.org/object/2021bnv}.
}, DC21bkrj. 
DC21bkrj has been analyzed by~\citet{Graham2023}. Using a different pipeline, we obtain consistent results with reasonable error bars.

\subsection{Stellar flares in the DWF data}\label{sec:flare}
\begin{figure*}[htb]
    \plottwo{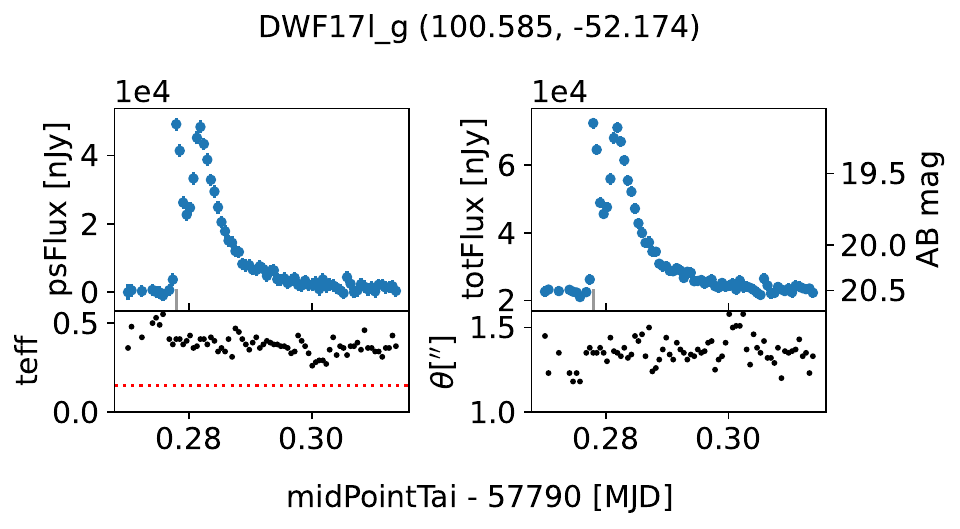}{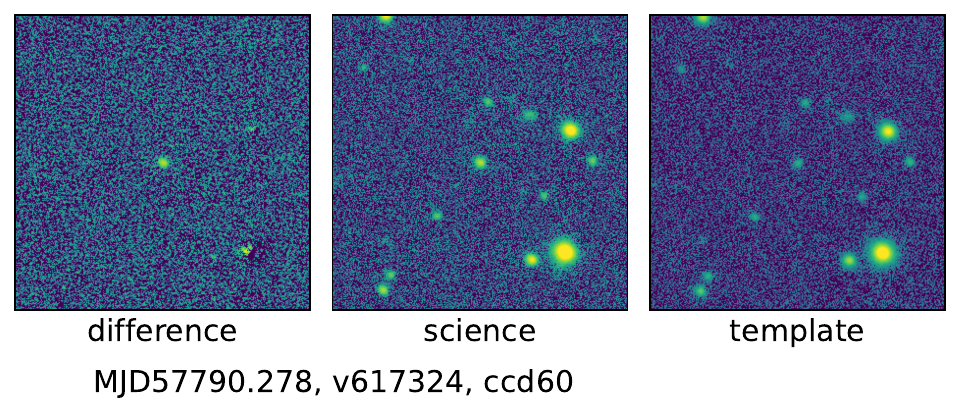}

    \caption{Flares of an M-type star in $g$ band observed in the DWF program, with the difference image at the first flux peak (visit 617324; MJD 57790.278).}
    \label{fig:dwf_lc}
\end{figure*}

We test another type of transient, the stellar flares of M-type dwarf stars. 
We reprocess the observational data of the Deeper, Wider, Faster program~\citep[DWF;][]{Andreoni2020} using the LSST Science Pipelines software. Those are 20-sec $g$-band exposures with $\sim1$ min gaps. We use SkyMapper DR1 for photometric calibration (using DR1 produces a slightly better light curve than DR2 here) and use the first exposure to build a template. Figure~\ref{fig:dwf_lc} shows the light curve of flares emitted from an M5 dwarf. 

\subsection{Variable stars in the diffuse galaxy Crater II}\label{sec:craterii}
\begin{figure*}[htb]
    \centering
    \plottwo{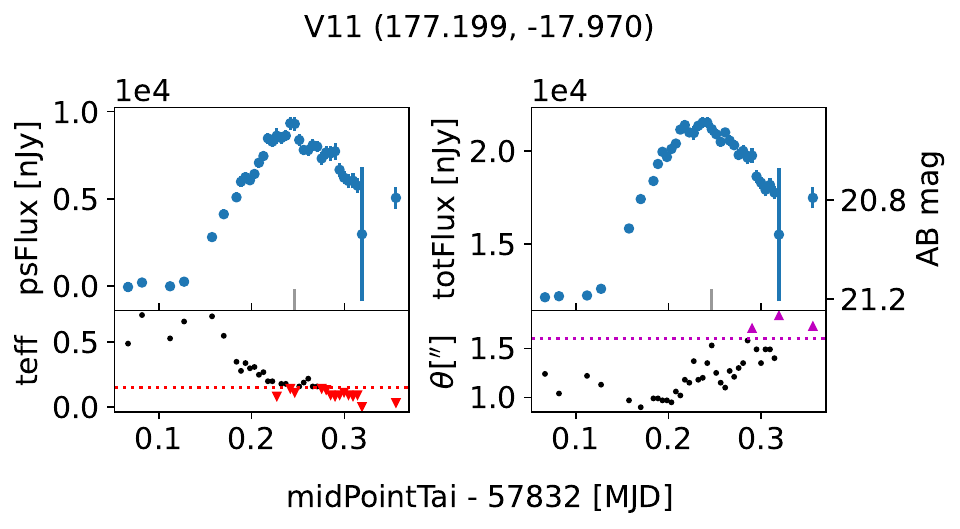}{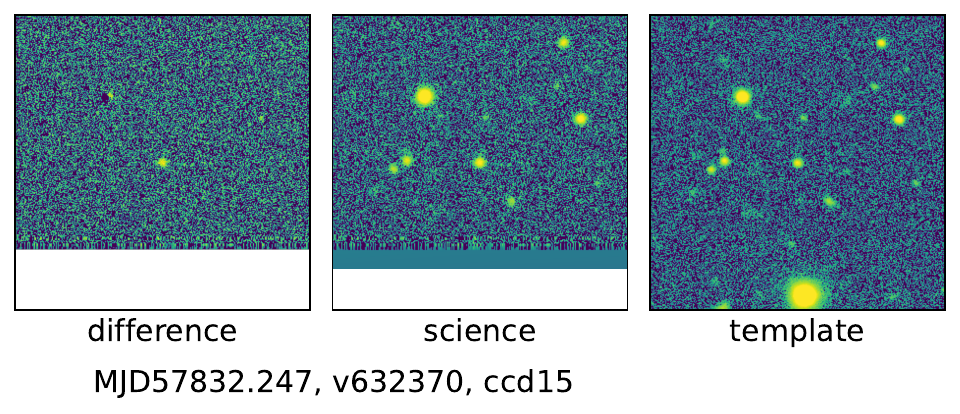}
    \plottwo{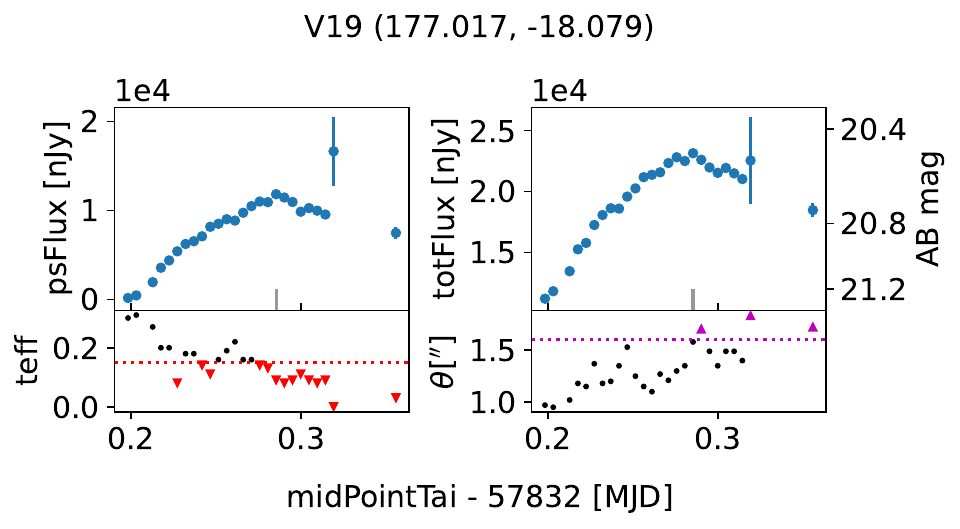}{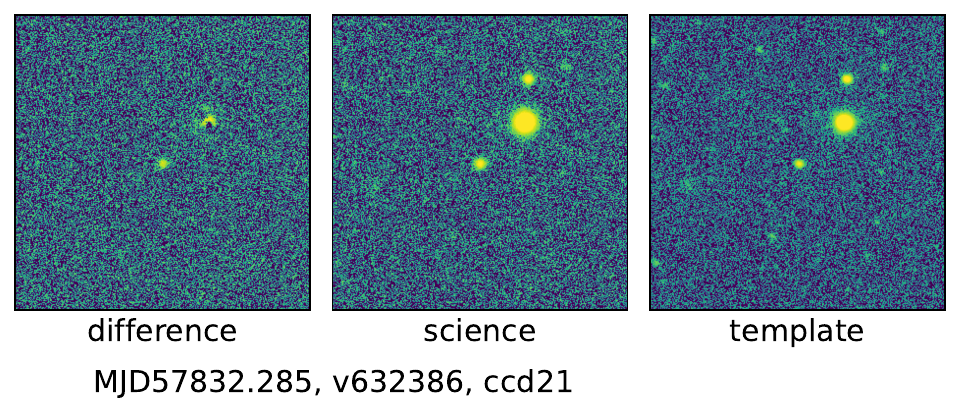}
    
    \caption{
    Variable stars in the diffuse dwarf galaxy  Crater II (both in $g$ band): V11 (\textit{top}), V19 (\textit{bottom}). 
    }
    \label{fig:craterii_lc}
\end{figure*}

We switch to variable stars and start with two RR Lyrae stars in a diffuse dwarf galaxy Crater II~\citep{Vivas2020}. 
We take the exposure before the first visit in the light curve as the template and reprocess the $g$-band data captured during the first night
($\sim$ 7 hrs) of the program. We use PS1 for photometric calibration as the field is not covered by DES (or DECaLS DR9). 
Our light curves are presented in Figure~\ref{fig:craterii_lc} and match the literature results. 

\subsection{Variable stars in the Galactic Bulge}\label{sec:bulge}
\begin{figure*}[htb]
   \plottwo{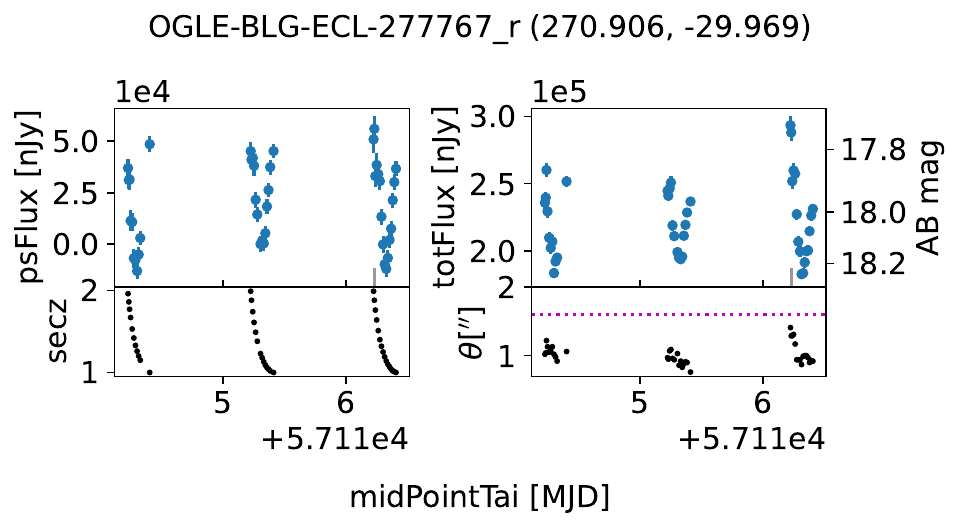}{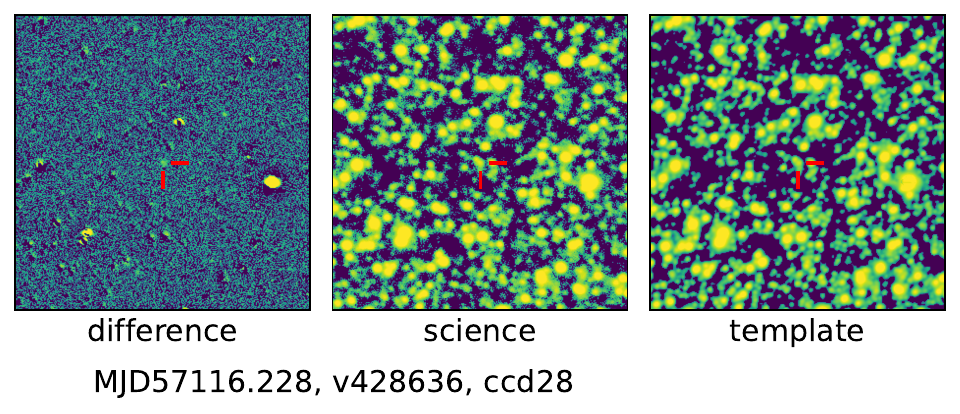}
    \plottwo{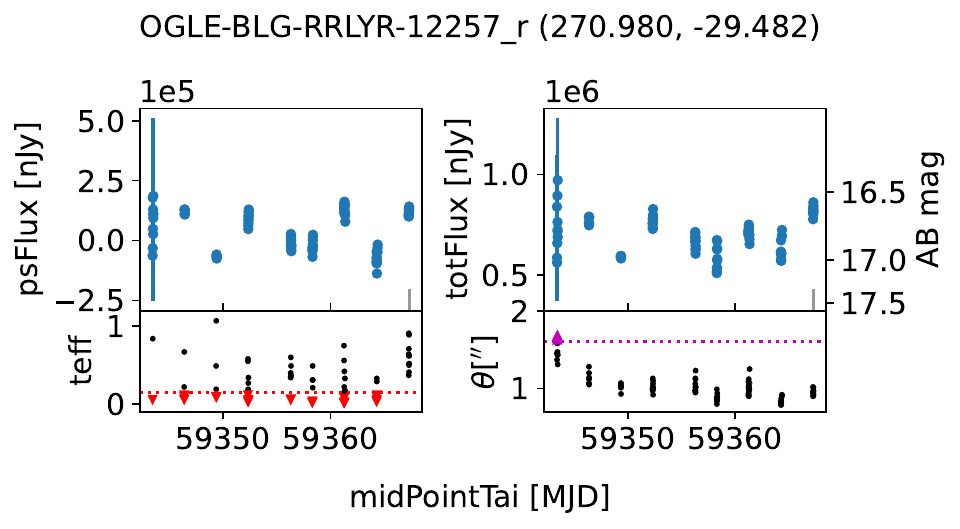}{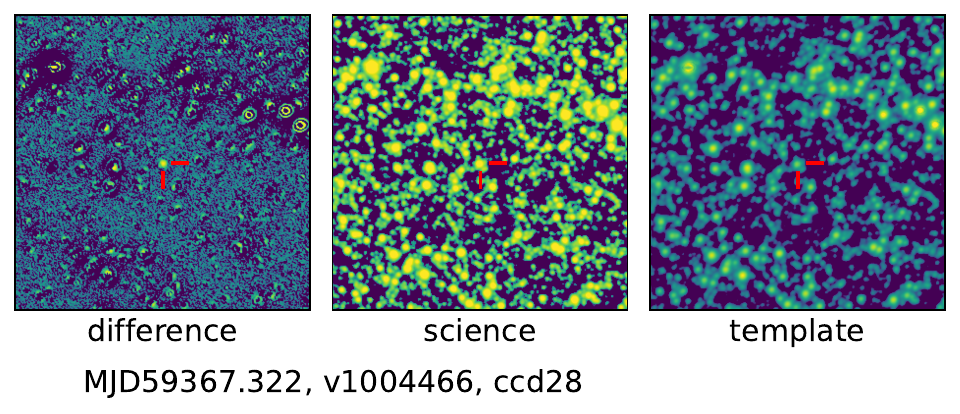}
    \plottwo{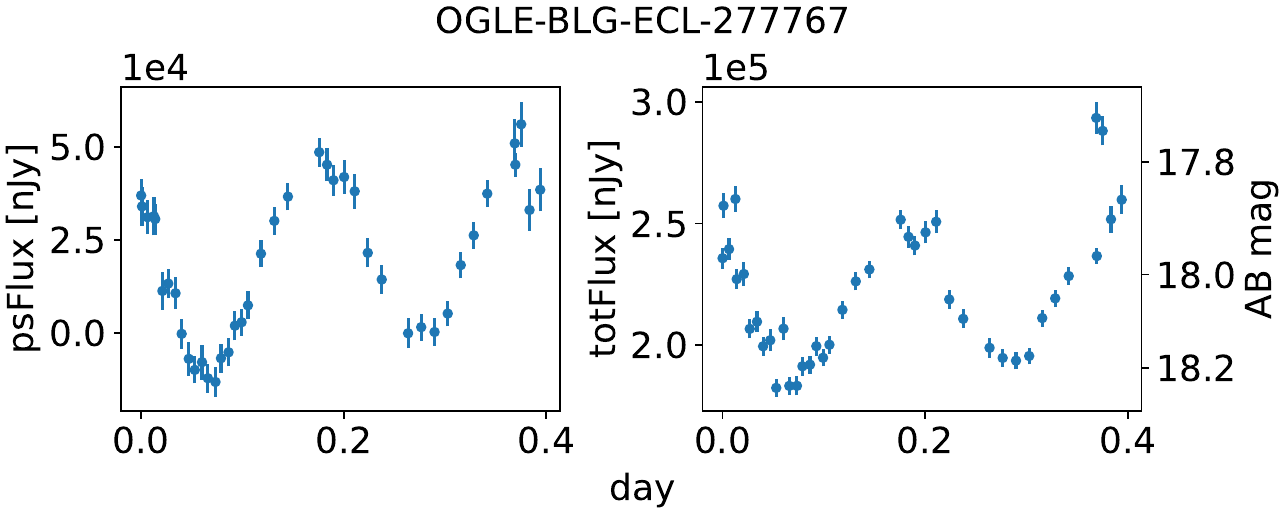}{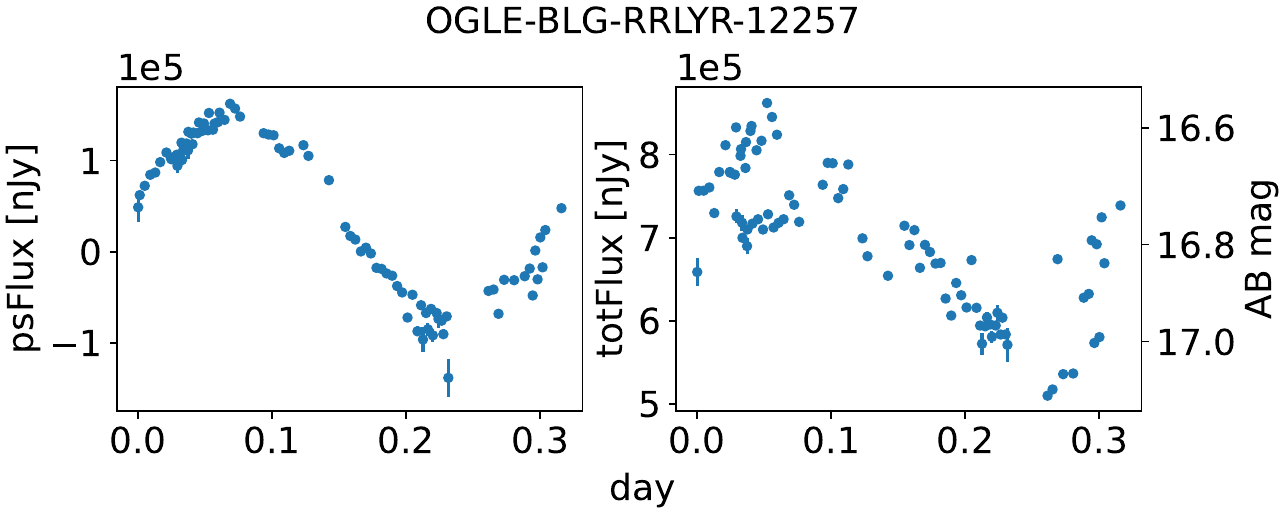}
     
    \caption{OGLE variables in the Galactic Bulge in $r$ band: Eclipsing variable OGLE-BLG-ECL-277767 (\textit{top};  $P=0.406$d; 5-sec exposures) and RR Lyrae variable OGLE-BLG-RRLYR-12257 (\textit{middle}; $P=0.318$d; 50-sec exposures) with the original light curves (\textit{top, middle}) and the phase folded version (\textit{bottom}; we use the period value from OGLE to shift the data points). 
    For the RR Lyrae variable, after the direct image flux (\texttt{totFlux}) cut mentioned earlier and a cut on significant outliers, in the bottom row we also remove outliers in the difference image flux (\texttt{psFlux}) by using a $3\sigma$ cut around the median, and remove data points that have significantly large errorbars (1/10 of the light curve amplitude) to produce light curves cleaner than the top row. 
    In the difference images, the annular rings around bright stars are \rr1{mainly} caused by saturation, but they do not affect our targets. 
    The observing condition metrics are estimated from the header information of \texttt{instcal} as these are not included in \texttt{qcInv}. The guider transparency is not available for the \textit{top} panel exposures, and we give the airmass $\sec(z)$ instead; the sky brightness shows a similar trend to the airmass. 
    }
    \label{fig:bulge_lc}
\end{figure*}

We test the photometry of variable stars in the 
high star density fields of the Galactic Bulge using the observations of~\citet{Saha2019} and~\citet{Graham2023}. 
Those images were taken in $r$ band with 5-sec and 50-sec exposures respectively.  
We use Gaia for astrometric calibration and 
SkyMapper for photometric calibration (PS1 does not have enough coverage). We note that Gaia DR1 has better coverage than DR2 in that region and thus use DR1. 
We stack 9 high-quality archival exposures to make a template.\footnote{We thank Melissa Graham for the exposure search tools based on the Astro Data Lab tutorial notebook~\citep{Fitzpatrick2014,Nikutta2020,Juneau2021} \url{https://github.com/astro-datalab/notebooks-latest/blob/master/04_HowTos/SiaService/How_to_use_the_Simple_Image_Access_service.ipynb}. } 
The high star density in this field requires more processing time and memory than those in the previous fields. 
\rr1{Given the observation window (observing the field for several hours and coming back a few days later)}, we consider variable stars that have a period of $\sim0.5$ day so that the full light curve can be well captured by the DECam observations.  
Figure \ref{fig:bulge_lc} shows our results. 
Because the weather metrics are not available in \texttt{qcInv} \rr1{(caused by the high number density of stars in these fields)}, they are estimated with the information from \texttt{instcal} \rr1{images}\rr1{\footnote{The \texttt{instcal} images are instrumentally calibrated exposures generated by the DECam Community Pipeline~\citep{Valdes2014} and provided by NOIRLab. More details can be found at \url{https://noirlab.edu/science/documents/scidoc1203}. }}: we use the guider sky transparency and the medians of the sky brightness and FWHM among CCDs\footnote{See tutorials at \url{https://github.com/NOAO/nat-nb/tree/master}. }, and we compute the \texttt{teff} using Eq.~\ref{eq:teff} and the fiducial seeing and sky brightness~\citep{Neilsen2016,Morganson2018}.

We compare our light curves with those from the Optical Gravitational Lensing Experiment survey~\citep[OGLE;][]{Udalski2015}\footnote{\url{https://ogledb.astrouw.edu.pl/~ogle/OCVS/catalog_query.php}}. 
Our photometry is close to OGLE, especially the flux variability amplitude. 
Compared to the methods in previous work~\citep[e.g.,][]{Saha2019}, in the Bulge region the image subtraction of the LSST Science Pipelines software still produces  clean results. Therefore, we utilize  the source catalogs derived from the difference imaging, rather than comparing a catalog obtained from a direct image with a baseline/reference catalog (Section~\ref{sec:intro}).

\subsection{Summary of the light curves}
We compared our light curves against previous results and found consistent shapes and amplitudes with slightly different  magnitude zero points ($\sim0.1$ mag for reported DECam results), which could be caused by the color terms between the reference instrument and DECam. The outliers in the light curves could be caused by weather or instrumental artifacts. 

Our results also show that in general \texttt{psFlux} (measured on difference images) is less sensitive to weather and produces cleaner light curves than \texttt{totFlux} (from direct images), 
which is important for transient studies (to remove the flux from the host/environment). 
On the other hand, the direct image photometry gives the magnitude of an object, which is useful for variable star analysis.


\section{Real/Bogus classification algorithm}\label{sec:rb}
Given the fast brightness change and the large search area of KN, we carry out R/B on individual CCDs of single exposures; we do not rely on the collective information from multiple exposures, e.g., the light curve or the color of an object. We leave further classification based on the time evolution or the multi-band information of an object to downstream brokers. 

To perform R/B, we first clean the sources detected on difference images using the flags provided by the LSST Science Pipelines software, because those flags remove most artifacts~\citep{Liu2024}. 
Then, we apply Principal Component Analysis (PCA) to several quantities to select sources. 
We give the details and examples below. 

\subsection{Flags}\label{sec:flags}
We consider a list of flags (Table~\ref{tab:flags}) in the source catalog based on their definitions and their performance on real objects in the archival \rr1{observational} data. \rr1{We note that these flags give high completeness of real sources in simulations as well~\citep{Liu2024}.} 

We adopt a ``tight'' cut by using multiple flags; using fewer flags can still generate a list of MMA candidates, but there would be more false positives. As the sky search area and the number of exposures are both large for DECam MMA detections, we decide to use a narrow cut to reduce false positives and noise. 
On the other hand, we note that when the source has low S/N or is in a dense region, we may have to skip some flags, e.g., \texttt{base\_PixelFlags\_flag\_suspect}. 

In the following text, we study the time evolution of flags in some representative cases as robustness tests. The corresponding light curves have been presented in Section~\ref{sec:examples}.

\paragraph{GW170817}
We consider  the $g$-band exposures of the KN (Section~\ref{sec:kn}). 
The KN is only detected in the individual difference images of the first \textit{four} exposures, because \rr1{the star rapidly lost its brightness,} and the observing conditions were non-optimal (but note the forced photometry still captured it in later exposures; Figure~\ref{fig:kn_lc}). We find that all flags considered in Table~\ref{tab:flags} are \texttt{False} in those four exposures. 

\paragraph{DES15C2eaz}
We examine the time evolution of the flags and compare that with the \texttt{teff} and the SN flux. 
Though the \texttt{teff} has  large variations during the observations (Figure~\ref{fig:des_lc}), the considered flags are fixed at \texttt{False} (except at the visit 508832, MJD 57394.148,  \verb|base_SdssCentroid_flag|, \rr1{which is the general failure flag of the centroid  measurement algorithm}, turns \texttt{True} as the SN is faint). This is similar to the case of GW170817 above. 

\paragraph{DC21bwbfe/SN2021bnv \& DES16X3jj}
Sometimes the constructed template under satisfactory weather conditions may already contain the transient, and we test the performance of image subtraction in such cases. 
For example, DC21bwbfe shows negative flux in about half of the difference images (Figure~\ref{fig:decat_lc_bwbfe}), while DES16X3jj (Figure~\ref{fig:des_lc}) shows negative flux in almost all difference images. 
We note that the SDSS-style shape/centroid flags give \texttt{True} when the source is negative, and thus we skip those flags to keep the SN. Dropping those flags causes more sources to be retained, and the result can be noisier with more false positives. However, we expect this situation to be rare in MMA -- in difference images, a KN should generally have positive pixels, because it gets faint quickly and the template can hardly capture it. 
The situation would be more common for SN and variable stars. 
In a new version of the LSST Science Pipelines software, those flags will not automatically be set to \texttt{True} for negative sources. 

In summary, the results above indicate that for a bright transient, the flags that we consider will not filter it out, unless it is below (i.e., undetected) or close to the detection limit.

\begin{table*}[htb]
    \centering
    \begin{tabular}{l l l}
    \hline
    \hline
    Type & Flag index \& Name & Definition \\
    \hline
    Flux & 0. \verb|base_PsfFlux_flag| & General \rr1{failure} flag for PSF flux. \\
    & 1. \verb|base_CircularApertureFlux_12_0_flag| & General \rr1{failure flag for 12-pix radius aperture flux}.\\
    & 2. \verb|base_PixelFlags_flag_saturated| & Whether the source has a saturated pixel. \\
    & 3. \verb|base_PixelFlags_flag_bad| &  Whether the source has a bad pixel. \\
    & 4. \verb|base_PixelFlags_flag_edge| &  Whether the source \rr1{is} near the CCD edge. \\
    & 5. \verb|base_PixelFlags_flag_suspect| &  Whether the source has a suspicious pixel. \\
    & 6. \verb|ip_diffim_DipoleFit_flag_classification| & Whether the source is recognized as a dipole. \\
    & 7. \verb|ip_diffim_NaiveDipoleFlux_flag| & General \rr1{failure} flag for dipole flux. \\
    & 8. \verb|ip_diffim_forced_PsfFlux_flag| & \rr1{Failure} flag for the direct image forced photometry. \\
    \hline
    Geometry & 9. \verb|base_SdssCentroid_flag| & General \rr1{failure} flag for SDSS-style centroid coordinates. \\
    & 10. \verb|base_SdssShape_flag| & General \rr1{failure} flag for SDSS-style shapes. \\
    \hline
    \end{tabular}
    \caption{LSST Science Pipelines software flags for filtering the source catalog \texttt{diaSrc} of a difference image. We group the flags by types and assign them integer index values to simplify retrieving their information. 
    Note, \texttt{base\_SdssShape\_flag} and \texttt{base\_SdssCentroid\_flag} may be set to \texttt{True} for a source that contains negative pixel values (Section~\ref{sec:pca_test}). 
    However, for a KN/MMA source we expect it to have positive pixel values in general. 
    }
    \label{tab:flags}
\end{table*}

\subsection{\rr1{PCA workflow}}\label{sec:pca_theory}

PCA uses the covariances between the features of different objects to find a direction in the multi-dimension feature space, along which the data is maximally ``stretched'', and therefore cutting the data on that direction gives the best split~\citep[e.g.,][]{Jolliffe16}. 
We define a list of features so that real objects are expected to have larger feature values than bogus ones (Section~\ref{sec:features}).

After flag filtering (Section~\ref{sec:flags}), we ``standardize'' each feature of remaining sources (subtract the mean and then divide it by the standard deviation); this centers the values around the origin and rescales the values so that they have similar scatter ranges.  

Next, we diagonalize the covariance matrix (which is a real symmetric matrix) and sort the eigenvalues. The unit eigenvector (as a linear combination of different features) that has the highest eigenvalue indicates the direction in which the data has the greatest spread. This direction (i.e., the first principal component) generally represents how the data points are separated, especially when the scatters along other eigenvectors are much smaller. The corresponding eigenvalue shows how much the data variance is explained by this component.  
We implement the above process using \texttt{scikit-learn}~\citep{scikit-learn}. 
Note that the sign-flipped version of an eigenvector is still an eigenvector; for the first principal component, we choose the sign when the first feature (the S/N of \texttt{|psFlux|}; Table~\ref{tab:features}) coefficient is positive, because a higher S/N would suggest a more valid source. 

Finally, we make a cut on the first component direction and select sources that have high first component values (PC1); \rr1{in general, we consider sources that have PC1 greater than or equal to their median as candidates  (but users can adjust the threshold depending on the science cases)}.
Each selected source has a weight based on the explained variance ratio (the fraction of the data variance captured by the first PCA component) and the \texttt{teff} of that exposure, and the PC1 value. We present the details of our algorithm in the following subsections.

The advantages of using PCA for R/B are as follows. 
PCA is mathematically simple and can run with minimal computational resource utilization. 
It does not require training and thus is flexible to work on different datasets. 
Also, the features are correlated -- a real object tends to be bright (have large S/N) and to have a size comparable to that of the PSF and a small ellipticity, and thus those quantities are associated naturally. 
PCA considers different features simultaneously -- though a single feature may separate sources, running R/B based on multiple features can give a cleaner result.

\subsection{Features for PCA}\label{sec:features}

For each source detected on the difference image (of a single CCD exposure), we consider the following features. 
After tests on known transients, we find these features can well separate legitimate candidates from bogus ones.
Some features are inspired by ZTF~\citep{Mahabal2019}.
Here ``\verb|_instFlux|'' means the instrumental flux that is directly given in the LSST Science Pipelines software source catalogs stored in FITS tables (\texttt{diaSrc}), which can be easily and quickly obtained. 
A further correction to the flux can be made to deal with the color-terms between DECam and reference catalogs; we skip this correction because it is small and relative flux carries enough information about photometry.  
Also, the flux-related features we consider in R/B are flux ratios, and we expect the correction factor to be reduced in the ratios. 
Table~\ref{tab:features} summarizes the features that we adopt, and we present PCA examples in Section~\ref{sec:pca_test}. 
More details about the LSST Science Pipelines software cataloged quantities are described in the LSST Data Products Definition Document (DPDD)\footnote{\url{https://lse-163.lsst.io}}.

\begin{table*} 
    \centering
    \begin{tabular}{l l l}
    \hline
    \hline
    Type & Feature index \& Name & Definition  \\
    \hline
    S/N  &
    
    0. $\texttt{|base\_PsfFlux\_instFlux|/..Err}$ & PSF flux S/N.  \\ & 1. $\texttt{|base\_CircularApertureFlux\_12\_0\_instFlux|/..Err}$ & Flux S/N within a 12-pix (radius) aperture. \\
    \hline
    Ratio &  2. $\texttt{|base\_PsfFlux\_instFlux|}/f_\textrm{tot}$ & Fractional PSF flux. \\
    & 3. $\texttt{|base\_CircularApertureFlux\_12\_0\_instFlux|}/f_\textrm{tot}$ &
    Fractional flux within a 12-pix aperture ($d=6\farcs3$).\\
    \hline
    Geometry & 4. \verb|ip_diffim_NaiveDipoleFlux_npos+..nneg| & Total number of positive and negative pixels.\\
    & 5. $|\verb|ip_diffim_NaiveDipoleFlux_npos-..nneg||$ & Difference in number of  positive and negative pixels.\\
    & 6. $-\verb|base_SdssCentroid_xErr|$ & Uncertainty of the centroid X-coordinate (opposite).\\
    & 7. $-\verb|base_SdssCentroid_yErr|$ & Uncertainty of the centroid Y-coordinate (..).\\
    & 8. $-\verb|base_SdssShape_xxErr|$ & Uncertainty of moment $I_{xx}$ (..).\\
    & 9. $-\verb|base_SdssShape_xyErr|$ & Uncertainty of moment $I_{xy}$ (..).\\
    & 10. $-\verb|base_SdssShape_yyErr|$ & Uncertainty of moment $I_{yy}$ (..).\\
    & 11. $-\sqrt{e_1^2+e_2^2}$ & Ellipticity derived from second moments (..).  \\
    & 12. $-|\sigma-\sigma_{\rm PSF}|$ & Size derived from second moments (..).  \\
    & 13. \verb|base_SdssShape_instFlux_xx_Cov| & Uncertainty  covariance between \verb|instFlux| and \verb|xx|.\\
    & 14. \verb|base_SdssShape_instFlux_xy_Cov| & Uncertainty  covariance between \verb|instFlux| and \verb|xy|.\\
    \hline
    \end{tabular}
    \caption{
    Features for R/B; they are derived from the difference image unless otherwise stated. 
    Here $e_1=(I_{xx}-I_{yy})/(I_{xx}+I_{yy})$ and  $e_2=2I_{xy}/(I_{xx}+I_{yy})$. 
    $I_{xx}$, $I_{xy}$, and $I_{yy}$ are elliptical Gaussian adaptive moments  \texttt{base\_SdssShape\_xx}, 
    \texttt{base\_SdssShape\_xy}, and 
    \texttt{base\_SdssShape\_yy}, 
    respectively. 
    The total flux $f_\textrm{tot}$ is \texttt{ip\_diffim\_forced\_PsfFlux\_instFlux}.  
    The index numbers will be used in the PCA example images (Section~\ref{sec:pca_test}). 
    The size $\sigma=I_{xx} I_{yy} - I_{xy}^2$. Another common way to describe size is $\sigma=I_{xx} + I_{yy}$, and we find the PCA performance is similar. 
    The PSF second moments are \texttt{base\_SdssShape\_psf\_xx}, \texttt{base\_SdssShape\_psf\_xy}, and \texttt{base\_SdssShape\_psf\_yy}. 
    The ``..'' refers to repeated text. 
    We skip \texttt{base\_SdssShape\_instFlux\_yy\_Cov} because it does not well separate our target from other sources (Appendix~\ref{sec:ineffective}). 
    We note that the SDSS-style shape/centroid quantities have \texttt{nan} for negative sources, and thus we skip them in those cases (from feature \textnumero6 onward). 
    }
    \label{tab:features}
\end{table*}

\paragraph{Flux S/N}

We expect a real transient to have a large flux S/N  on the difference image. We consider both PSF flux and aperture flux (default 12-pix radius).

\paragraph{Fractional flux} 

We consider a ratio between the difference image flux and the direct image (total) flux \\  ($f_\textrm{tot}=\verb|ip_diffim_forced_PsfFlux_instFlux|$). 
\rr1{In this ratio, the difference image flux is scaled by the total flux measured on the direct image; this gives a relative (instead of absolute) change of the flux.}
We expect the fractional fluxes to be large for real transients.

\paragraph{Geometric features} This group of features includes  the uncertainties of positions/moments, the number of pixels in the positive and negative lobes of the dipole, the size/shape information, and the covariance between the flux and moments. 
Compared to spurious detections, a real transient source is expected to have e.g., small measurement errors, a small ellipticity, and an area similar to that of the PSF.

\paragraph{Other features}
After testing real objects, we find that a few features make no significant contribution to separating sources by PCA. For example, the ratio between the PSF flux and aperture flux is expected to be $\sim1$ for a real (stellar) source, while spurious sources may have this value strongly departing from 1. However, our tests show that this ratio does not yield a clear separation between real versus spurious sources. 
The reason could be that the previous flag filtering step has already removed most artifacts that have anomalous photometry.   
We present further details in Appendix~\ref{sec:ineffective}.

\subsection{PCA tests}\label{sec:pca_test}

In R/B, we filter the source catalog of each difference image based on the LSST Science Pipelines software flags (Section~\ref{sec:flags}), standardize the flag-filtered catalogs, and then run PCA using the features extracted from the catalog (Section~\ref{sec:pca_theory} and \ref{sec:features}). 
We adopt a 2-component PCA; we find that higher components do not significantly improve our candidate selection.

\begin{figure*}[htb]
    \plotone{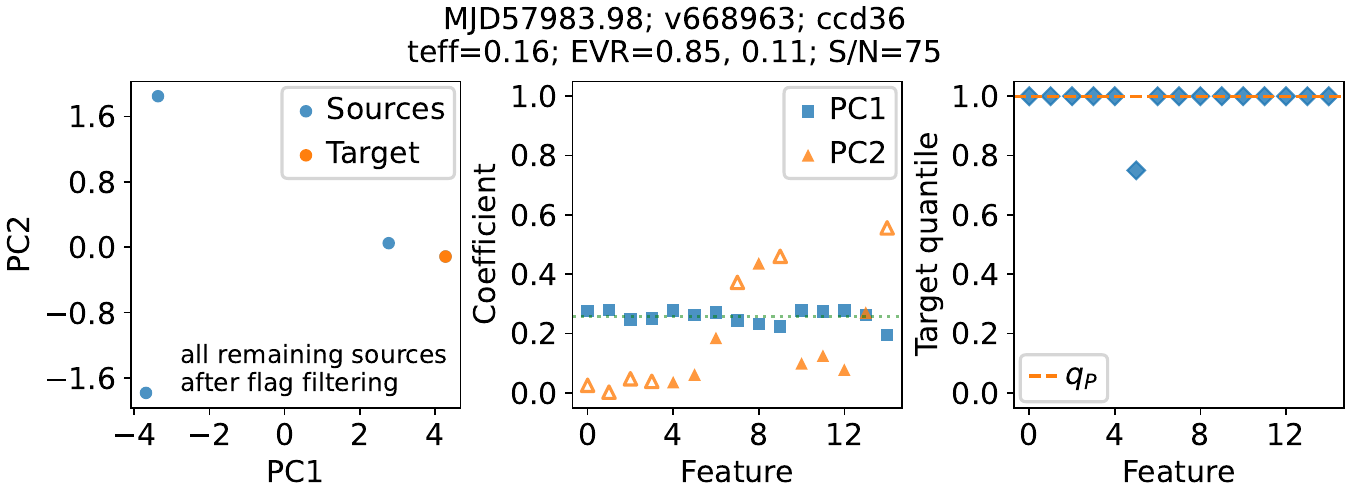}
    
    \caption{Example of running PCA on the DECam data set for GW170817 KN. \textit{Top}: \rr1{We give the observation time, visit/exposure number, and CCD number. Then, we show the effective exposure time, and }EVR means explained variance ratios of the first and second principal components. The title also lists the PSF flux S/N on the difference image. 
    \textit{Left}: The first (PC1) and second (PC2) principal component  values of sources in the transformed space. Those \rr1{points are all remaining sources after flag filtering (Section~\ref{sec:flags})}. The target (KN) is marked by an orange point. 
    \textit{Middle}: The coefficient of each feature along each principal component direction (i.e., the projection on the principal component vector). 
    Hollow markers indicate the negative values  (sign flipped). The reference horizontal  dotted line denotes one divided by the square root of the number of features ($\sqrt{1/15}\sim0.26$), showing the effectiveness of each  feature. 
    \textit{Right}: The feature quantiles of the target (i.e., the fraction of  sources that have a value for this feature smaller than or equal to that of the target).   
    The horizontal  dashed line gives the PC1 quantile of the target among sources, showing the effectiveness of candidate selection (R/B). 
    }
    \label{fig:pca_visit_kn}
\end{figure*}

\begin{figure*}[htb]
    \plotone{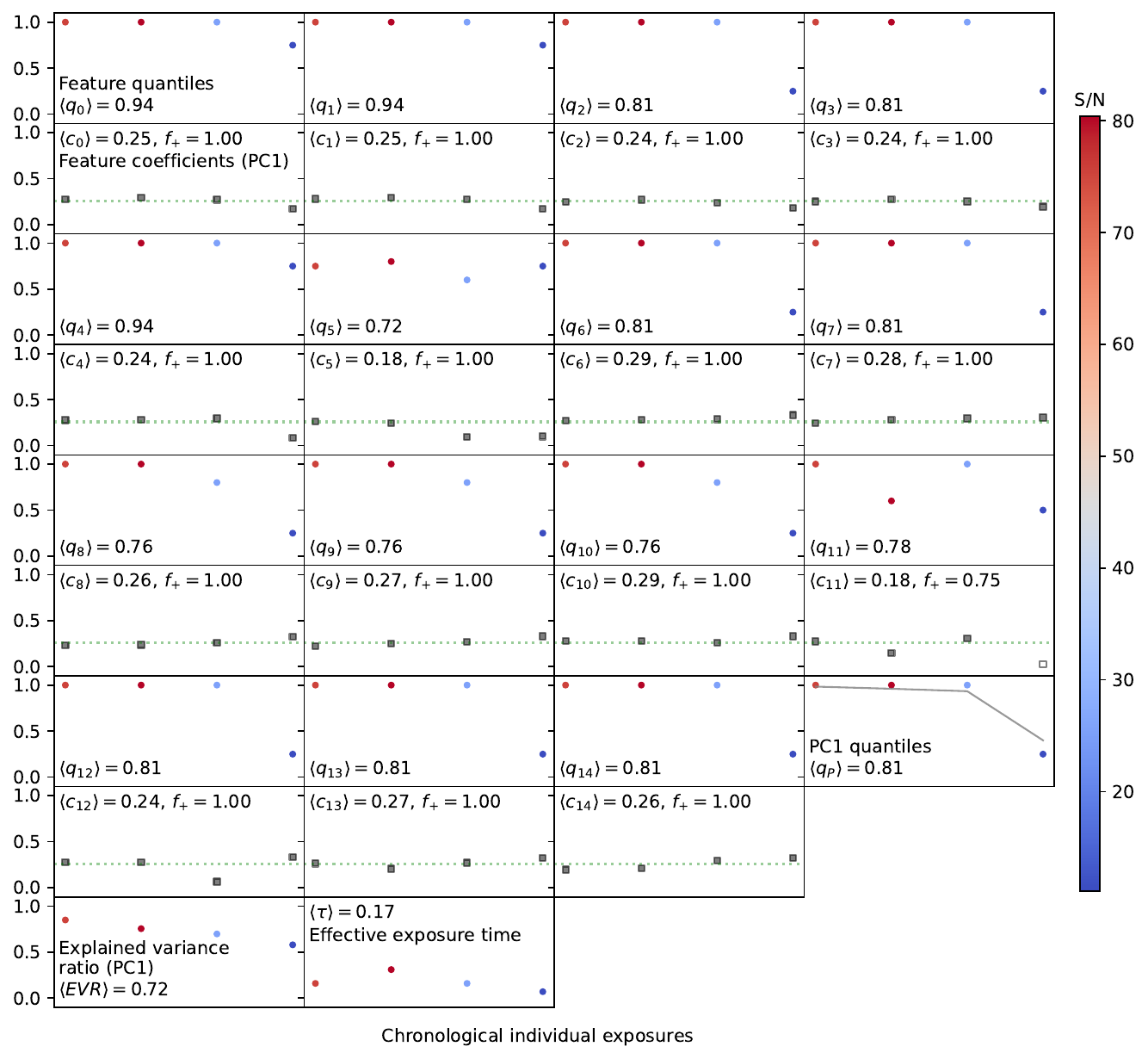}
    \caption{Evolution of the feature quantiles ($q_i$, where $i$ is the feature index) and the feature coefficients ($c_i$, on the first principal component) of the target (KN of GW170817; matched by coordinate within $1''$). 
    The X-axis corresponds to \rr1{individual exposures in chronological order (four exposures here)}, and we omit the exact \rr1{exposure} number \rr1{(or observation time)} for clarity. 
    The circular marker color shows the  PSF flux S/N (a redder color corresponds to a higher S/N) on the difference image. 
    The gray squares give the coefficients of a feature on the \textit{first} principal component for different exposures; the hollow squares give the (sign flipped) negative values; the horizontal dotted line shows one divided by the square root of the number of features. 
    We also present the time evolution of the \textit{first} component EVR, the quantile of the \textit{first} component value of the target compared to other sources ($q_{\rm P}$), and the \texttt{teff} ($\tau$). 
    Additionally, we give the mean coefficient value $\langle c_i \rangle$, with the fraction of positive values ($f_+$),  and the mean quantile value $\langle q_i \rangle$ among exposures.  
    The gray curve in the $q_{\rm P}$ diagram gives the mean of all feature quantiles of the target, in order to test the effectiveness of PCA candidate selection. 
    }
    \label{fig:quantile_evolution_kn}
\end{figure*}

\begin{figure*}[htb]
    \plotone{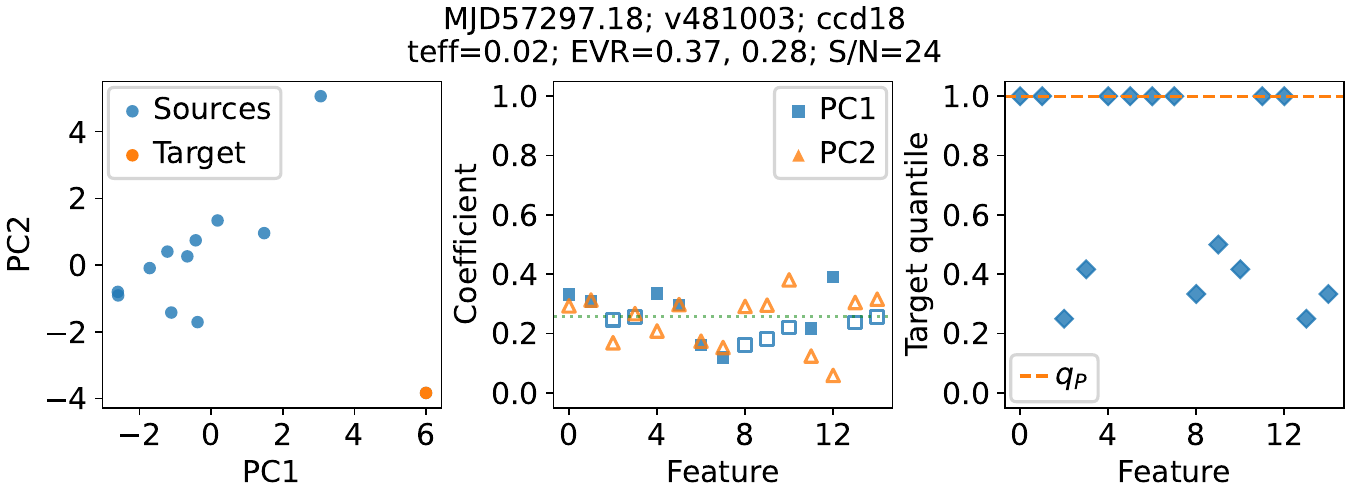}
    \plotone{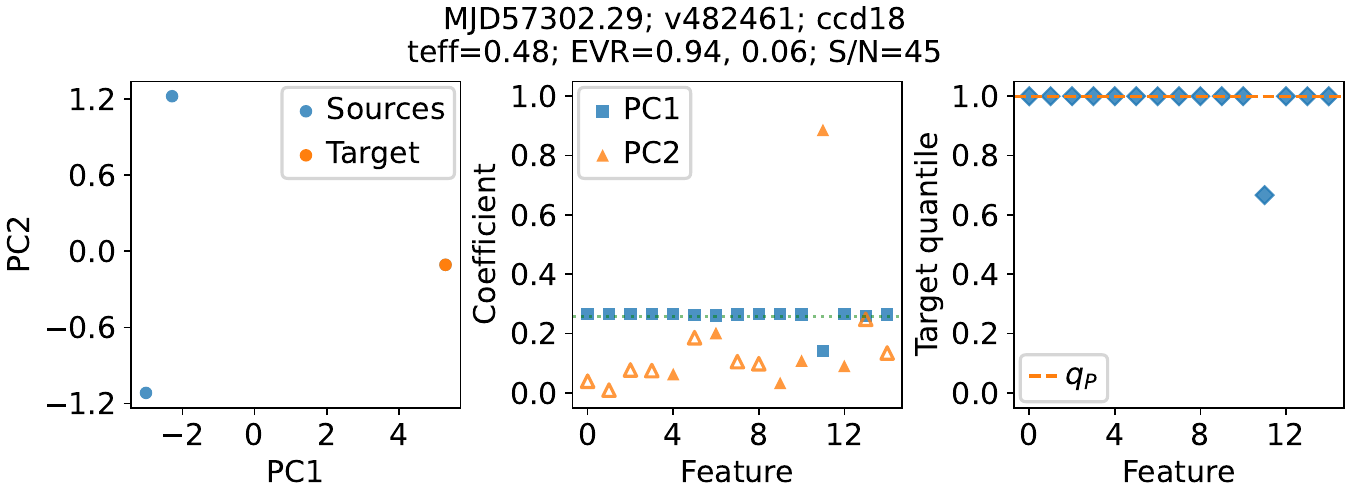}
    
    \caption{Examples of running PCA on the DES15C2eaz data (Type II SN). We use the same color scheme as Figure~\ref{fig:pca_visit_kn}. Note, even though the \texttt{teff} is low for the \textit{top} exposure, PCA still separates the target well from other sources ($q_{\rm P}=1$). 
    }
    \label{fig:pca_visit_c2}
\end{figure*}

\begin{figure*}[htb]
   \plotone{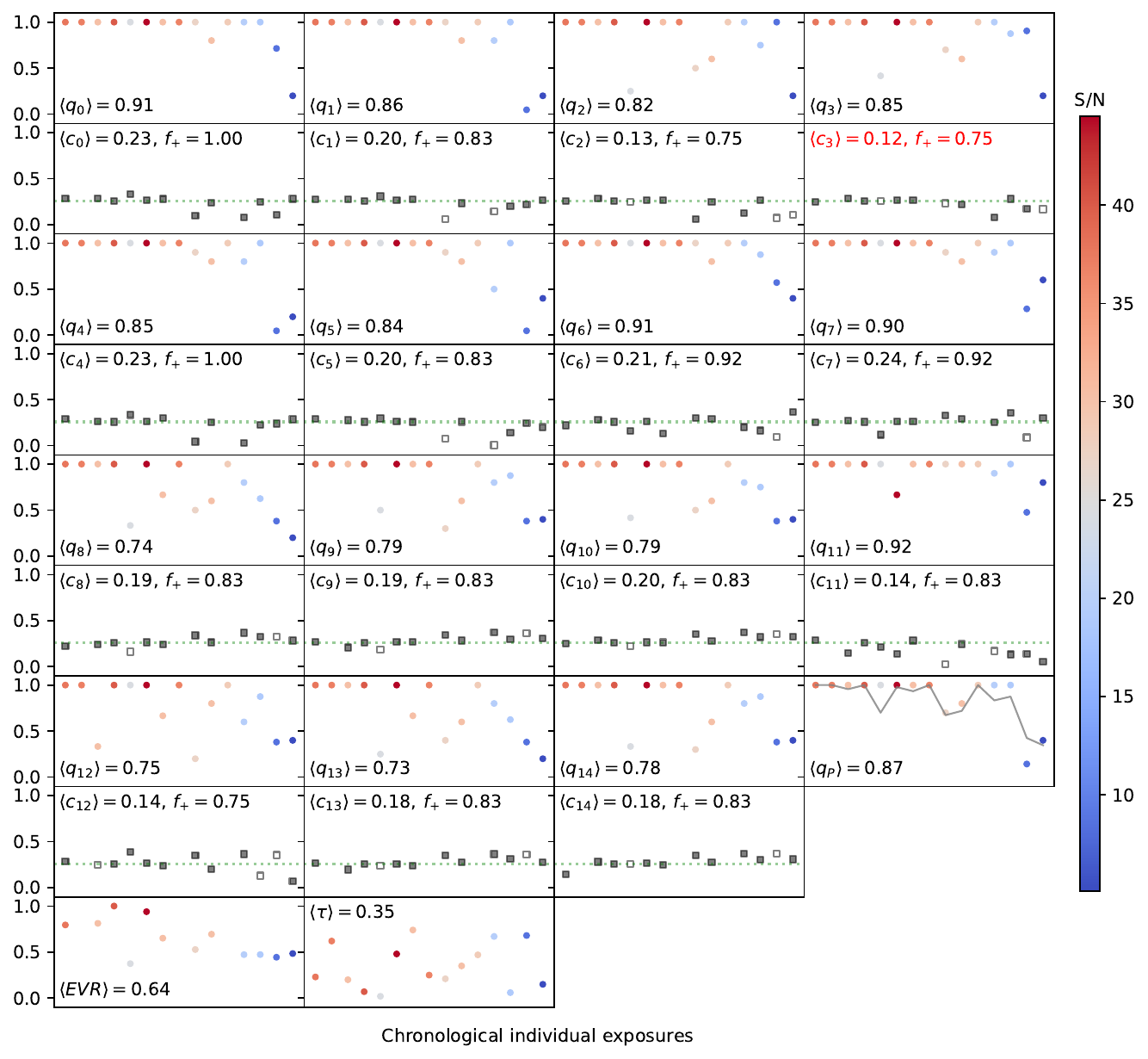}
    \caption{Time evolution of the features and PCA performance for the target (DES15C2eaz, Type II SN). We use the same color scheme as Figure~\ref{fig:quantile_evolution_kn}. 
    The exposures shown in Figure~\ref{fig:pca_visit_c2} correspond to the 4th (low \texttt{teff}) and 5th (regular \texttt{teff}) exposures here. 
    We note that for some exposures the number of sources drops to one after the flag filtering, and accordingly we set the quantiles to 1 and skip the coefficients if that source is the target. 
    All feature quantiles drop when the source is faint. 
    The features  \textnumero2, 3, 8, 9, 10, 13, 14 are sensitive to the weather. 
    Feature \textnumero3 is less effective than the others, as it has smaller $\langle c_i \rangle$ (we use red text when $\langle c_i \rangle<0.5/\sqrt{N_f}$). 
    The EVR drops as the star gets dimmer. 
    Similar to Figure~\ref{fig:quantile_evolution_kn}, the $q_{\rm P}$ is higher than the average quantile (gray curve) except at low S/N. Also, at the last point $q_{\rm P}>q_0$, indicating that PCA has the potential to perform better selection than just using the flux S/N only. 
    }
    \label{fig:quantile_evolution_eaz}
\end{figure*}

\begin{figure*}[htb]
    \plotone{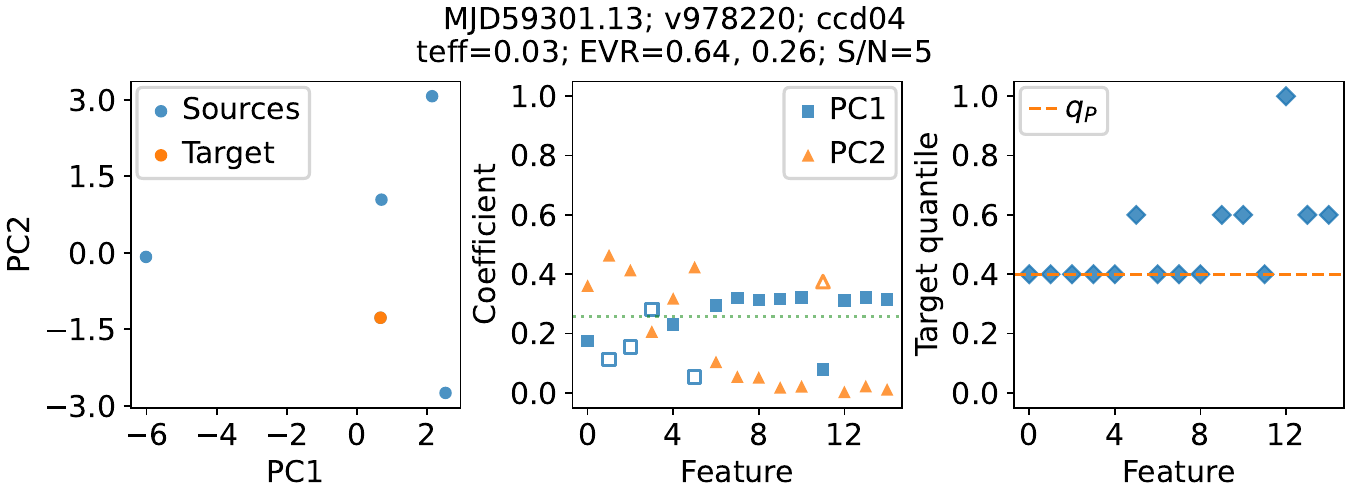}
    \plotone{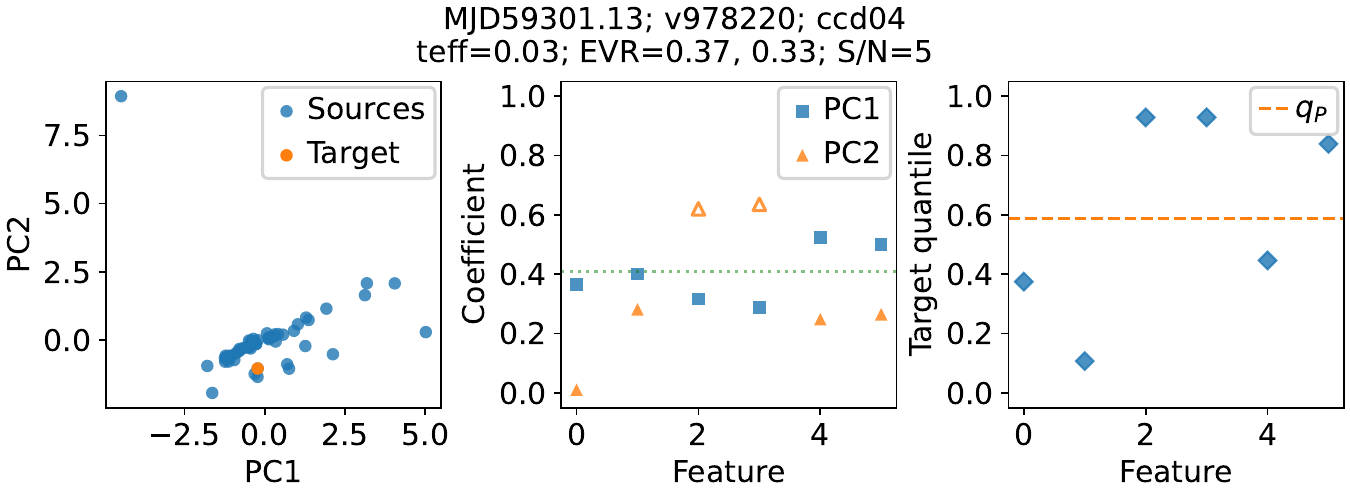}
    \caption{Examples of running PCA on the DC21bwbfe/SN2021bnv data (Type Ia SN; $r$ band). 
    We show the instances where the S/N is low (approaching the detection limit) but other features still have \textit{high} quantiles to assist the candidate  selection. At visit 978220, the low S/N is caused by poor weather, and the S/N increases afterwards even though the transient is getting fainter. Here the results are strongly contaminated by noise, and the PCA R/B shows mediocre but acceptable performance. 
    \textit{Top}: PCA result using 15 features (Table~\ref{tab:features}, Figure~\ref{fig:quantile_evolution_decat0}). 
    \textit{Bottom}: PCA result using 6 features (Figure~\ref{fig:quantile_evolution_decat}).  
    }
    \label{fig:pca_visit_dc}
\end{figure*}

\begin{figure*}
    \plotone{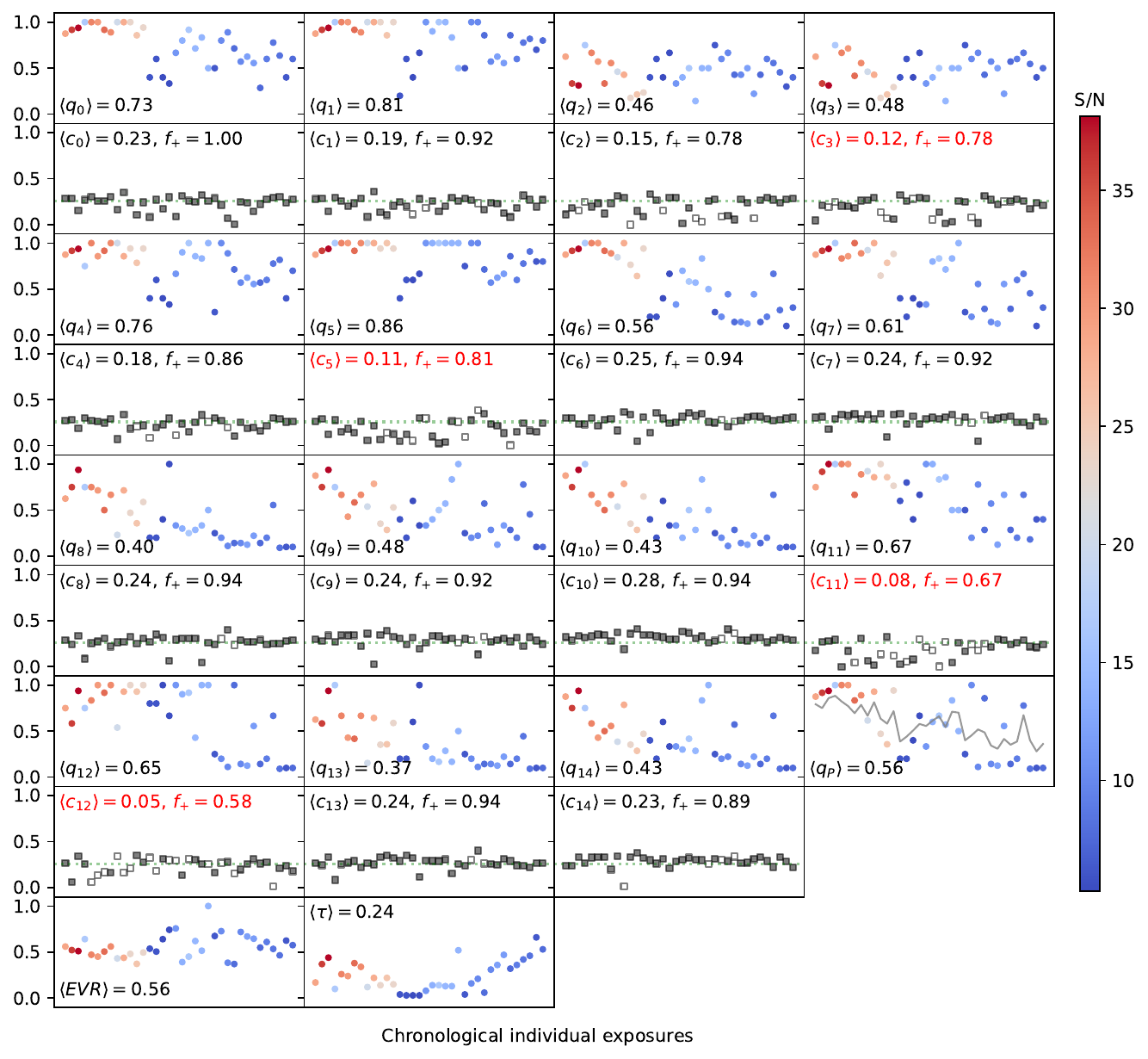}
    \caption{
    Evolution of the features and PCA performance for the target (DC21bwbfe/SN2021bnv, Type Ia SN; $r$ band); positive flux cases only. The color scheme is the same as previous figures. 
    Here features \textnumero3, 5, 11, 12 are less effective than the others. 
    Compared to Figure~\ref{fig:qe_dc_app} that uses a larger set of features, here PCA generally has better separation (higher EVR).  However, when the poor weather leads to low S/N  (blue points in the middle of the exposure sequence with low \texttt{teff}),  using more features as in Figure~\ref{fig:qe_dc_app} gives slightly better results (higher $q_{\rm P}$). 
    }
    \label{fig:quantile_evolution_decat0}
\end{figure*}

\begin{figure*}
    \plotone{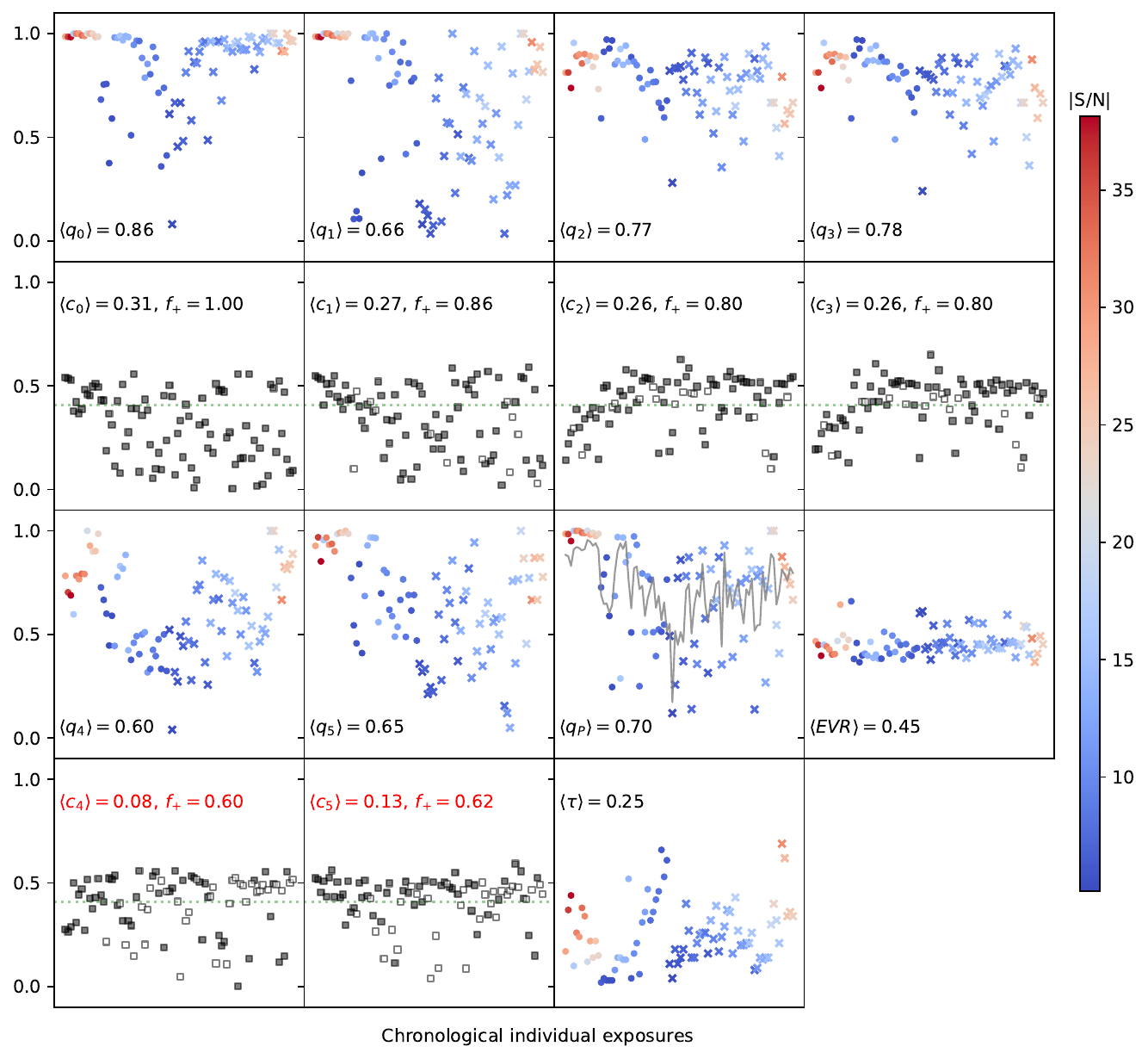}
    
    \caption{Exposures including both the positive flux cases (``point'' markers) and the negative  flux cases (``cross'' markers) on the difference images for DC21bwbfe/SN2021bnv (Type Ia SN; $r$ band). 
    Since we include the negative flux cases, we skip the SDSS-style shape and centroid flags and quantities, and the top 6 features in Table~\ref{tab:features} remain. 
    This example illustrates what happens when the features from the SDSS algorithm are unavailable. Here features \textnumero4, 5 are less effective than the others. 
    In both positive and negative flux cases, when the source is sufficiently bright (high S/N in absolute value), the PSF flux S/N itself is sufficient for the candidate selection. However, when it is faint in the positive flux cases, using more features can give better candidate selection (see also Figure~\ref{fig:pca_visit_dc}).   
    }
    \label{fig:quantile_evolution_decat}
\end{figure*}

\begin{figure*}[htb]
    \plotone{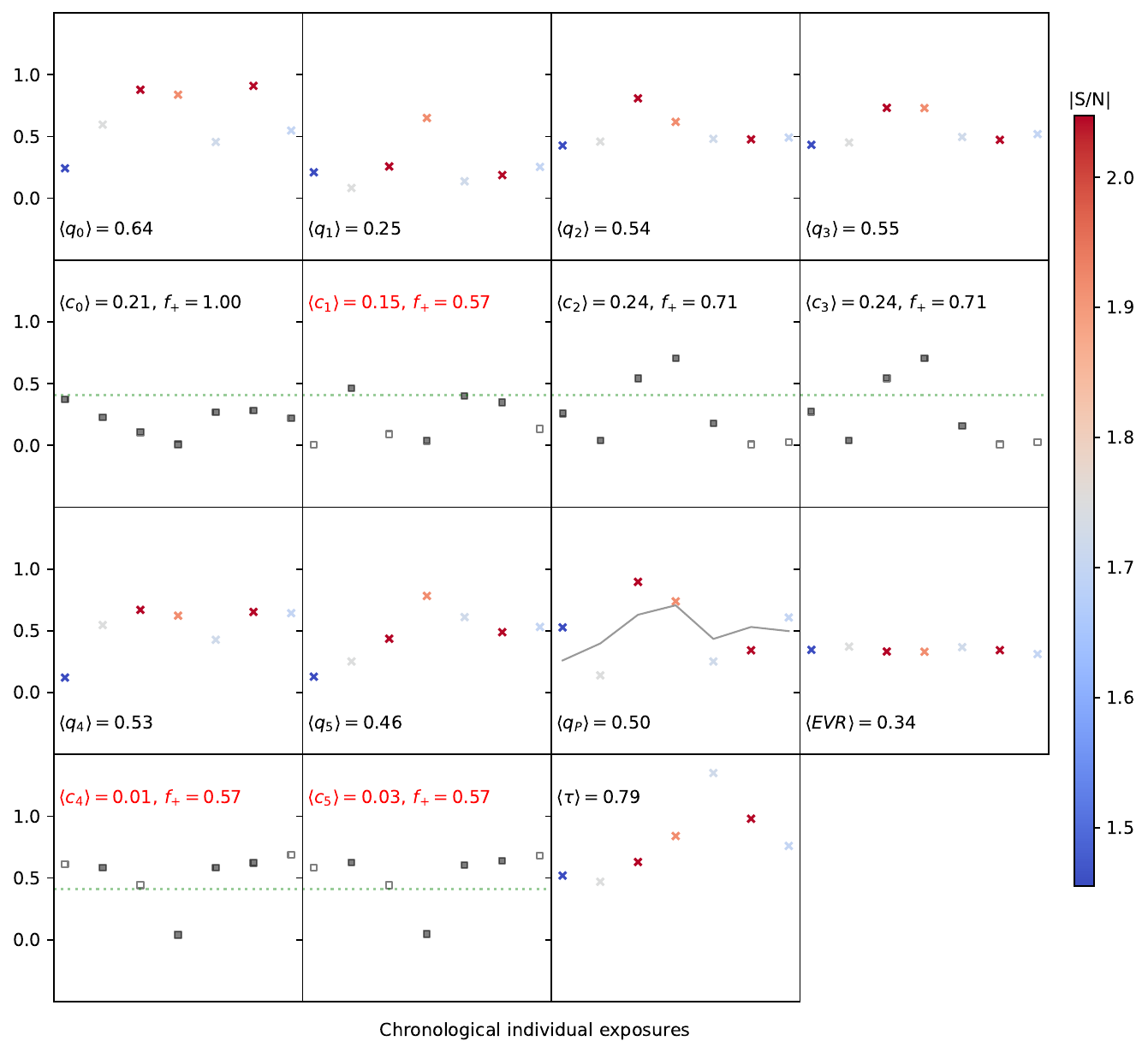}
    \caption{Evolution of the features and PCA performance for the target (DES16X3jj, possible Type II SN; $i$ band).  This object was captured by short exposures (10 sec) and has low brightness. It  generally has negative flux in our difference images. We thus only use 6 features, same as Figure~\ref{fig:quantile_evolution_decat}. The corresponding light curve is shown in Figure~\ref{fig:des_lc}. Here features \textnumero1, 4, and 5 are less effective than the others. The first point has low S/N, but the feature quantiles \textnumero2 and 3 (and $q_{\rm P}$) are still high. 
    }
    \label{fig:quantile_evolution_x3}
\end{figure*}

Here, we examine the PCA component values (Figure~\ref{fig:pca_visit_kn},~\ref{fig:pca_visit_c2},~\ref{fig:pca_visit_dc}) and the feature quantiles of real objects among all sources in the archival exposures, with the goal of checking how well the real objects ``stand out'' from the others.  
Additionally, we study how those feature quantiles evolve from exposure to exposure to test the feature robustness (Figure~\ref{fig:quantile_evolution_kn},~\ref{fig:quantile_evolution_eaz},~\ref{fig:quantile_evolution_decat0},~\ref{fig:quantile_evolution_decat},~\ref{fig:quantile_evolution_x3}).  
We present the details below. 

\paragraph{GW170817}
First, we consider the $g$-band exposures of the KN (Section~\ref{sec:kn}). 
Figure~\ref{fig:pca_visit_kn} shows an example of the PCA classification for the first exposure at the first night. 
We match our flag-filtered catalog with the known coordinates of the KN to locate it (same in the following tests); if a  bright transient  with known coordinates is not in the catalog, it is likely caused by poor weather. 
Here, the KN (target) shows a much larger first component value compared to other sources (the \textit{left} panel). 

In the title, the explained variance ratios (EVR) indicate how well the sources are separated -- a larger difference between those ratios means a better split of the data (and thus a better selection of the candidates), as the data variance can be more captured by the first component.  

In the \textit{middle} panel, 
the linear coefficient of each feature for the PCA component (i.e., its proportion along the PCA component direction) indicates how important the feature is and whether the feature has a positive/negative effect on the PCA component. 
We find that most features do affect the first component value (PC1) positively and uniformly -- their coefficients are $\sim1/\sqrt{N_f}\sim0.26$, where $N_f=15$ is the number of features. 
For the second component (PC2), about half of the features have negative effects and most features have small absolute coefficient values (i.e., they are ineffective). 

The \textit{right} panel shows the quantile of each feature of the target (compared to other sources) -- the quantile values are generally high, which makes the target distinct from other sources. 
Note, we define the quantile value as a normalized \textit{rank} for $N_s$ sources: for a set of $N_s$ elements $\{x_1, x_2, ..., x_{N_s}\}$, the quantile value of $x_j$ is the number of elements smaller than or equal to it then divided by $N_s$, so that it is straightforward to compute this value without interpolation.   

In Figure~\ref{fig:quantile_evolution_kn}, we study the time evolution of feature quantiles of the target, similar to our previous tests on flags.  
The $q_{\rm P}$ diagram indicates that the PC1 rank (points) provides better candidate selection than the mean of feature quantiles (gray curve) when the target is sufficiently bright. 
In addition, a candidate selection can be made by adding a cut at $q_{\rm P}\gtrsim0.5$.  

\paragraph{DES15C2eaz}
 
Compared to the KN, the DES15C2eaz (Type II SN) exposure sequence has very low \texttt{teff} when the transient was still bright (Figure~\ref{fig:des_lc}), and Figure~\ref{fig:pca_visit_c2} shows the transition from low \texttt{teff} to high \texttt{teff}.   
The \textit{top} panel of Figure~\ref{fig:pca_visit_c2} shows an example of running PCA on the observational data captured under poor weather conditions ($\texttt{teff}=0.02$). The first principal component still successfully separates the target from others in the positive direction (\textit{top left}), mainly because most of its features are larger than other sources (\textit{top right}).  
The EVR of the first component (and the sum of the EVR of the two components) is lower than other exposures, which means the sources are less separated. Also, even in the first component, some features exhibit  negative effects (\textit{top middle}). 
These results indicate that a weather metric, such as \texttt{teff}, is essential and worth being included as a \textit{weight} for the candidate selection. 
The \texttt{bottom} panel, on the other hand, shows an opposite case which happened a few days later. Though the target is less bright (Figure~\ref{fig:des_lc}), the PCA separation is much cleaner. Compared to the the \textit{top} panel, fewer sources are included after the flag filtering, indicating that many sources in the \textit{top left} diagram are likely spurious. 
The target now has the largest values for almost all features compared to other sources, leading to positive and nearly equal coefficients in the first component. 

Figure~\ref{fig:quantile_evolution_eaz} shows the feature quantiles of the target among sources. 
We note that the PCA first component quantile ($q_{\rm P}$) has better performance than the average of feature quantiles for selecting out the target. Also, when the target is faint, using multiple features seems to select out the target more easily than only using the S/N.

\paragraph{DC21bwbfe (SN2021bnv)}
The DECam exposures of DC21bwbfe (Type Ia SN) were taken after the explosion. Its flux drops in the exposures, and on the difference images its flux changes from positive to negative (Figure~\ref{fig:decat_lc_bwbfe}). It is useful for testing the transition of the R/B performance. 

Noting that the SDSS-style flags and quantities do not work for the cases of negative flux, we run R/B in two approaches. First, we use all features but only on positive flux exposures (Figure~\ref{fig:quantile_evolution_decat0}). Second, we use the features without the SDSS-style ones on all exposures (Figure~\ref{fig:quantile_evolution_decat}). 

Similar to the previous target (Figure~\ref{fig:pca_visit_c2}), here we also study cases when difficult observing conditions lead to poor source detection and measurement. 
In Figure~\ref{fig:pca_visit_dc}, we show examples when weather is poor ($\texttt{teff}=0.03$) and the (measured) S/N of the target is close to the detection limit, but the candidate selection still has acceptable performance ($q_{\rm P}\sim0.5$). The true brightness of the SN is likely still high at that time according to the light curve trend (Figure~\ref{fig:decat_lc_bwbfe}). Also, though the flux S/N of the target  has low quantiles (the first 2 features), some other features have high quantiles. 
This example shows that using multiple features is more robust for candidate selection than using a single feature, especially at low S/N. 

In Figure~\ref{fig:quantile_evolution_decat0} and Figure~\ref{fig:quantile_evolution_decat}, we study the time evolution of the features and the PCA performance. 
In Figure~\ref{fig:quantile_evolution_decat}, we skip the SDSS-style shape/centroid flags and quantities because of the negative flux cases on the  difference images. 
Again, from those examples we find that when the target has high S/N under high \texttt{teff}, the flux S/N quantile is sufficient for candidate selection; when the target has low S/N due to low \texttt{teff}, using more features gives better selection.

\paragraph{DES16X3jj}
In our last example, we consider an extreme case -- a faint SN that is decreasing the brightness and has mostly negative flux values on the difference images. 
Figure~\ref{fig:quantile_evolution_x3} shows similar behavior to the previous ones -- when the S/N is low (especially under low \texttt{teff}), using multiple features other than the flux S/N only (the first data point) yields better selection. 
The features we use generally have positive effects on selecting the target (positive coefficients). Also, PCA gives slightly better selection than using the mean of quantile values only -- the mean of the average quantile (gray curve) is $0.49<\langle q_{\rm P}\rangle=0.5$. 

\paragraph{Summary}
From the quantile evolution tests above we conclude that real transients can be well selected ($q_{\rm P}\gtrsim0.5$) when they are sufficiently bright (S/N $\gtrsim15$, with the default detection limit $5\sigma$). 
Additionally, those tests show that when a source is bright, only using a cut on the quantile of PSF flux S/N can select the source successfully. However, when the source is faint (and under poor weather), using multiple features  would be able to select the source more successfully. 
The effectiveness of each feature depends on the target. 
At low S/N, using the mean (or median) of feature quantiles instead of ML results may also give acceptable candidate selection. 
Though we have only tested PCA in this work, our conclusion about features may be applicable to other ML methods. In the future, we will test other  R/B algorithms and evaluate their performance using artificial source injection~\citep[e.g.,][]{Brennan2022,Everett2022} with some known input flux and customized host/environment (Section~\ref{sec:source_injection}).

\subsection{Serendipitous source detection}\label{sec:serendipitous}

In the PCA run of a single exposure, 
what are the sources that have high first principal component values (PC1) but are not the target (Figure~\ref{fig:pca_visit_kn}, ~\ref{fig:pca_visit_c2}, ~\ref{fig:pca_visit_dc}; the left panel)? 
We inspect those sources, and we find that they correspond to real objects (we regard them as serendipitous detections) or noise. 

\begin{figure*}
    \plotone{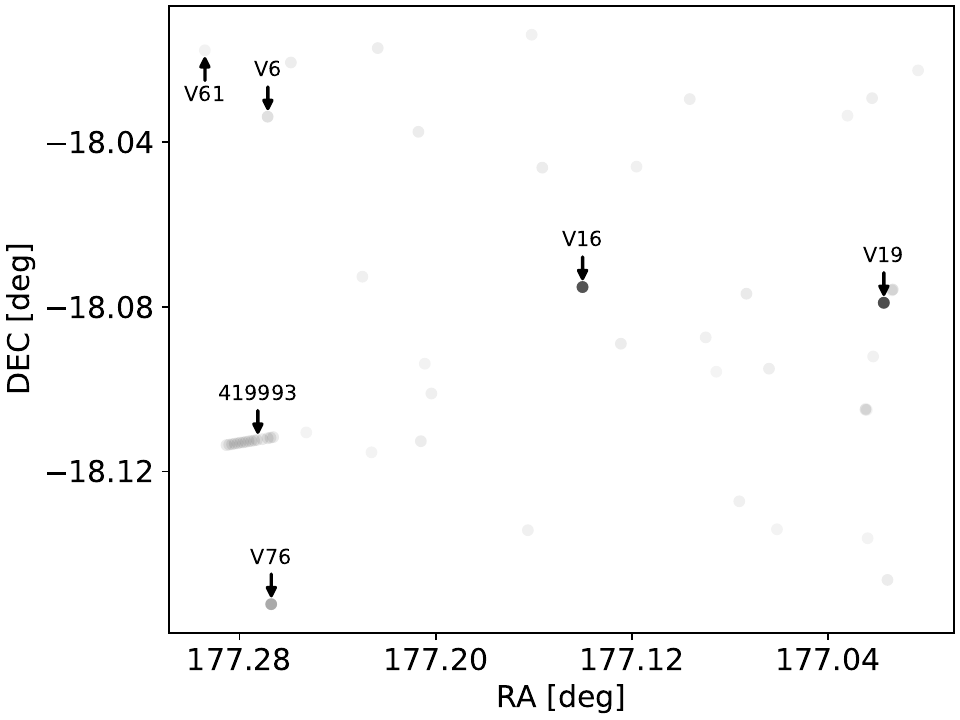}
    \caption{PCA results  in the Crater II field, including an asteroid (419993/2011CW32). We plot the sources with PC1 quantile larger than 0.5 and combine the results of 18 exposures (visit 632364 to 632398 on CCD 21, 3 min exposures in $g$ band each). The corresponding MJD range is from 57832.232 to 57832.314, and the \texttt{teff} range is from 0.08 to 0.22. 
    We annotate the variables~\citep{Vivas2020} and mark the asteroid (moving from SE to NW).  
    The marker \rr1{opacity} is proportional to the sum of PC1 quantile of each source in individual exposures. 
    V6 and V61 have relatively low brightness during the observation window. 
    }
    \label{fig:asteroid}
\end{figure*}

In the first example, we inspect the Crater II field data. We use the default 15 features for PCA (Table~\ref{tab:features}). In the PCA result (Figure~\ref{fig:asteroid}), the variable stars stand out as expected. For the remaining sources with high PC1, clearly, an object is moving with a pattern. We find that it corresponds to an asteroid (419993/2011CW32); during the exposures it is $\sim20$ mag.\footnote{\url{https://minorplanetcenter.net/db_search/show_object?object_id=419993}} 
Note, some extra flags/quantities related to dipoles in the LSST Science Pipelines software might also be useful for detecting asteroids.

\begin{figure*}
    \plottwo{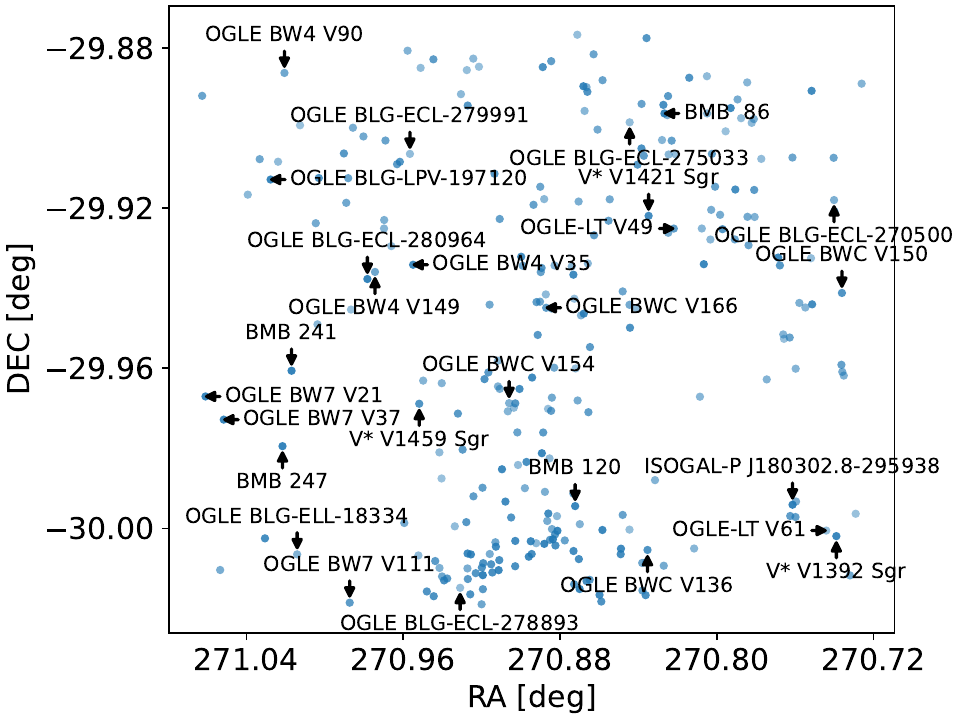}{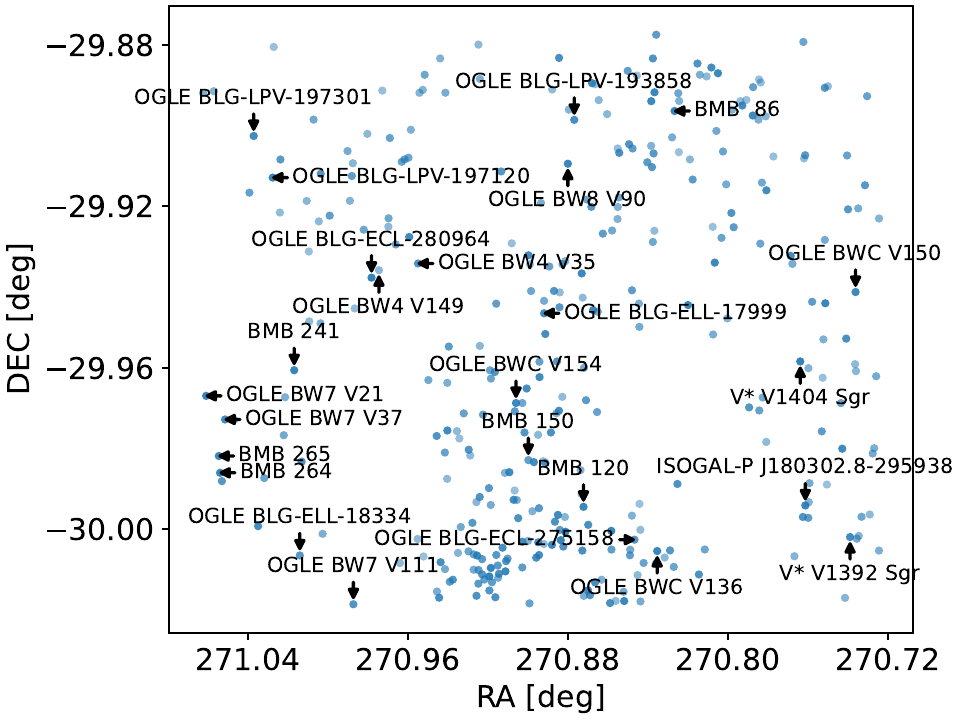}
    \caption{PCA results in the Bulge field~\citep{Saha2019}, which includes many 
    variable stars.   
    We show visits 428030 (\textit{left}) and 429041 (\textit{right}) on CCD 28 in $r$ band, and plot the sources \rr1{that have the} PC1 quantile larger than 0.5; \rr1{the marker opacity is proportional to the PC1 quantile of each source}.    
    The corresponding exposure times in MJD are 57115.264 and 57116.403 (the interval is $\sim27$~hr). 
    We annotate the objects recorded in SIMBAD~\citep{Wenger2000}. 
    Their stellar types can be identified by name. 
    ``OGLE BWC V154'' corresponds to ``OGLE-BLG-ECL-27776'' studied earlier (light curves in Figure~\ref{fig:bulge_lc}). 
    ``BMB'' indicates AGB stars~\citep{Blanco1984}, and ``ISOGAL-P J180302.8-295938'' is an AGB star~\citep{Ojha2003} as well. 
    Some short period stars were also captured because of the timing of their phases, e.g., those names beginning with ``OGLE BLG-ECL''. 
    The source detection depends on weather and instrumental conditions.  
    The detected sources without annotation could be noise (especially near CCD edges or saturated stars), or previously uncataloged real objects. 
    }
    \label{fig:agb}
\end{figure*}

In the second example, we inspect the Bulge field. 
We tested both the exposures from~\citet{Saha2019} and~\citet{Graham2023}, which have 5-sec and 50-sec exposure times respectively, and find the results are similar. Since long exposure time may cause stars to be affected by saturation more easily, and previous studies (and their catalogs for our comparison) focus more on bright stars, we prioritize the 5-sec exposures here.    
Noting that flag \texttt{base\_PixelFlags\_flag\_suspect} is often \texttt{True} due to high source density in this region, we skip this flag here. In addition, we only keep the first 6 features in Table~\ref{tab:features}, because they generally have higher quantiles compared to others here. 
In the PCA result (Figure~\ref{fig:agb}), short period variable stars -- the objects we are interested in and presented in Figure~\ref{fig:bulge_lc}, stand out. Some long period variable stars, such as asymptotic giant branch (AGB or AB) stars stand out as well. The reason is that the exposures constituting the template are much earlier than the new exposure (by 1--2 years). This example shows the potential of our algorithm to detect long period variables or AGN. 

Going further, if a source repeatedly shows up in different exposures (at the same coordinates or moving in a clear pattern), it probably belongs to a real object, because noise would appear at random locations. This method can be used to identify transients spanning multiple exposures (Section~\ref{sec:weight_count}).

The number of serendipitous sources may grow when the exposure depth improves, because the number of sources with sufficient S/N has increased. 
In the future, we will test detecting faint transients using the image subtraction between deep coadded images of DECam (instead of the image subtraction between a single exposure and a coadded image), and we expect more serendipitous sources to appear. 
This analysis will also be useful for studying future deep surveys such as LSST. 

\subsection{Grouping per-exposure candidates on the sky}\label{sec:weight_count}

The goal of our MMA pipeline is to select candidates in each exposure and send them to a broker. The broker analyzes those candidates captured in different exposures and performs a further classification, and then distributes the information to other telescopes. 
Though our pipeline is designed to examine individual exposures, here we consider an algorithm (as a ``mini-broker'') which collects and summarizes our candidates from multiple exposures. 

First, we divide the sky into super-pixels (e.g., $\lesssim1'$).  
Then in each super-pixel, we count the number of candidates over all exposures. 
This gives a cumulative distribution (in celestial coordinates) of candidates over time -- a 2D histogram over the plane of the sky.  
Next, we expect that the super-pixels with high counts correspond to real objects (like ``signals'' on the 2D histogram), and bogus/artifacts may show up at random super-pixels (like ``noise'' on the histogram); objects with high peculiar motions should present distinctive patterns on the histogram. 
Finally, we can look into those special super-pixels to locate the objects of interest. 
Therefore, we can use this algorithm to test R/B and search for transients.  

Taking this further, we can apply weights to those number counts, and then the final 2D histogram shows the weighted sum of counts from individual exposures. 
For one exposure, we can use the \texttt{teff} or the first component EVR as the (same) weight of all candidates in that exposure.  Also, we can use the PC1 quantiles (normalized ranks) of the candidates in that exposure as alternative weights. 
We expect the candidates to be more credible when those weights have larger values. 
If available, external information about the host galaxies (how likely they contain transients) or the stars (how likely they are variable) can also be used as weights. Although we do not consider multi-band information here, it may improve the grouping of single-exposure candidates as well.

\begin{figure*}[htb]
    \plotone{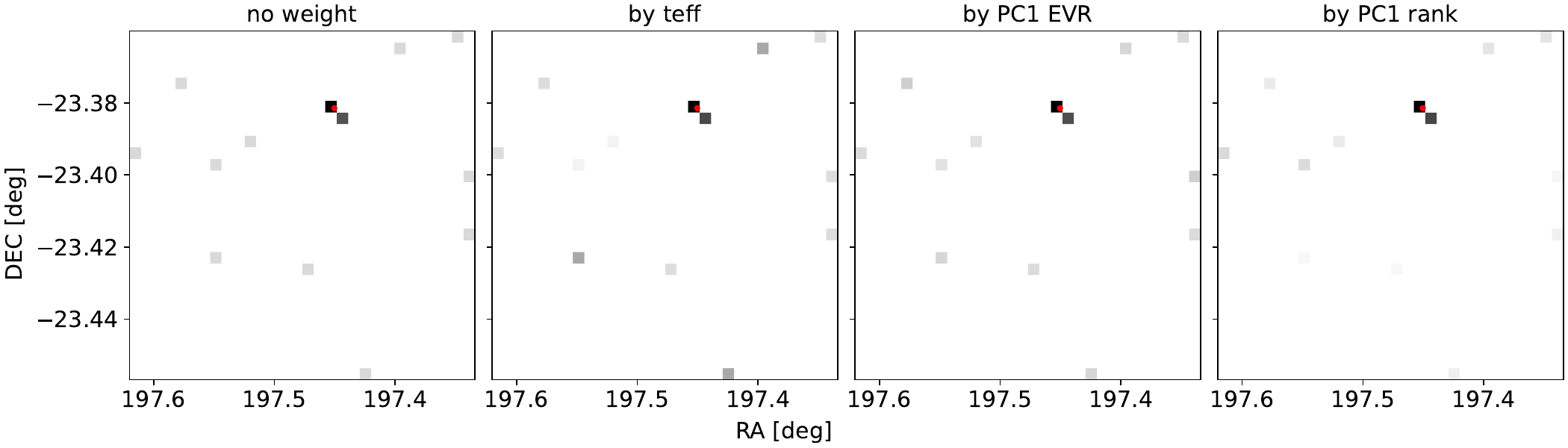}
    \plotone{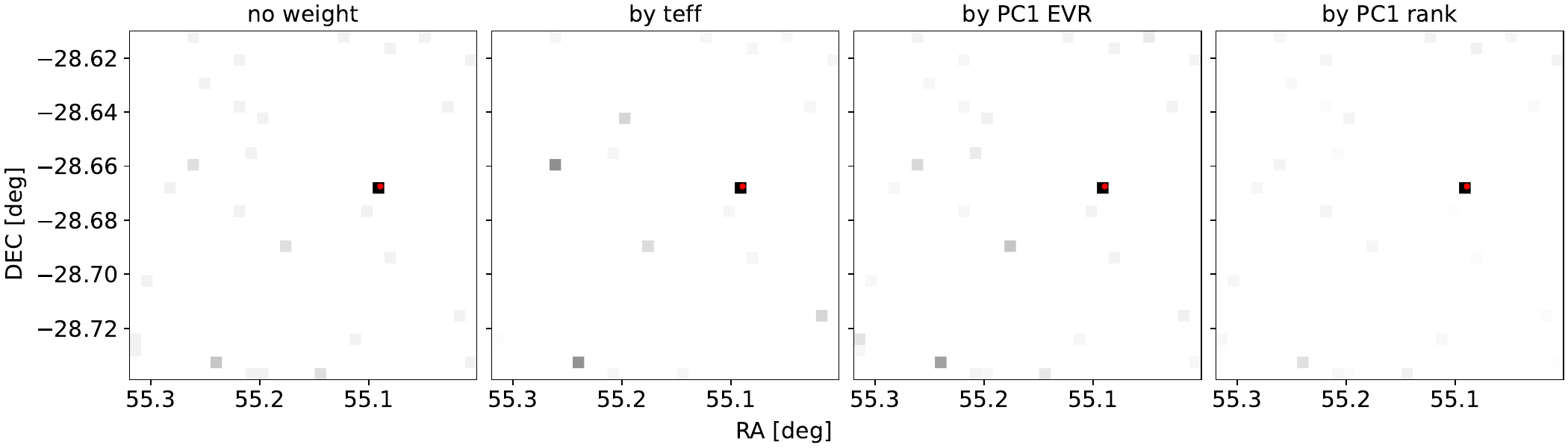}
    \caption{2D histograms for PCA selected sources from multiple exposures of the same sky region. The targets are the KN of GW170817 (\textit{top} row; combining the first 4 exposures at CCD 36 in $g$ band) and the DES15C2eaz SN (\textit{bottom} row; the first 9 valid exposures at CCD 18 in $r$ band). 
    We show the counts with no weight (\textit{first} column), weighted by \texttt{teff} (\textit{second} column), by PCA first component EVR  (\textit{third} column), and by first component quantile/rank  (\textit{last} column). 
    In each diagram, a darker color corresponds to a higher count, and the red dot marks the (true) location of the target. Note that there is a possible variable star near the KN. 
    For the SN, if there is only one source after the flag filtering, we set the EVR weight to be zero.  
    \label{fig:count_2d}
    }
\end{figure*}

We present examples in Figure~\ref{fig:count_2d}. The super-pixel with the highest count corresponds to our target, and the PC1 rank weighting gives the cleanest result. 

\section{Discussion}\label{sec:discussion}

\subsection{Input weights for sources and features}

In PCA, we standardize feature values across different sources, compute the covariances between the (standardized) features, and then diagonalize the covariance matrix to get eigenvalues and eigenvectors; the eigenvector with the highest eigenvalue gives the first principal component. 
When we compute the covariances between features, we may consider Eq.~\ref{eq:cov} and~\ref{eq:mean}, where $\{s_i\}=\{s_1,s_2,..., s_{N_s}\}$ are $N_s$ sources, $w_i$ is an input weight of the $i$th source, $\alpha$ and $\beta$ are arbitrary indices of features, and $v_\alpha$ is an input weight of the corresponding feature $f_\alpha$. Note, those weights are different from the weight derived from \texttt{teff}, EVR, and PC1 mentioned earlier. 
By default, here we set $w_i=v_\alpha=v_\beta=1$, but we may adjust them to emphasize the importance of some sources or features. For instance, a source is more likely to be a real transient if it has high brightness, or it is close to a possible host; the input weight of a feature may be determined by archival observations or image simulations. 

\begin{equation}
    \mathrm{Cov}[f_\alpha,f_\beta] = \frac{1}{N_s}\sum_{i}^{N_s} w_i v_\alpha v_\beta [f_\alpha (s_i)-\bar{f}_\alpha][f_\beta (s_i)-\bar{f}_\beta]
    \label{eq:cov}
\end{equation}

\begin{equation}
    \bar{f}_\alpha = \frac{1}{N_s}\sum_{i}^{N_s} w_i f_\alpha (s_i)
    \label{eq:mean}
\end{equation}

\subsection{Classifiers}\label{sec:classifiers}
In this work, we employ a straightforward unsupervised learning method to perform Real/Bogus classification. 
Our algorithm can easily be applied to other instruments (especially the upcoming LSST), and the flags and features we selected may be useful for other classifiers. 
The advantages of using unsupervised learning are that it does not rely on the observing or instrument conditions, and it does not require a specific training set. 
Those benefits speed up the construction of the classifier.  
Nonetheless, we do need to use known real objects in observations to determine flags for removing artifacts and to determine classifier features, and to use the weather information as a weight for the R/B result.

On the other hand, since the unsupervised learning method is more generic and not specific to exposure conditions, in some cases it might be less accurate compared to a more sophisticated supervised learning method. 
In the future we will test (and possibly train) more advanced classifiers for the data taken under specific circumstances. 
To achieve that, we are currently building infrastructure to perform large-scale data processing with our DECam MMA pipeline, so that we can efficiently test different algorithms on various data sets. 
The observing conditions that could affect the new classifier may include \texttt{teff}, seeing, sky brightness, airmass, and clouds, and the instrument conditions may include exposure time, saturation, CCD edges/response. Blendedness and source number density may also be important dimensions to explore.
Additionally, testing templates constructed from exposures taken under different conditions requires future large-scale data processing. 
\rr1{It is also possible to reanalyze the processed images generated by the LSST Science Pipelines, and to study the detailed features of sources and classify sources at the image level (instead of the catalog level presented in this work), using Neural Networks (Section~\ref{sec:intro}) or analytical methods~\citep[e.g.,][]{Ackley2019}.}
We have started some initial studies for those tests and will present more details in future publications.

\subsection{Source injection}\label{sec:source_injection}

In this paper, we use real objects that change flux to validate the performance of source detection, photometry, and selection, and assess these pipeline components as a function of exposure conditions. 
An alternative approach could be injecting sources with known fluxes/positions, as we do not generally know the true fluxes of our real test sources in the DECam observations (only the measured flux), though the injected sources might not be realistic enough. 
An injected source can either be synthetic (such as a model PSF) or the stamp of a source cropped from a real exposure, or even be an artifact (for testing R/B).  
The input flux of the injected source needs to be scaled to match the observing conditions and the CCD response of the real exposure which is being injected \rr1{into}. 
\rr1{The injected stars may have different sizes/shapes imitating representative PSFs derived from real observations. Also, we could categorize observed sources into groups based on their PCA results, and then ingest (the stamps of) representative sources from those groups. }
It is also possible to simulate a DECam exposure directly and run image processing on that, but injecting sources \rr1{into} real exposures can provide more realistic results. 

Source injection may help to determine the detection limit and completeness to find out the ``true-negative'' cases of MMA sources under different conditions. However, the corresponding source injection tests would require the large-scale processing mentioned above,  because various types of conditions and sources in real exposures need to be considered, and therefore we leave that study to future work. 

The source injection will also help us to determine the ``false-positive'' (FP) rate in R/B.   
Some selected candidates could be artifacts/noise, but it might not be easy to find all of them in real observations, because there are uncataloged real objects and sometimes it is difficult to recognize artifacts by eye. 
In addition, fast transients (e.g., KN, stellar flares) can be hard to recognize if captured in very few exposures. 
A better estimation of FP thus requires source injection -- this is beyond the scope of this work and will be studied in the future.   
On the other hand, for variable stars that are included as candidates, we expect them to be easily classified by the broker, because they repeatedly show up in difference images and are usually cataloged by previous studies.

\section{Summary}\label{sec:summary}

This paper presents the framework of an MMA pipeline for DECam (Section~\ref{sec:pipeline}). We describe using the LSST Science Pipelines to process raw DECam data and produce catalogs, visualizing difference images and light curves for photometry validation (Section~\ref{sec:examples}), classifying single-exposure sources as Real/Bogus via a new algorithm (with 11 flags and 15 features), and generating candidates that will be sent to downstream brokers (Section~\ref{sec:rb}). We show examples of running the pipeline on  archival DECam data for various types of transients/variables (13 objects in 4 types, Table~\ref{tab:target_list}), and discuss potential future improvements to our pipeline (Section~\ref{sec:discussion}).

In Section~\ref{sec:examples}, we plot the weather metrics together with corresponding light curves, providing some intuition for the uncertainties and fluctuations seen in these light curves.
Additionally, using archival DECam data, we show clean results of image subtraction and difference imaging photometry generated by the LSST Science Pipelines software -- these encouraging results underscore the great potential of the LSST survey and software in the future. 

Our R/B algorithm operates  on single exposures and uses an unsupervised classifier to deal with various conditions in real observations. 
The algorithm does not strongly depend on the source type and thus may be applicable to other user cases, such as discovering early-stage SN to enable fast alerts and rapid follow-up observations. 
Our flags and features (Section~\ref{sec:rb}) are comparable and complementary to the ones derived from simulations~\citep{Liu2024}, and can be used as a reference for other instruments including LSST and other R/B algorithms. Future generations of the DECam MMA pipeline will be built upon the flags and features presented in this work.

Future work includes experimenting with more advanced LSST Science Pipelines software and classifiers, different approaches to selecting exposures for making templates, and MMA source injection.

\section*{Acknowledgements}
\rr1{We thank the anonymous reviewer for constructive comments that have helped improve the paper.}
We thank Abhijit Saha, Melissa Graham, Monika Soraisam, Shu Liu, Meredith Rawls, Lee Kelvin, Kathy Vivas, Armin Rest, Tod Lauer, Alejandro Clocchiatti Garc\'ia, John Parejko, Anthony Englert, 
Kenneth Herner, and the LSST Data Management (DM) Team for comments and support. 

This work is supported by NOIRLab, which is managed by the Association of Universities for Research in Astronomy (AURA) under a cooperative agreement with the U.S. National Science Foundation. 

This paper makes use of LSST Science Pipelines software developed by the Vera C. Rubin Observatory. We thank the Rubin Observatory for making their code available as free software at \url{https://pipelines.lsst.io}.

Part of this research was conducted using computational resources and services at the Center for Computation and Visualization, Brown University. 

This work has made use of data from the European Space Agency (ESA) mission
{\it Gaia} (\url{https://www.cosmos.esa.int/gaia}), processed by the {\it Gaia}
Data Processing and Analysis Consortium (DPAC,
\url{https://www.cosmos.esa.int/web/gaia/dpac/consortium}). Funding for the DPAC
has been provided by national institutions, in particular the institutions
participating in the {\it Gaia} Multilateral Agreement. 

The Pan-STARRS1 Surveys (PS1) and the PS1 public science archive have been made possible through contributions by the Institute for Astronomy, the University of Hawaii, the Pan-STARRS Project Office, the Max-Planck Society and its participating institutes, the Max Planck Institute for Astronomy, Heidelberg and the Max Planck Institute for Extraterrestrial Physics, Garching, The Johns Hopkins University, Durham University, the University of Edinburgh, the Queen's University Belfast, the Harvard-Smithsonian Center for Astrophysics, the Las Cumbres Observatory Global Telescope Network Incorporated, the National Central University of Taiwan, the Space Telescope Science Institute, the National Aeronautics and Space Administration under Grant No. NNX08AR22G issued through the Planetary Science Division of the NASA Science Mission Directorate, the National Science Foundation Grant No. AST-1238877, the University of Maryland, Eotvos Lorand University (ELTE), the Los Alamos National Laboratory, and the Gordon and Betty Moore Foundation. 
All the {\it PS1} data used in this paper can be found in MAST\rr1{~\citep{PS1-DB}}.

The national facility capability for SkyMapper has been funded through ARC LIEF grant LE130100104 from the Australian Research Council, awarded to the University of Sydney, the Australian National University, Swinburne University of Technology, the University of Queensland, the University of Western Australia, the University of Melbourne, Curtin University of Technology, Monash University and the Australian Astronomical Observatory. SkyMapper is owned and operated by The Australian National University's Research School of Astronomy and Astrophysics. The survey data were processed and provided by the SkyMapper Team at ANU. The SkyMapper node of the All-Sky Virtual Observatory (ASVO) is hosted at the National Computational Infrastructure (NCI). Development and support of the SkyMapper node of the ASVO has been funded in part by Astronomy Australia Limited (AAL) and the Australian Government through the Commonwealth's Education Investment Fund (EIF) and National Collaborative Research Infrastructure Strategy (NCRIS), particularly the National eResearch Collaboration Tools and Resources (NeCTAR) and the Australian National Data Service Projects (ANDS).

Funding for the Sloan Digital Sky Survey V has been provided by the Alfred P. Sloan Foundation, the Heising-Simons Foundation, the National Science Foundation, and the Participating Institutions. SDSS acknowledges support and resources from the Center for High-Performance Computing at the University of Utah. SDSS telescopes are located at Apache Point Observatory, funded by the Astrophysical Research Consortium and operated by New Mexico State University, and at Las Campanas Observatory, operated by the Carnegie Institution for Science. The SDSS web site is \url{www.sdss.org}.
SDSS is managed by the Astrophysical Research Consortium for the Participating Institutions of the SDSS Collaboration, including Caltech, The Carnegie Institution for Science, Chilean National Time Allocation Committee (CNTAC) ratified researchers, The Flatiron Institute, the Gotham Participation Group, Harvard University, Heidelberg University, The Johns Hopkins University, L’Ecole polytechnique f\'{e}d\'{e}rale de Lausanne (EPFL), Leibniz-Institut f\"{u}r Astrophysik Potsdam (AIP), Max-Planck-Institut f\"{u}r Astronomie (MPIA Heidelberg), Max-Planck-Institut f\"{u}r Extraterrestrische Physik (MPE), Nanjing University, National Astronomical Observatories of China (NAOC), New Mexico State University, The Ohio State University, Pennsylvania State University, Smithsonian Astrophysical Observatory, Space Telescope Science Institute (STScI), the Stellar Astrophysics Participation Group, Universidad Nacional Aut\'{o}noma de M\'{e}xico, University of Arizona, University of Colorado Boulder, University of Illinois at Urbana-Champaign, University of Toronto, University of Utah, University of Virginia, Yale University, and Yunnan University.

This research has made use of the SIMBAD database, operated at CDS, Strasbourg, France.

This project used data obtained with the Dark Energy Camera (DECam), which was constructed by the Dark Energy Survey (DES) collaboration. This project used public archival data from the Dark Energy Survey (DES). Funding for the DES Projects has been provided by the U.S. Department of Energy, the U.S. National Science Foundation, the Ministry of Science and Education of Spain, the Science and Technology Facilities Council of the United Kingdom, the Higher Education Funding Council for England, the National Center for Supercomputing Applications at the University of Illinois at Urbana-Champaign, the Kavli Institute of Cosmological Physics at the University of Chicago, the Center for Cosmology and Astro-Particle Physics at the Ohio State University, the Mitchell Institute for Fundamental Physics and Astronomy at Texas A\&M University, Financiadora de Estudos e Projetos, Funda{\c c}{\~a}o Carlos Chagas Filho de Amparo {\`a} Pesquisa do Estado do Rio de Janeiro, Conselho Nacional de Desenvolvimento Cient{\'i}fico e Tecnol{\'o}gico and the Minist{\'e}rio da Ci{\^e}ncia, Tecnologia e Inova{\c c}{\~a}o, the Deutsche Forschungsgemeinschaft, and the Collaborating Institutions in the Dark Energy Survey.
The Collaborating Institutions are Argonne National Laboratory, the University of California at Santa Cruz, the University of Cambridge, Centro de Investigaciones Energ{\'e}ticas, Medioambientales y Tecnol{\'o}gicas-Madrid, the University of Chicago, University College London, the DES-Brazil Consortium, the University of Edinburgh, the Eidgen{\"o}ssische Technische Hochschule (ETH) Z{\"u}rich,  Fermi National Accelerator Laboratory, the University of Illinois at Urbana-Champaign, the Institut de Ci{\`e}ncies de l'Espai (IEEC/CSIC), the Institut de F{\'i}sica d'Altes Energies, Lawrence Berkeley National Laboratory, the Ludwig-Maximilians Universit{\"a}t M{\"u}nchen and the associated Excellence Cluster Universe, the University of Michigan, the National Optical Astronomy Observatory, the University of Nottingham, The Ohio State University, the OzDES Membership Consortium, the University of Pennsylvania, the University of Portsmouth, SLAC National Accelerator Laboratory, Stanford University, the University of Sussex, and Texas A\&M University.
Based on observations at Cerro Tololo Inter-American Observatory, a programme of NOIRLab (NOIRLab Prop. 2012B-0001; PI J. Frieman).

The Legacy Surveys consist of three individual and complementary projects: the Dark Energy Camera Legacy Survey (DECaLS; Proposal ID \#2014B-0404; PIs: David Schlegel and Arjun Dey), the Beijing-Arizona Sky Survey (BASS; NOAO Prop. ID \#2015A-0801; PIs: Zhou Xu and Xiaohui Fan), and the Mayall z-band Legacy Survey (MzLS; Prop. ID \#2016A-0453; PI: Arjun Dey). DECaLS, BASS and MzLS together include data obtained, respectively, at the NSF V\'ictor M. Blanco 4-meter Telescope, Cerro Tololo Inter-American Observatory, NSF NOIRLab; the Bok telescope, Steward Observatory, University of Arizona; and the NSF Nicholas U. Mayall 4-meter Telescope, Kitt Peak National Observatory, NOIRLab. The Legacy Surveys project is honored to be permitted to conduct astronomical research on I'oligam Du'ag (Kitt Peak), a mountain with particular significance to the Tohono O'odham Nation.
BASS is a key project of the Telescope Access Programme (TAP), which has been funded by the National Astronomical Observatories of China, the Chinese Academy of Sciences (the Strategic Priority Research Programme `The Emergence of Cosmological Structures' Grant \# XDB09000000), and the Special Fund for Astronomy from the Ministry of Finance. The BASS is also supported by the External Cooperation Programme of Chinese Academy of Sciences (Grant \# 114A11KYSB20160057), and Chinese National Natural Science Foundation (Grant \# 11433005).
The Legacy Survey team makes use of data products from the Near-Earth Object Wide-field Infrared Survey Explorer (\textit{NEOWISE}), which is a project of the Jet Propulsion Laboratory/California Institute of Technology. \textit{NEOWISE} is funded by the National Aeronautics and Space Administration.
The Legacy Surveys imaging of the DESI footprint is supported by the Director, Office of Science, Office of High Energy Physics of the U.S. Department of Energy under Contract No. DE-AC02-05CH1123, by the National Energy Research Scientific Computing Center, a DOE Office of Science User Facility under the same contract; and by the U.S. National Science Foundation, Division of Astronomical Sciences under Contract No. AST-0950945 to NOAO.

This research uses services or data provided by the Astro Data Lab, which is part of the Community Science and Data Center (CSDC) Program of NSF NOIRLab. 

This research uses services or data provided by the Astro Data Archive at NSF NOIRLab. These data are associated with observing programs 2012B-0001 (PI J. Frieman), 2013A-0719 (PI A. Saha), 2015A-0608 (PI F. F\"orster), 2017A-0210 (PI A. Walker), 2017A-0909 (PI J. Cooke), 2017B-0110 (PI E. Berger), 2021A-0113 (PI M. Graham).

\facilities{
CTIO:4m(Blanco/DECam), Astro Data Lab, Astro Data Archive
}

\software{
Astropy~\citep{AstropyCollaboration2013,AstropyCollaboration2018,AstropyCollaboration2022},
LSST Science Pipelines~\citep{Juric2017,Bosch2019},
Matplotlib~\citep{Hunter2007}, 
Numpy~\citep{Harris2020}, 
Scikit-learn~\citep{scikit-learn},
Scipy~\citep{Virtanen2020}
}

\appendix

\section{Target list}\label{sec:target_list}
In Table~\ref{tab:target_list} we present a list of targets studied in this work. 

\begin{table*}[htb]
\centering
\begin{tabular}{lrrr}
\hline
\hline
Name & R.A. (deg) & Decl. (deg) & Type \\
\hline
GW170817 & 197.450 & -23.381 & KN \\
SNHiTS15H & 139.044 & -0.393 & SN \\
SNHiTS15bd & 142.364 & -3.489 & SN \\
DES15C2eaz & 55.089 & -28.667 & SN \\
DES16X3jj & 35.696 & -4.359 & SN \\
DC21cove & 150.205 & 3.743 & SN \\
DC21bwbfe/SN2021bnv & 149.505 & 3.810 & SN \\
DC21bkrj & 150.044 & 3.477 & SN \\
DWF17l & 100.585 & -52.174 & Flare \\
CraterII-V11 & 177.199 & -17.970 & Var* \\
CraterII-V19 & 177.017 & -18.079 & Var* \\
OGLE-BLG-ECL-277767 & 270.906 & -29.969 & Var* \\
OGLE-BLG-RRLYR-12257 & 270.980 & -29.482 & Var* \\
\hline
\end{tabular}
\caption{List of real objects tested in this work. ``Var*'' means ``variable star''. 
}
\label{tab:target_list}
\end{table*}

\section{SN in the COSMOS field of DECaLS}
\label{sec:cosmos_gen3}

\begin{figure*}
    \plottwo{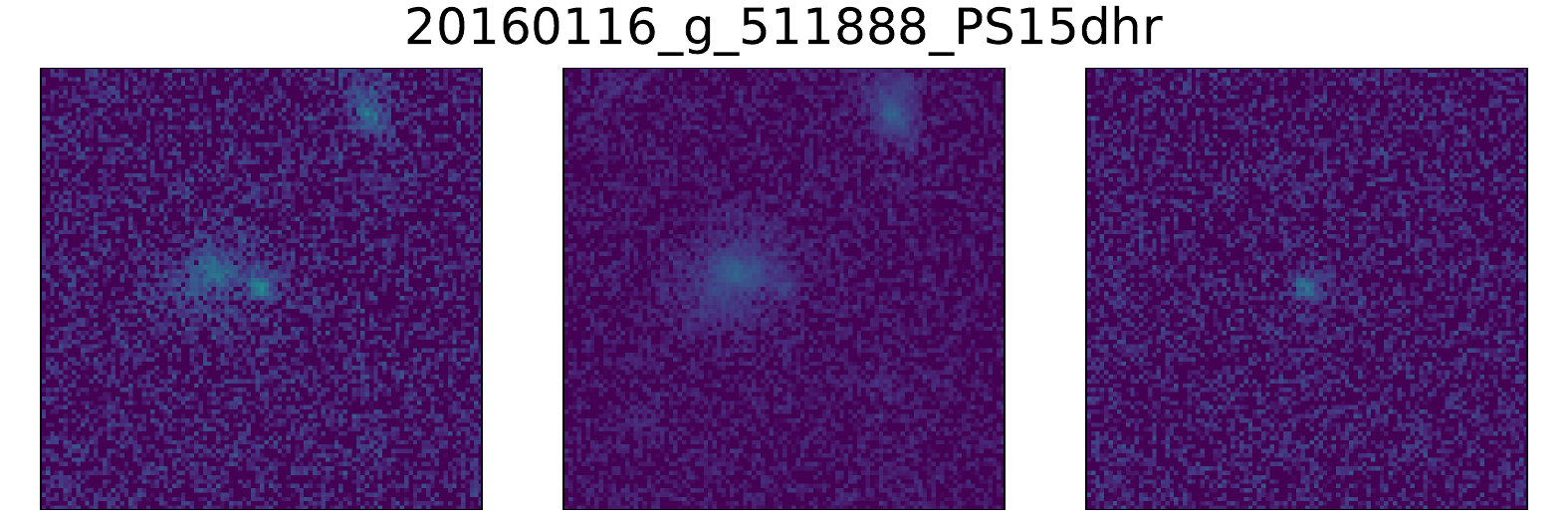}{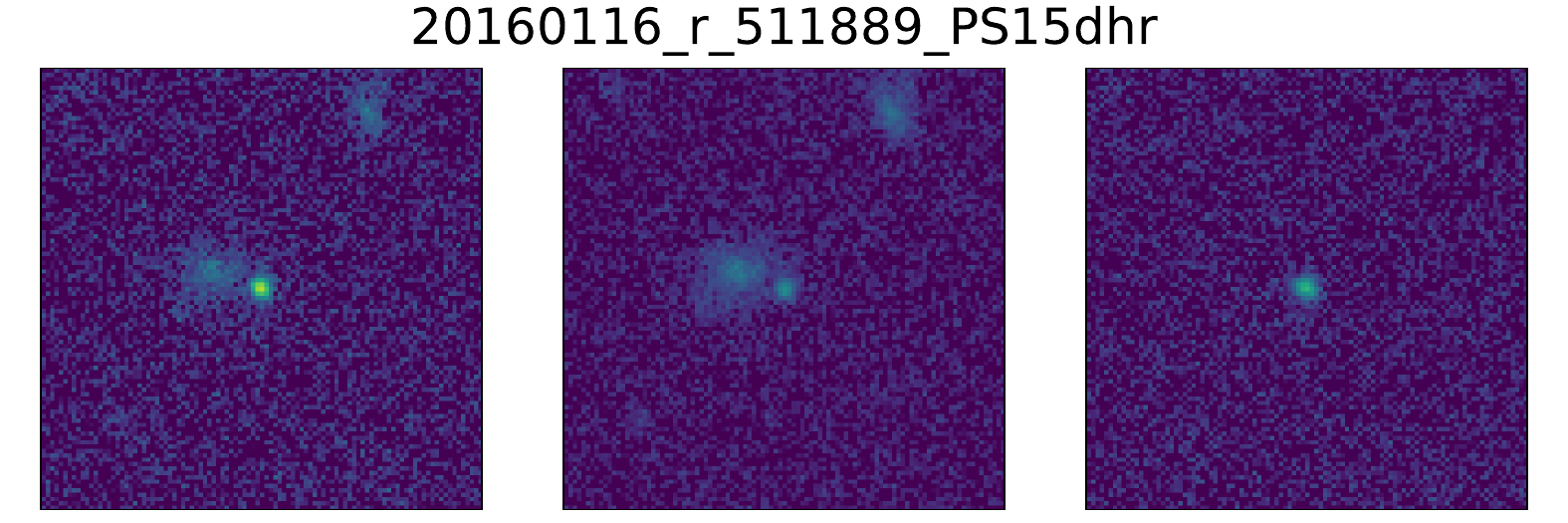}

    \plottwo{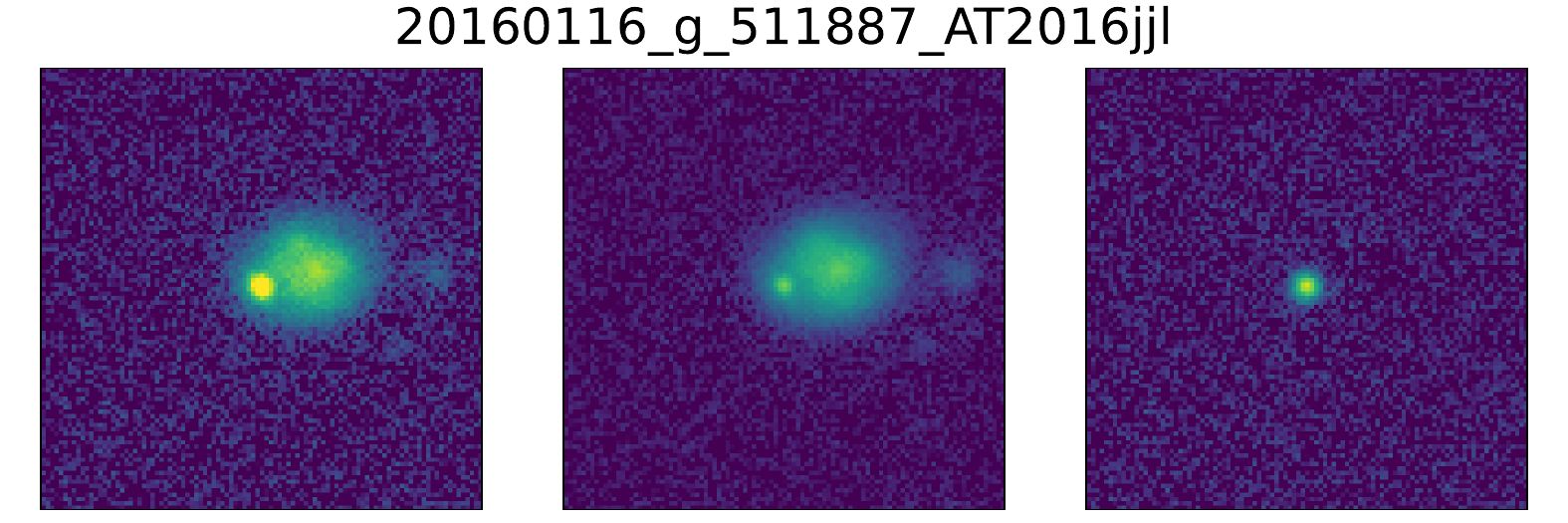}{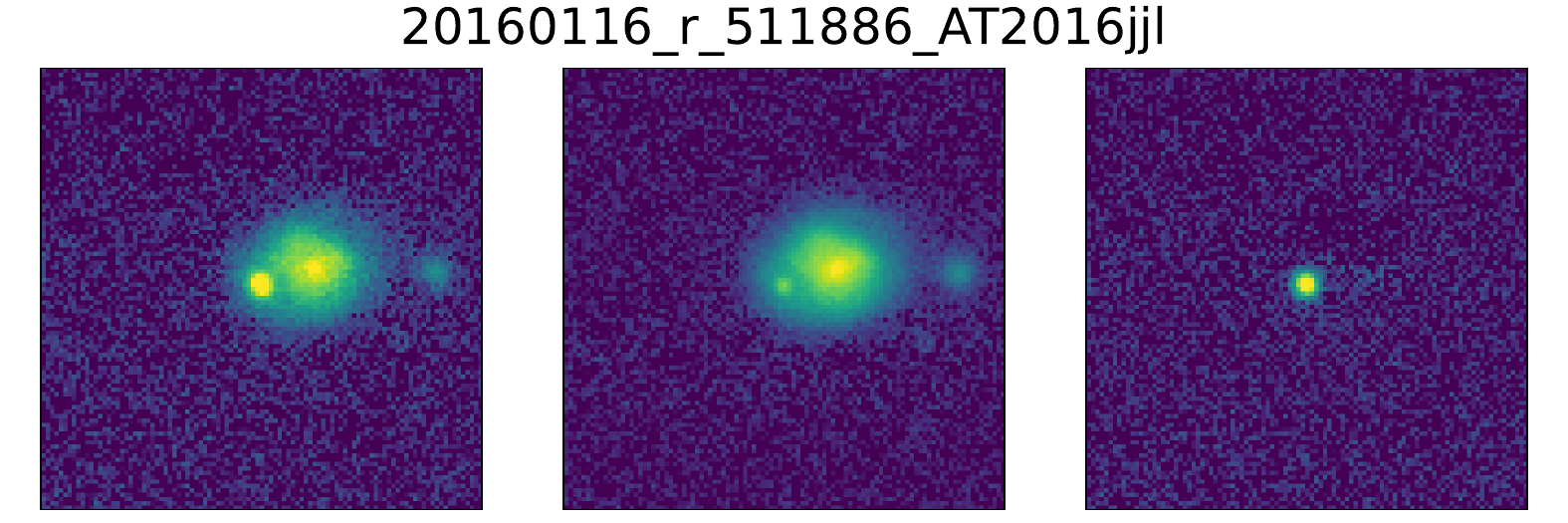}
    \plottwo{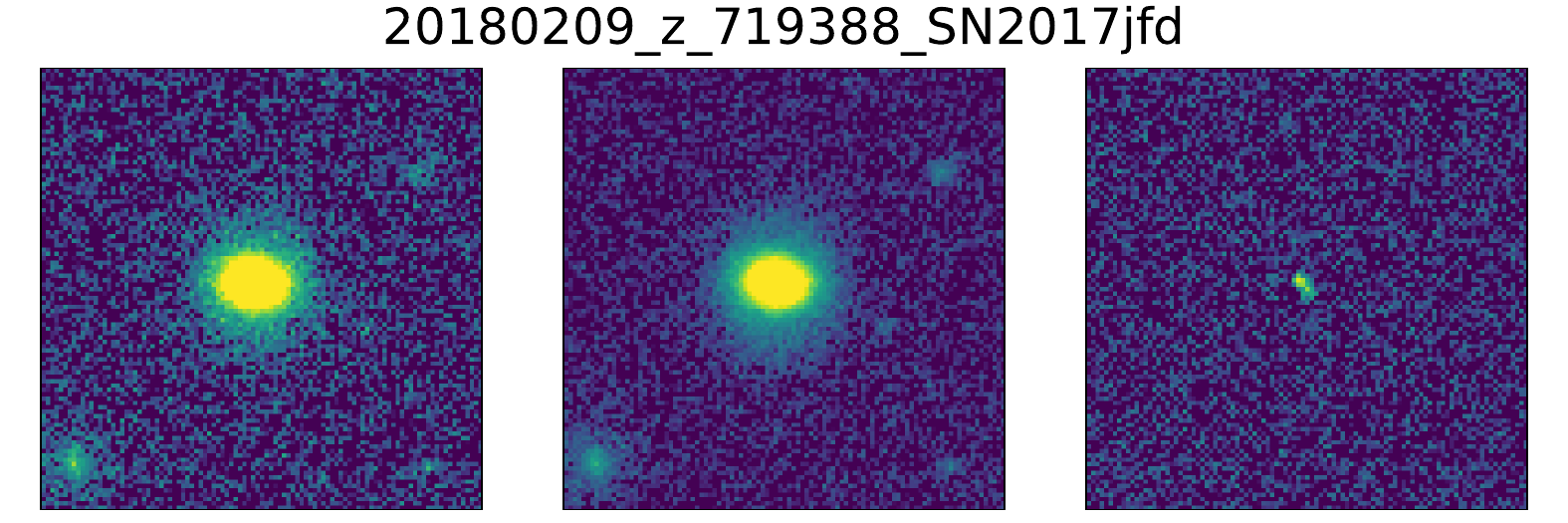}{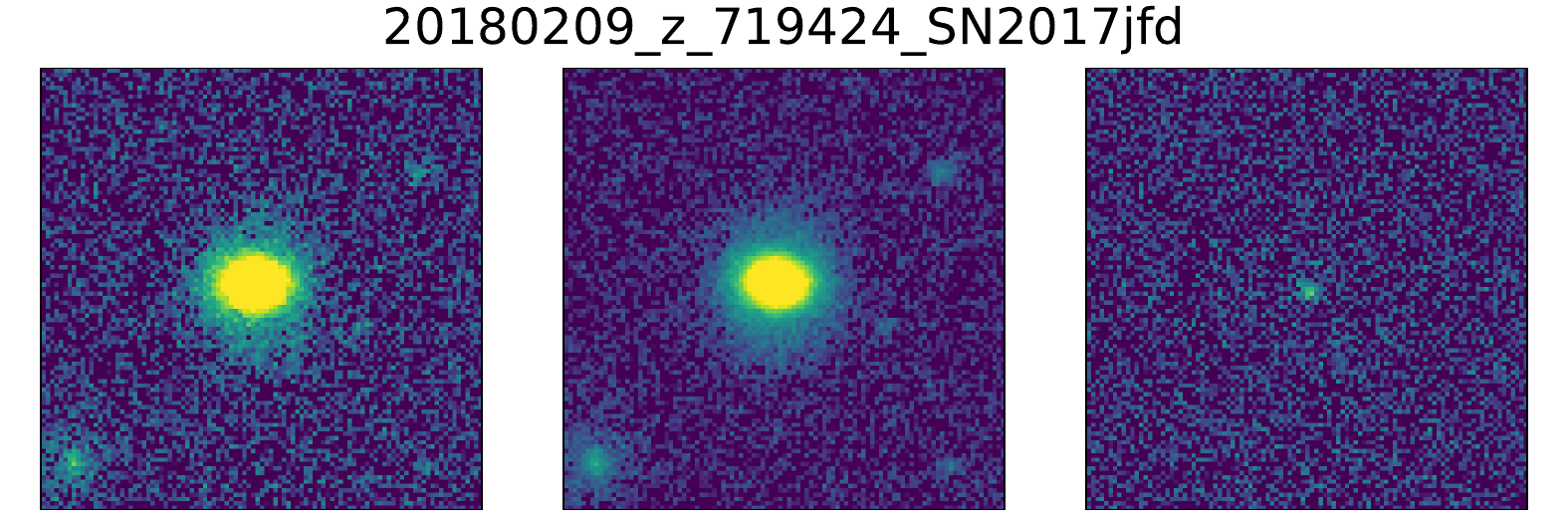}
    
    \caption{COSMOS SN in the \texttt{calexp}, template, and difference images. The exposure date, band, exposure number, and name  of each SN are given in the title. Each row shows the cutouts of a SN. Cutouts are $100\times100$ pix with North towards top and East towards left. 
    }
    \label{fig:cosmos_sn}
\end{figure*}

We reprocessed DECaLS DR9 exposures using the recent Gen3 version 23.0.1 of the LSST Science Pipelines software. 
Here we present all covered COSMOS SN (PS15dhr, AT2016jjl, SN2017jfd)\footnote{Searched in the Open SN catalog~\citep{Guillochon2017}. } in the reprocessed data set. 
Figure~\ref{fig:cosmos_sn} shows the calibrated exposure \texttt{calexp}, the template, and the difference image of each SN. Other exposures are far from their brightness peaks. 
The templates are the coadded images of good-seeing exposures in DECaLS observations, and some of them captured the SN.

\section{Ineffective  features}\label{sec:ineffective}
Here we comment on features that were ultimately not effective for our candidate selection and thus were not included in our analysis in the main text.

\begin{figure*}[htb]
    \plotone{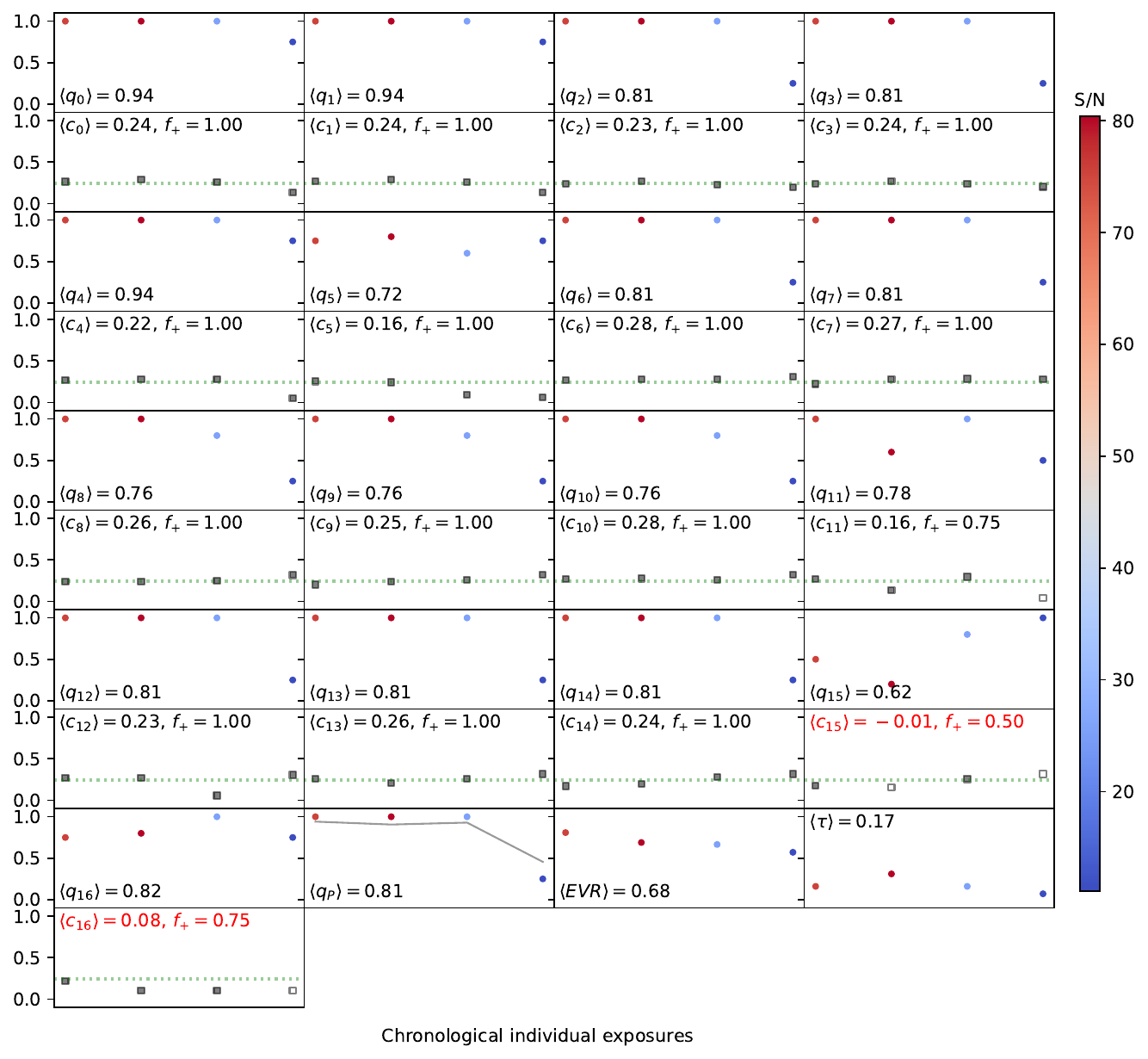}

    \caption{Evolution of \textit{all} features and PCA performance for the KN of GW170817. The labels/tags have the same meanings as Figure~\ref{fig:quantile_evolution_kn}. The mean coefficients of features \textnumero15 and 16 are close to zero. }
    \label{fig:qe_gw_app}
\end{figure*}

\begin{figure*}[htb]
\plotone{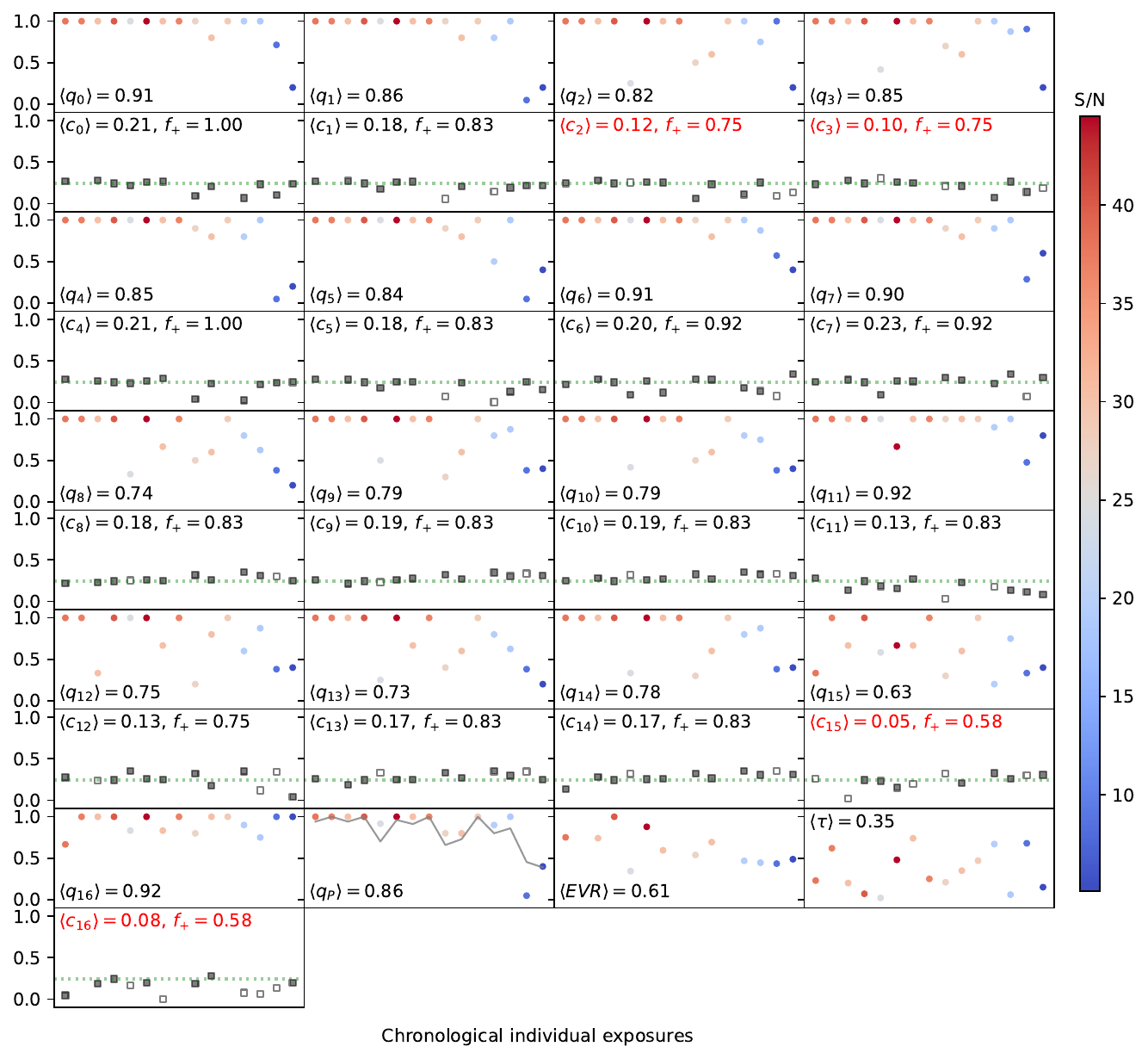}

    \caption{Evolution of all features and PCA performance for DES15C2eaz. Again, the mean coefficients of features \textnumero15 and 16 are close to zero,  indicating that they are not effective (nearly half of the coefficients are negative). 
    }
    \label{fig:qe_c2_app}
\end{figure*}

\begin{figure*}[htb]
    \plotone{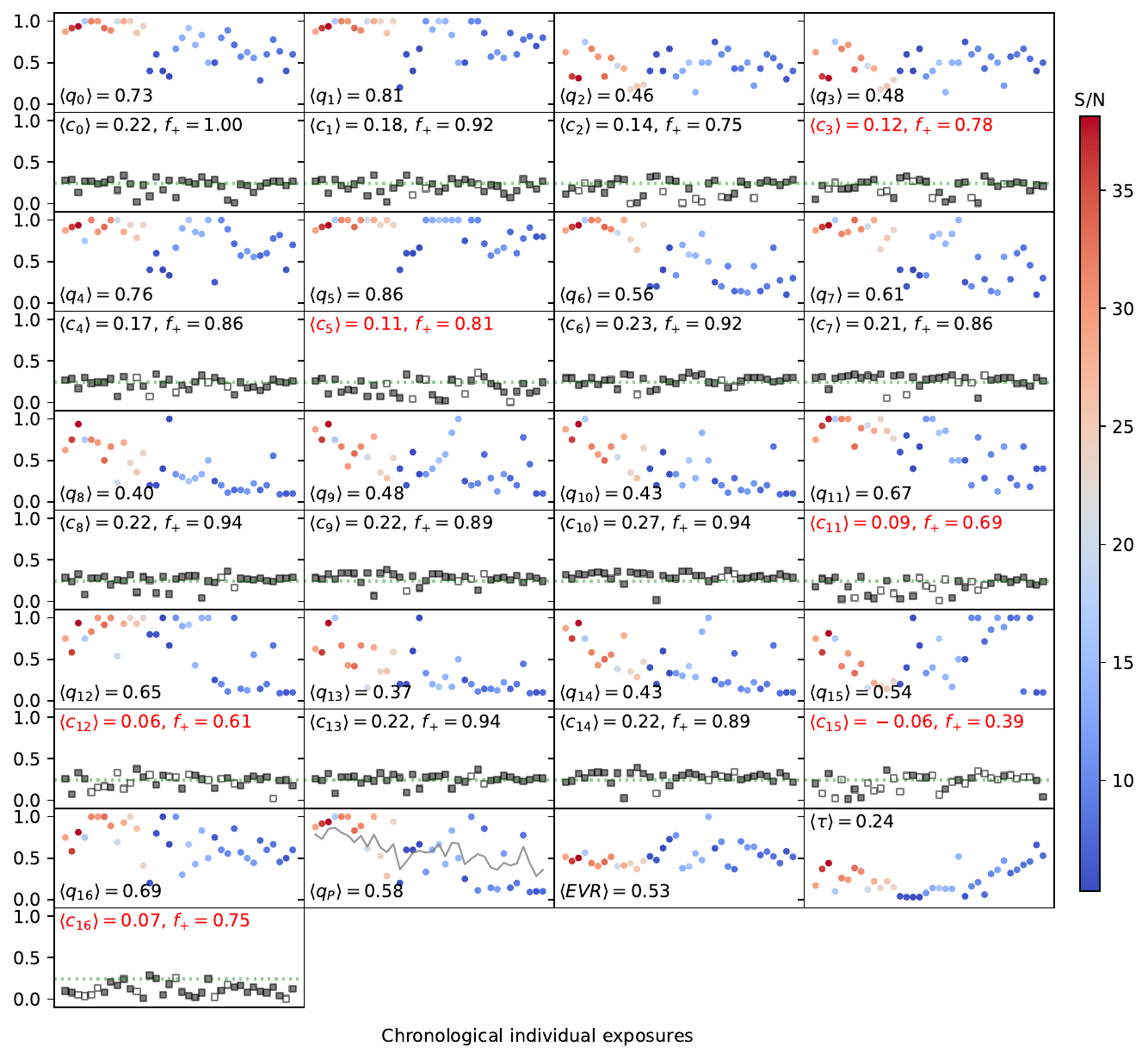}
    \caption{Evolution of all features and PCA performance for DC21bwbfe/SN2021bn (positive flux cases).  The mean coefficients of features \textnumero15 and 16 are small. 
    }
    \label{fig:qe_dc_app}
\end{figure*}

As mentioned in Section~\ref{sec:features}, 
we test including extra features (Table~\ref{tab:features_extra}) in PCA and then remake the figures of quantile evolution  (Figure~\ref{fig:qe_gw_app},~\ref{fig:qe_c2_app},~\ref{fig:qe_dc_app}). 
The results show that those newly added features are not useful for selecting candidates. 
We note that the \verb|base_SdssShape_instFlux_yy_Cov| term does not separate our target well from other sources, and we thus skip it in the main text (but the covariances with \texttt{xx} and \texttt{xy} work better for candidation selection). 
The reason of this asymmetry among \texttt{xx}, \texttt{xy}, and \texttt{yy} could come from the instrument instead of the sky -- along the Y-direction (CCD column) a source can be more easily affected by the instrumental defects nearby (e.g., bad columns and bleeding trails). This was pointed out by John Parejko from the LSST DM team, and we are grateful for his helpful comments.\footnote{\url{https://community.lsst.org/t/question-about-base-sdssshape-instflux-yy-cov-in-the-pipeline-source-catalog}} 

It is possible that more advanced ML algorithms such as Neural Network and Random Forest can make use of these extra features -- instead of requiring the real source to have much larger values than the others as in PCA -- but potentially that can cause redundancy in the feature space. We will further explore this in the future.

\begin{table*}[htb]
    \centering
    \begin{tabular}{l l l}
    \hline
    \hline
    Type & Feature index \& Name & Definition  \\
    \hline
    Others & 15. \verb|base_SdssShape_instFlux_yy_Cov| & Uncertainty  covariance between \verb|instFlux| and \verb|yy|.\\
    & 16. $-(f-1)^2$ & Difference between PSF flux and aperture flux. \\ 
    \hline
    \end{tabular}
    \caption{
    Extra features for R/B. 
    Here the ratio  $f=\texttt{base\_PsfFlux\_instFlux}/\texttt{base\_CircularApertureFlux\_12\_0\_instFlux}$. 
    }
    \label{tab:features_extra}
\end{table*}

\newpage


\bibliography{main}{}
\bibliographystyle{aasjournal}

\end{document}